\documentclass[11pt,a4paper]{article}
\pdfoutput=1
\usepackage{jcappub}
\usepackage{aas_macros_JCAP}
\usepackage{blindtext}
\usepackage{bm}
\usepackage{verbatim}

\newcommand{\hMpc}{\textrm{Mpc}/h}
\newcommand{\hkpc}{\textrm{kpc}/h}
\newcommand{\kms}{\textrm{km/s}}

\newcommand{\Msun}{M_{\odot}/h}
\newcommand{\Mpch}{{\textrm{Mpc}}/h}
\newcommand{\void}{\mathrm{v}}
\newcommand{\halo}{\mathrm{h}}

\newcommand{\tracer}{\mathrm{t}}
\newcommand{\matter}{\mathrm{m}}

\newcommand{\coreDens}{\hat{n}_\mathrm{min}}

\newcommand{\hydro}{\texttt{hydro}}
\newcommand{\DMo}{\texttt{DMo}}

\newcommand{\mr}{\texttt{mr}}

\newcommand{\hr}{\texttt{hr}}
\newcommand{\MR}{\texttt{midres}}
\newcommand{\HR}{\texttt{highres}}
\newcommand{\UHR}{\texttt{ultra-hr}}
\newcommand{\uhr}{\texttt{uhr}}
\newcommand{\Mag}{\texttt{Magneticum}}

\title{Why Cosmic Voids Matter: Mitigation of Baryonic Physics}

\author[a,b]{Nico Schuster,}
\author[a,b]{Nico Hamaus,}
\author[a,c]{Klaus Dolag,}
\author[a,b,d]{Jochen Weller}

\affiliation[a]{Universit\"ats-Sternwarte M\"unchen, Fakult\"at f\"ur Physik, Ludwig-Maximilians-Universit\"at, Scheinerstr. 1, 81679 M\"unchen, Germany}
\affiliation[b]{Excellence Cluster ORIGINS, Bolzmannstr. 2, 85748 Garching, Germany}
\affiliation[c]{Max-Planck-Institut f\"ur Astrophysik, Karl-Schwarzschild-Str. 1, 85748 Garching, Germany }
\affiliation[d]{Max-Planck-Institut f\"ur extraterrestrische Physik, Giessenbachstr. 1, 85748 Garching, Germany}

\emailAdd{Nico.Schuster@physik.lmu.de}
\emailAdd{n.hamaus@physik.lmu.de}
\emailAdd{kdolag@mpa-garching.mpg.de}
\emailAdd{jochen.weller@lmu.de}

\abstract{ We utilize the Magneticum suite of state-of-the-art hydrodynamical, as well as dark-matter-only simulations to investigate the effects of baryonic physics on cosmic voids in the highest-resolution study of its kind. This includes the size, shape and inner density distributions of voids, as well as their radial density and velocity profiles traced by halos, baryonic and cold dark matter particles. Our results reveal observationally insignificant effects that slightly increase with the inner densities of voids and are exclusively relevant on scales of only a few Mpc. Most notably, we identify deviations in the distributions of baryons and cold dark matter around halo-defined voids, relevant for weak lensing studies. In contrast, we find that voids identified in cold dark matter, as well as in halos of fixed tracer density exhibit nearly indistinguishable distributions and profiles between hydrodynamical and dark-matter-only simulations, consolidating the universality and robustness of the latter for comparisons of void statistics with observations in upcoming surveys. This corroborates that voids are the components of the cosmic web that are least affected by baryonic physics, further enhancing their use as cosmological probes.}

\date{\today}

\keywords{cosmological simulations, hydrodynamical simulations, cosmic web, galaxy clustering}

\begin{document}
\maketitle

\newpage

\section{Introduction\label{sec:intro}}

The formation of cosmic voids as the underdense regions of large-scale structure is governed by the gravitational interaction of initially tiny fluctuations in density at the time of the cosmic microwave background (CMB), back when the Universe was first observable in its infancy~\cite{Bennett2003,Planck2020}. As these fluctuations were amplified by gravity and clustered into gravitationally bound structures, such as sheets, filaments and dense nodes~\cite{Zeldovich1970}, most of the volume in our Universe became devoid of almost all matter and formed expanding voids~\cite{Gregory1978,Joeveer1978,Kirshner1981,Zeldovich1982,Bertschinger1985,vdWeygaert1993}. This combination of structures, known as the \emph{cosmic web}, is assumed to be largely composed of \emph{cold dark matter} (CDM), a form of matter that interacts solely through gravity and whose exact particle nature remains one of the big questions in cosmology. In contrast, visible structures are made up of baryonic matter, which, apart from gravity, interacts through the other fundamental forces of nature.

Moreover, a late-time accelerated expansion of space has been observed and is modelled with \emph{dark energy}, which is so far consistent with the model of a cosmological constant $\Lambda$~\cite{Riess1998,Perlmutter1999}. The evolution of CDM in combination with dark energy is determined by general relativity (GR) and together they form the basis of the $\Lambda$CDM model. Even though its constituents are mostly unknown, simulations that follow the formation and evolution of structures allow a detailed study of a $\Lambda$CDM universe~\cite[e.g.,][]{Springel2001a,Dolag2016}.

Since their initial discovery in the 1970s~\cite{Gregory1978,Joeveer1978}, the study of cosmic voids evolved significantly, and due to recent surveys, such as BOSS~\cite{Dawson2013}, DES~\cite{DarkEnergySurvey2005}, eBOSS~\cite{Dawson2016} and SDSS~\cite{Eisenstein2011}, their use as efficient cosmological probes was established~\cite{Biswas2010,Pan2012,Sutter2012b,Hamaus2014a,Pisani2015a,Hamaus2015,Pisani2019,Correa2021a,Moresco2022}. Many studies rely on the radial density profile of voids~\cite{Hamaus2014b,Ricciardelli2014} and their observable shape in redshift space~\cite[e.g.,][]{Paz2013,Hamaus2014c,Cai2016,Hawken2017,Hamaus2017,Massara2018,Achitouv2019,Correa2021b,Massara2022}. The latter is affected by redshift-space distortions (RSD), originating from the contribution of the peculiar velocity of tracers to the observed redshift. These RSD have to be modelled for a correct shape reconstruction, which is also relevant for the Alcock-Paczynski effect, as it tests cosmology through the use of voids as standard spheres~\cite{Alcock1979,Ryden1995,Lavaux2012,Hamaus2016}. While tracer velocities can hardly be measured directly and voids themselves are nonlinear objects with substantial density variations with respect to the mean value at their boundary, the interplay between density and velocity is remarkably well described by the linearized continuity equation down to scales of only few Mpc~\cite{Hamaus2014b,Schuster2023}. The use of void density profiles ranges from studies of the $\Lambda$CDM model~\cite[e.g.,][]{Chan2014,Leclercq2015,Cautun2016,SanchezC2017,Pollina2017,Chantavat2017,Fang2019,Stopyra2021,Shim2021,Tavasoli2021} to modifications to GR~\cite[e.g.,][]{Zivick2015,Cai2015,Barreira2015,Falck2018,Baker2018,Paillas2019,Davies2019,Perico2019,Wilson2021,Tamosiunas2022,Fiorini2022}, inflation~\cite{Chan2020}, as well as in the context of exploring particle dark matter models~\cite[e.g.,][]{Yang2015,Reed2015,Baldi2018,Lester2021,Arcari2022} and massive neutrinos~\cite[e.g.,][]{Massara2015,Banerjee2016,Kreisch2019,Schuster2019,Zhang2020,Contarini2021,Bayer2021,Kreisch2022,Thiele2023}.

However, the impact of baryonic physics poses a major obstacle in cosmological experiments, as processes such as the formation and death of stars and feedback from active galactic nuclei (AGN) affect the large-scale matter distribution and its power spectrum~\cite[e.g.,][]{Schneider2015,Schneider2019,Chisari2019}. In order to describe a $\Lambda$CDM universe in great detail down to small scales and to capture galaxy properties accurately, dark-matter-only simulations that solely implement gravitational interactions are not sufficient, and instead more complex hydrodynamical simulations are required. While effects from baryonic physics are immensely important on small scales, as well as for characteristics of dense structures such as galaxies and clusters, their impact on cosmic voids is largely unknown and has only been explored in few recent studies~\cite{Paillas2017,Habouzit2020,Panchal2020,Rodriguez2022,WangHe2023}. Since most of these studies investigated (spherical) voids at either low resolutions or in small cosmological volumes, we aim to study the impact of baryonic physics on watershed voids across a variety of resolutions and volumes. This includes their distribution in size, inner density and shape, as well as their radial density and velocity profiles. This is done with the help of hydrodynamical, as well as dark-matter-only counterpart simulations at identical initial conditions. We compare voids identified in various matter tracers, as well as biased halos between these simulations. Analysing void statistics from both setups can help determining the validity of void statistics derived from dark-matter-only simulations when these statistics are compared with observations. Furthermore, we examine differences in the distribution and movement of CDM, as well as baryons around halo-defined voids, which is relevant for weak lensing studies of voids.

The structure of this work is as follows: section~\ref{sec:Magneticum} provides details on the employed simulation suite and section~\ref{sec:methods} introduces the void finder and estimators for void profiles. Next, we describe the void catalogs of all tracers in section~\ref{sec:catalogs} and investigate baryonic effects in void profiles in section~\ref{sec:baryonic}, before summarizing our final conclusions in section~\ref{sec:conclusion}.

\section{The Magneticum simulations}
\label{sec:Magneticum}

\begin{table}[b]
\centering
\begin{tabular}{|c | c c c c c c|} 
 \hline
 Name & {\it Box} & $L_\mathrm{Box}$ & $N_\mathrm{particles}$ & $m_\mathrm{CDM}$ & $m_\mathrm{baryon}$ & $z$ \\ [0.5ex] 
 \hline
 \rule{0pt}{3ex}
 \MR{} (\mr{}) & {\it 0} & 2688 & $2 \times 4536^3$ & $1.3 \times 10^{10}$ & $2.6 \times 10^9$ & 0.00, 0.29  \\
 \HR{} (\hr{}) & {\it 2b} & 640 & $ 2 \times 2880^3$ & $6.9 \times 10^8$ & $1.4 \times 10^8$ & 0.25 \\
 \UHR{} (\uhr{}) & {\it 4} & 48 & $ 2 \times 576^3$ & $3.6 \times 10^7$ & $7.3 \times 10^6$ & 0.25 \\ [1ex]
 \hline
\end{tabular}
\caption{Properties of the \Mag{} simulations used in this work. The box length $L_\mathrm{Box}$ is in units of $\hMpc$ and particle masses ($m_\mathrm{CDM}$ \& $m_\mathrm{baryon}$) are in units of $\Msun$.}
\label{table_1}
\end{table}

This work utilizes simulations of the \Mag\footnote{\url{http://www.magneticum.org}} suite, a variety of state-of-the-art cosmological hydrodynamical, as well as dark-matter-only simulations, which cover a vast range of cosmological volumes at different mass resolutions. The hydrodynamical runs with included baryonic physics will henceforth be referred to as \hydro{} and the dark-matter-only ones as \DMo{}. The default will always be the \hydro{} runs and hence not always explicitly mentioned. We briefly describe the \Mag{} suite and for more details we refer to previous work using these simulations~\cite[e.g.,][]{Hirschmann2014,Dolag2015,Steinborn2015,Teklu2015,Bocquet2016,Dolag2016,Remus2017,Castro2018,Castro2021,Angelinelli2022}. All \Mag{} runs used in this analysis adopt a flat $\Lambda$CDM cosmology, with cosmological parameter values chosen according to the best fit values of WMAP7 (see~\cite{Komatsu2011}), hence $h = 0.704$, $\Omega_{\Lambda} = 0.728$, $\Omega_\mathrm{m} = 0.272$, $\Omega_\mathrm{b} = 0.0456$, $\sigma_8 = 0.809$ and the spectral index of the primordial power spectrum $n_s = 0.963$. These simulations have been performed using an advanced version of the tree particle mesh-smoothed particle hydrodynamics (TreePM-SPH) code \textsc{gadget3}~\cite{Springel2005}, which includes an improved SPH solver~\cite{Beck2016}. The code additionally implements a variety of processes that describe baryon physics and evolution, which includes the distribution of multiple metal species~\cite{Dolag2017}, as well as AGN feedback and black hole growth, based on previous work in~\cite{Springel2005a,DiMatteo2005,Fabjan2010}.

Our analysis on baryonic effects in and around voids focuses mostly on {\it Box0} and {\it Box2b}, as these are the largest \Mag{} boxes of medium ({\it Box0}) and high ({\it Box2b}) mass resolution, with volumes of large enough size to contain a sufficient number of cosmic voids, with typical radii on the order of $5-80\;\hMpc$. From now on we will refer to these boxes by their resolution, i.e. \MR{} (\mr{}) for {\it Box0}, as well as \HR{} (\hr{}) for {\it Box2b}. For tests of baryonic effects on even smaller scales, we make use of the ultra high resolution {\it Box4}, referred to as \UHR{} (\uhr{}), although this increased resolution comes with the caveat of fewer voids due to the smaller box size. Previous studies on the tracer bias around voids and tests on the validity of linear mass conservation around voids already made use of these \Mag{} simulations~\cite{Pollina2017,Schuster2023}. For a summary of details on the box sizes, the number of simulated particles and the mass resolution of baryons and CDM particles of the \hydro{} runs, we refer the reader to table~\ref{table_1}. The \DMo{} runs use the same initial conditions and box sizes as the \hydro{} simulations, but only contain CDM particles, with their particle masses simply given by $m_\mathrm{part} = m_\mathrm{CDM} + m_\mathrm{baryon}$ from table~\ref{table_1}. For both \HR{} and \UHR{} we analyze tracer catalogs at redshift $z = 0.25$, while \MR{} used $z = 0$ and $z = 0.29$.

For the analysis of voids in biased tracers, we identify subhalos and their properties via the \textsc{SubFind} algorithm~\cite{Springel2001b}, which was modified to additionally take baryonic components into account~\cite{Dolag2009}. From this point forward, we will refer to subhalos as `halos' for simplicity. In all the aforementioned simulations we compare voids found in halos from the hydrodynamical runs with the ones from \DMo{}, as well as with voids found in subsamplings of the underlying cold dark matter (CDM) and the baryons that formed the halos. As the simulations use different mass resolutions, we are able to select halos at different minimal masses for the void finding process and further analysis. In \MR{} a minimum halo mass of $10^{12} \Msun$ is used, while \HR{} uses $10^{11} \Msun$, which further reduces to $1.6 \times 10^{9} \Msun$ in \UHR{}. These values of mass cuts were chosen because at these particular values of a given resolution, the halos in both \hydro{} and \DMo{} still follow the halo mass function (see~\cite{Schuster2023}).

\section{Methodology \label{sec:methods}}

\subsection{VIDE void finding \label{subsec:void_finding}}

To identify voids in both the distributions of halos, as well as the underlying matter, we employ the commonly used Void IDentification and Examination toolkit {\textsc{vide}}\footnote{\url{https://bitbucket.org/cosmicvoids/vide_public/}} \cite{Sutter2015}. \textsc{vide} is based on an enhanced version of the ZOnes Bordering On Voidness algorithm \textsc{zobov}~\cite{Neyrinck2008}, which is a watershed algorithm (see~\cite{Platen2007}) that identifies local underdensities in the three-dimensional density field constructed from tracer particle positions. The density field is estimated via Voronoi tesselation, in which each tracer particle \textit{j} is assigned a unique Voronoi cell of volume $\textit{V}_j$, defined as the volume of space that is closer to its dedicated tracer particle than any other tracer. The density over the volume of a given cell is simply the inverse of the volume, i.e. $\rho_j = 1 / \textit{V}_j$. The watershed algorithm searches for extended density depressions by starting at the local minima in the density field and then monotonically increasing the density until it drops again, which defines the underdense cores and boundaries of our cosmic voids.

Furthermore, a density-based threshold within \textsc{zobov} can be imposed in \textsc{vide}, which is a free parameter. This density threshold is relevant for merging voids, as adjacent zones are added to a void only if the density in any part of their shared wall is less than this threshold value times the mean number density of tracers $\bar{n}_\tracer$. In the case where two neighbouring voids are added, the void catalog will contain two voids, where one encompasses the other one, creating a hierarchy of `parent' and `child' voids. Using only the very low default value of $10^{-9}$ prevents the merging of any voids and results in a catalog of purely \emph{isolated} voids. These \emph{isolated} voids will be the ones relevant in this work, as merging only creates a fraction of larger voids, where we have found no significant effects of the implemented baryonic physics. We therefore refrain from labeling voids as \emph{isolated} ones for the rest of this paper. For a more in-depth discussion on the merging thresholds and how merging affects different void statistics, we refer the reader to previous work~\cite{Schuster2023}.

The catalogs created with \textsc{vide} contain non-spherical voids with various properties. Their centers are defined as the volume-weighted barycenters of all member particles of a particular void:
\begin{equation}
\label{eq:void_barycenter}
\mathbf{X}_\mathrm{v} = \frac{\sum_j \, \bm{x}_j \, \textit{V}_j}{\sum_j \, \textit{V}_j} \,,
\end{equation}

where $\bm{x}_j$ are the comoving tracer positions. This barycenter can be thought of as the geometric center of a void, which is primarily constrained by its boundary, where most of the void's member particles are located. It does not necessarily coincide with the location of its Voronoi cell of lowest density. Additionally, by this definition, the centers of voids are robust against Poisson fluctuations in tracer positions. Due to the non-spherical nature of voids identified with \textsc{vide}, we can only define an  \emph{effective void radius} $r_\void$, which is given by the radius of a sphere with identical volume as the void, calculated as the sum over the volumes of all associated Voronoi cells:

\begin{equation}
\label{eq:void_radius}
r_\void = \left( \frac{3}{4 \pi} \sum_j  \, \textit{V}_j \right)^{1/3} .
\end{equation}

In order to quantify the shape of voids, \textsc{vide} first calculates the inertia tensor (see~\cite{Sutter2015,Schuster2023} for more details), and using its largest ($J_3$) and smallest ($J_1$) eigenvalues, the void ellipticity is given by:

\begin{equation}
\label{eq:void_ellipticity}
\varepsilon = 1 - \left( \frac{J_1}{J_3} \right)^{1/4} \,.
\end{equation}

Additional void properties of interest in this work include the compensation $\Delta_\tracer$ and core density $\coreDens$, where in the former the index `t' marks the tracers that were used for the void identification, i.e. `h' for halos and `m' for matter, respectively. The compensation is a measure of whether a void contains more or less member particles $N_\tracer$ than the average patch of the Universe of identical volume $V$ and therefore contains information about the environment a particular voids is located in, be it of lower or higher local average density. Voids with $\Delta_\tracer < 0$ are referred to as being undercompensated, whereas voids with $\Delta_\tracer > 0$ are referred to as being overcompensated~\cite{Hamaus2014a,Hamaus2014b}. The compensation is defined as:

\begin{equation}
\label{eq:void_compensation}
\Delta_\tracer \equiv \frac{N_\tracer/V}{\bar{n}} - 1 = \hat{n}_{\mathrm{avg}} - 1\,.
\end{equation}

Lastly, the core density is simply the density of the largest Voronoi cell of a void and therefore the cell of minimal density, expressed in units of the mean tracer density:

\begin{equation}
\label{eq:void_coreDensity}
\coreDens =   \frac{ n_{\mathrm{min}}   }{  \bar{n} }\,.
\end{equation}

\subsection{Void profiles \label{subsec:void_profiles}}

\begin{table}[t]
\centering
\begin{tabular}{|c  | c  |} 
 \hline
 Type of profile &  Formula \\
 \hline
  \rule{0pt}{4.5ex} 
density, individual & $n_\void^{(i)}(r) = \frac{3}{4\pi \, } \sum_{j } \frac{ \Theta(r_j)}{ \left( r + \delta r \right)^3 - \left( r - \delta r \right)^3 }$  \\
 \rule{0pt}{4.5ex} 
density, stacked & $n_\void(r) = \frac{1}{N_\void} \sum\limits_{i} n_\void^{(i)}(r)$  \\
 \rule{0pt}{4.5ex} 
velocity, individual & $u_\void^{(i)} (r) = \frac{  \sum_j \bm{u}_j \cdot \hat{\bm{r}}_j \,\, V_j \,\, \Theta(r_j)  }
{  \sum_j  V_j \,\, \Theta(r_j) }$ \\
  \rule{0pt}{4.5ex} 
velocity, stacked & $u_\void(r) = \frac{1}{N_\void} \sum\limits_{i} u_\void^{(i)}(r)$  \\ [2.5ex] 
 
 \hline
\end{tabular}
\caption{Formulas for calculating the void profiles presented in this work. For more details, as well as a discussion on weights in the density profiles and multiple ways of calculating velocity profiles with the respective biases in their methods, we refer the reader to sections 3.2 and 5.3 in reference~\cite{Schuster2023}. The meaning of the variables is explained in the text.}
\label{table_2}
\end{table}

As we investigate the impact of baryonic physics on the density, as well as the velocity profiles of both individual and stacked voids, we present a short summary of the formulas with which these profiles are calculated in table~\ref{table_2}. For a more in-depth discussion on the methods of calculating void profiles, we refer the reader to section 3.2 of our previous work using the Magneticum simulations~\cite{Schuster2023}. All of these profiles are averages of either density or velocity of tracers in spherical shells around the center of a void. The density profile is the spherically averaged density contrast from its mean value $\bar{n}$ in the Universe and the velocity profiles describes the radial movement of tracer particles around voids. In the latter, positive velocities correspond to an outflow of tracers from the perspective of the barycenter, whereas negative velocities describe inward movement of tracers. These profiles are calculated for individual voids using a constant radial bin size when expressed in units of their radius, $\delta r/r_\void = \mathrm{const}$. This ensures that characteristic features of void profiles are located at the same location from the void center when comparing both individual as well as stacked profiles of different sizes. One such feature is the compensation wall of voids, centered around $r = r_\void$.

In the formula for the individual density profiles in table~\ref{table_2}, the function $\Theta( r_j)$ is defined as $\Theta( r_j) \equiv \vartheta \left[ r_j - \left( r - \delta r \right) \right] \, \vartheta \left[- r_j + \left( r + \delta r \right) \right] $ and combines two Heaviside step functions $\vartheta$ in order to define the radial bins of the profile, with $r_j$ being the distance of tracer $j$ from the center of the void. The stacked density profiles are simply the average over the individual density profiles of voids, typically in different ranges of their radius $r_\void$, although stacks in other void properties are presented as well.

When calculating individual velocity profiles using the formula given in table~\ref{table_2}, $\bm{u}_j$ refers to the peculiar velocity vector of each tracer $j$ and $\hat{\bm{r}}_j = \bm{r}_j / r_j$ is the unit vector which connects the void center with the given tracer particle $j$. The individual velocity of each particle is weighted by its Voronoi volume $V_j$, which ensures that the velocity profile is a volumetric representation of the actual underlying velocity field~\cite{Hamaus2014b}. A uniform sampling of the velocity field without volume weights is biased due to higher sampling of tracers near the void boundaries and lower sampling near the center, equivalent to an estimate of the momentum profile instead.

For velocity profiles there are two options of stacking the individual profiles, namely \emph{individual} stacks, which simply average the individual velocity profiles in the same manner as it is done for density profiles, and \emph{global} stacks, which average denominator and numerator separately and then divide the averages afterwards. Both of these stacking methods have their own respective biases, namely \emph{individual} stacks are biased towards the profiles of the most numerous voids in a given stack, which are typically the smallest ones, while \emph{global} stacks are biased towards profiles of voids made up of the most member particles, typically the largest voids. For a more detailed discussion on these methods and their biases, we refer the reader to our previous work~\cite{Schuster2023}. In this work, we will only present results from \emph{individually} stacked velocity profiles, as we found that there are no significant differences in the impact of baryonic effects on the velocity profiles between \emph{individually} and \emph{globally} stacked profiles. This is reasonably expected, as the \emph{individually} stacked velocity profiles are more biased towards small voids, where the strongest effects are present, whilst the bias of \emph{globally} stacked profiles towards larger and better sampled voids might lead to additional biases in the effects of baryonic physics, as the number of voids in this regime is more sparse and fluctuations in void profiles might lead to spurious results that could be misinterpreted as baryonic effects in the velocity profiles.

\section{Magneticum catalogs \label{sec:catalogs}}

\subsection{Tracers \label{subsec:tracer_catalogs}}

\begin{table}[t]
\setlength{\tabcolsep}{5pt}
\centering
\begin{tabular}{|c | c c c c c c c|} 
 \hline
 Name &  \hydro{}/ & $M_\mathrm{cut}$ & $N_\halo $ & $\bar{n}_\halo $ &   $\bar{r}_\mathrm{t}$   & $N_\void$ in &  $N_\void$ in \\
 & \DMo{}  & $[\Msun]$ & $[\times 10^6]$ & $[(\hMpc)^{-3}]$ &$[\hMpc]$ &   halos &  matched halos  \\
 \hline
  \rule{0pt}{3ex} 
 \MR{} & \hydro{} & $1.0 \times 10^{12}$ & $65.5$ & $3.37 \cdot 10^{-3}$ & $6.67$ & $366\,709$ & -- \\
 
 \MR{} & \DMo{} & $1.0 \times 10^{12}$ & $66.4$ & $3.42 \cdot 10^{-3}$ & $6.64$ & $371\,943$ & $367\,791$ \\
 
 \HR{}  & \hydro{} & $1.0 \times 10^{11}$ & $8.20$ & $3.13 \cdot 10^{-2}$ & $3.17$ & $33\,254$ & $31\,139$ \\ 
 
 \HR{}  & \DMo{} & $1.0 \times 10^{11}$ & $7.53$ & $2.87 \cdot 10^{-2}$ & $3.27$ & $32\,050$ & -- \\ 
 
 \UHR{}  & \hydro{} & $1.6 \times 10^{9}$ & $0.113$ & $1.02 \cdot 10^{0}$ & $0.992$ & $281$ & -- \\
 
 \UHR{}  & \DMo{} & $1.6 \times 10^{9}$ & $0.143$ & $1.30 \cdot 10^{0}$ & $0.917$ & $348$ & $278$ \\ [1ex] 
 \hline
\end{tabular}
\caption{Summary of properties from the hydrodynamical (\hydro{}) and dark-matter-only (\DMo{}) simulations, like number of halos $N_\halo$ with $M_\halo \geq M_\mathrm{cut}$, mean halo density $\bar{n}_\halo$ and tracer separation $\bar{r}_\tracer$ after mass cuts, number of halo voids identified in the halo catalogs after applying the mass cut and matched halo densities (for simulation with larger $N_\halo$) in the different \Mag{} runs, all at the redshifts $z$ given in table~\ref{table_1}. }
\label{table_3}
\end{table}

Since the void finding with \textsc{vide} only requires the positions of any kind of tracer particles, we will identify voids in both the distributions of halos, as well as the different matter tracers. In case of the \hydro{} simulations, we identify voids in both CDM and baryon particles and will refer to those voids as \emph{CDM voids} and \emph{baryon voids}, respectively. For the \DMo{} simulations, only CDM particles are present and therein identified voids will be referred to as \emph{CDMo voids}. When using halos as tracers instead, voids identified in the halos of the \hydro{} simulation are referred to as \emph{halo voids}, while voids identified in halos from the \DMo{} simulations are referred to as \emph{\DMo{}-halo voids}. 

The halos utilized for void identification are usually selected after applying different mass cuts (see table~\ref{table_3}), which lie above the resolution limit of each box, which is explained in more detail in section 2 of~\cite{Schuster2023}. This approach is valid for halo voids in both \hydro{} and \DMo{} runs. However, as the number of halos above any given mass cut can differ between \hydro{} and \DMo{} runs, it can yield significant differences in void numbers, which subsequently affects common void statistics, such as the void size function and void profiles. These changes could already be attributed to baryonic effects, although in real observations we only observe a given number of halos and galaxies. We therefore choose to match the halo densities in both \hydro{} and \DMo{} runs to additionally identify voids in these matched catalogs, next to the mass cut catalogs. This is done in order to pinpoint where baryonic physics affect void statistics and to examine if void statistics from other dark-matter-only simulations at a given tracer density would have to take into account these effects when comparing observations with statistics from simulations. For this matching, we select the most massive halos of those simulations that have the higher number of halos above a given mass cut. Therefore, when investigating baryonic effects around halo voids, we will present void statistics in both catalogs obtained from matched halo densities and from mass cuts.

In the \MR{} and \HR{} simulations we additionally identify voids in subsamplings of the underlying matter particles. These values of subsampling were chosen to approximately match the number of matter tracers to the total number of halos used for the void finding, in order to obtain matter and halo-defined voids of roughly similar ranges in $r_\void$. This is done for both CDM and baryons from the \hydro{} simulations, as well as for CDM from the \DMo{} simulations (CDMo). The subsampling fractions compared to the total number of matter tracers of each kind and the resulting tracer numbers are given in table~\ref{table_4}. While voids identified in the distribution of matter are not directly accessible through observations, we nevertheless study them in order to identify effects from baryonic physics.

\begin{table}[t]
\centering
\begin{tabular}{|c | c c c c c|} 
 \hline
 Name &    subsampling & $N_\mathrm{t}$ after & $N_\void$ in   & $N_\void$ in &  $N_\void$ in \\
  &  fraction & subsampling $[\times 10^6]$ & CDM &   baryons &  CDMo  \\
 \hline
  \rule{0pt}{3ex} 
 \MR{}  &  0.0665 $\%$ & $62.1$ & $538\,255$ & $539\,284$ & $537\,951$ \\
 
 \HR{}  &  0.0344 $\%$ & $8.21$ & $51\,741$ & $52\,712$ & $51\,772$ \\ [1ex] 
 
 \hline
\end{tabular}
\caption{Subsampling fraction from total number of particles of each kind and resulting number of subsampled matter tracers used for the void identification, number of voids found in the CDM and baryons of the \hydro{} runs, as well as in the CDM of the \DMo{} runs (CDMo).}
\label{table_4}
\end{table}

\subsection{Voids \label{subsec:void_catalogs}}

As we have detected no significant baryonic effects between the \DMo{} and  \hydro{} runs of the \MR{} simulations in all matter and halo-defined voids, we refrain from presenting our analysis from these and instead focus on voids identified in the \HR{} simulations. Moreover, the absence of relevant baryonic effects in the \MR{} simulations is the reason as to why we do not investigate baryonic effects in alternative cosmologies from other \Mag{} runs, as they are only available at identical resolution to \MR{}, but in a smaller box with substantially fewer voids. However, the numbers of \MR{} voids, as well as tracer densities and simulation details are still presented in tables~\ref{table_1},~\ref{table_3} and~\ref{table_4} for completeness. Furthermore, we analyzed both \emph{isolated} and \emph{merged} void catalogs, but as relevant effects were only identified in voids of small $r_\void$, we refrain from presenting \emph{merged} void catalogs and only focus on \emph{isolated} voids, as the small voids and their statistics are typically identical in both options for merging~\cite{Schuster2023}. Void profiles from the \UHR{} simulation will only be presented in a resolution study in section~\ref{subsec:uhr} to further support our results. Common void statistics in \UHR{} will not be presented in other sections of this work due to the extremely low number of voids identified therein.

Table~\ref{table_3} contains the number of halo and \DMo{} halo voids identified in the tracers after applying mass cuts, as well as for the matched halo density catalogs. In the case of \MR{} there are more halos and correspondingly more halo voids in the \DMo{} run after applying mass cuts. This behavior changes for the \HR{} simulation, with a higher halo number density in the \hydro{} run, which reverses again in \UHR{}. As expected, after matching the halo number densities, the number of voids identified in halos becomes more similar.

The number of voids found in baryons and CDM are presented in table~\ref{table_4}. We see that at a fixed matter tracer density after subsampling (small variations in $N_\tracer$ exist), there are generally slightly more voids identified in baryons than in the CDM of the \hydro{} simulations due to the more uniform distribution of baryons, while the void numbers in CDM of both \hydro{} and \DMo{} simulations are almost identical.

\begin{figure}[t]

               \centering

               \resizebox{\hsize}{!}{

                               \includegraphics[trim=7 50 0 5, clip]{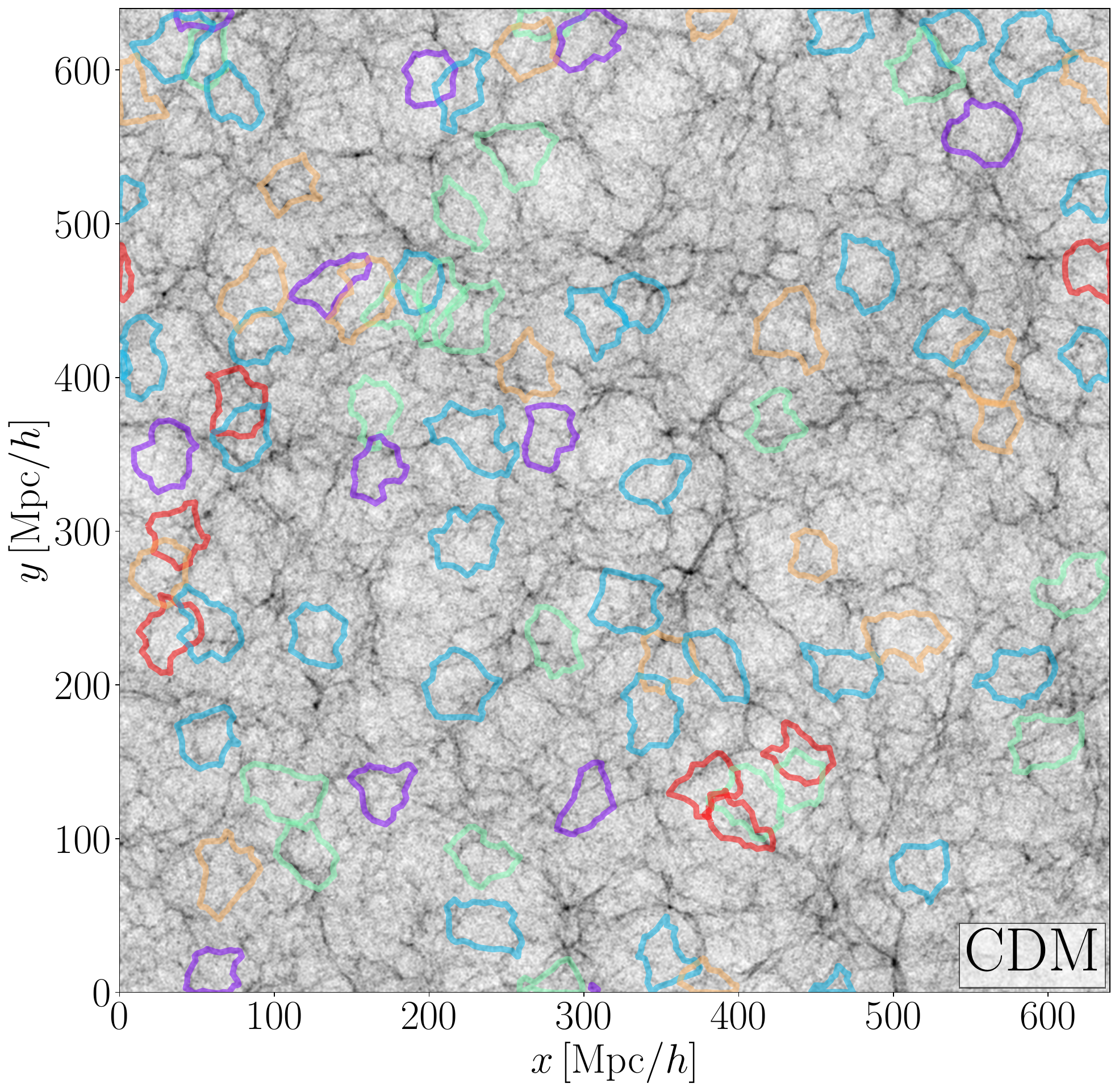}

                               \includegraphics[trim=45 50 0 5, clip]{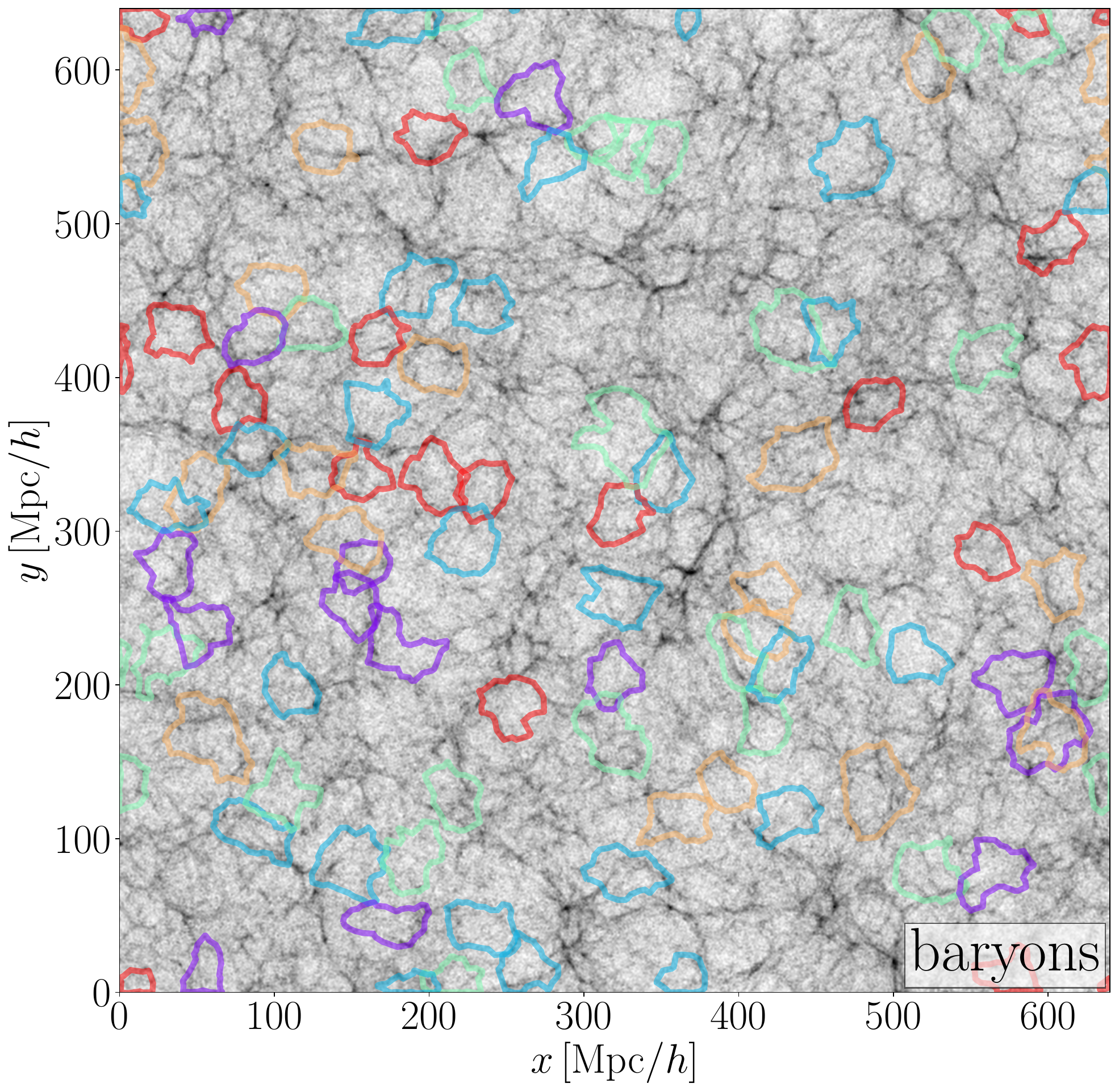}}

               \resizebox{\hsize}{!}{

                               \includegraphics[trim=0 10 0 5, clip]{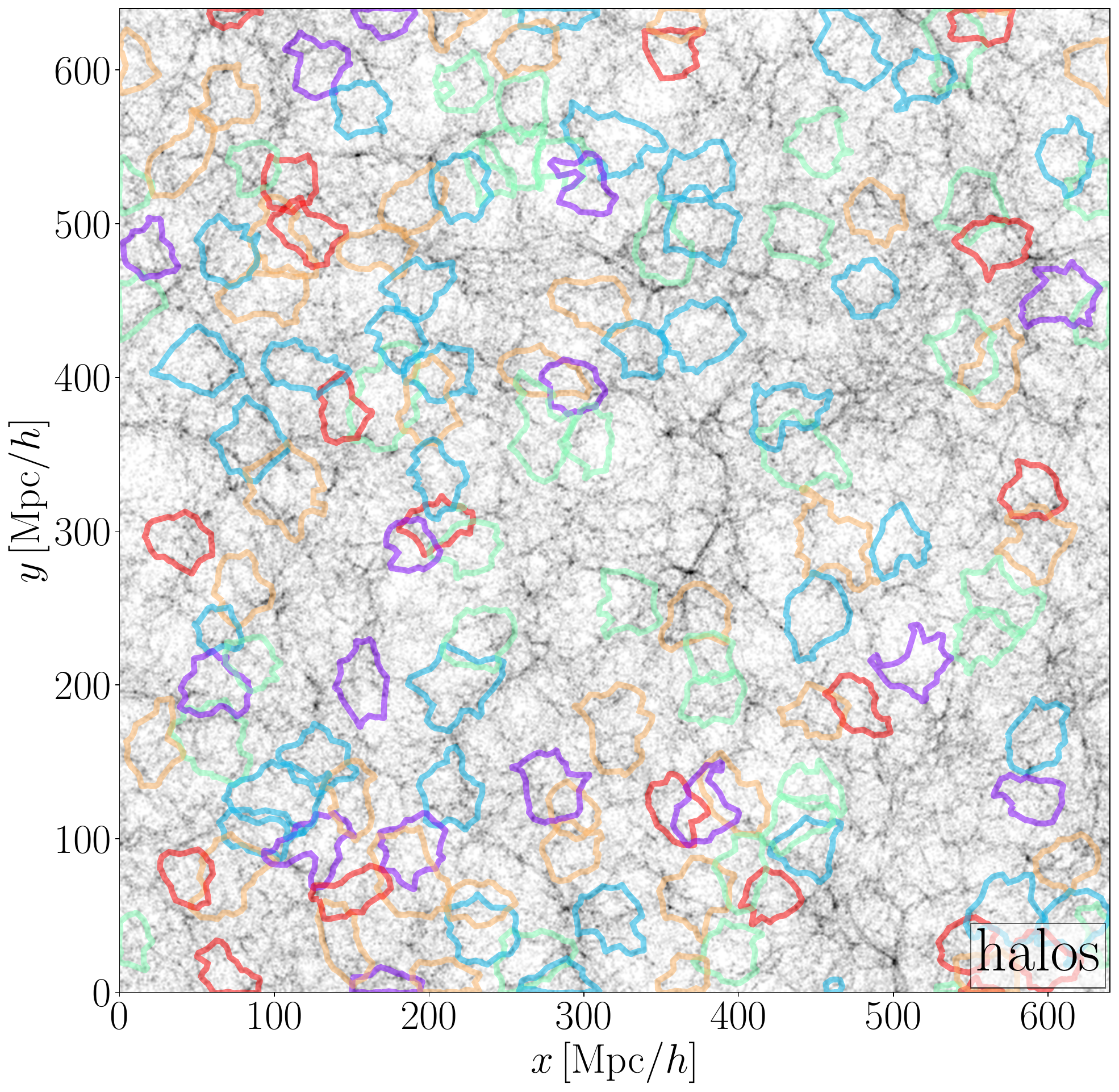}

                               \includegraphics[trim=45 10 0 5, clip]{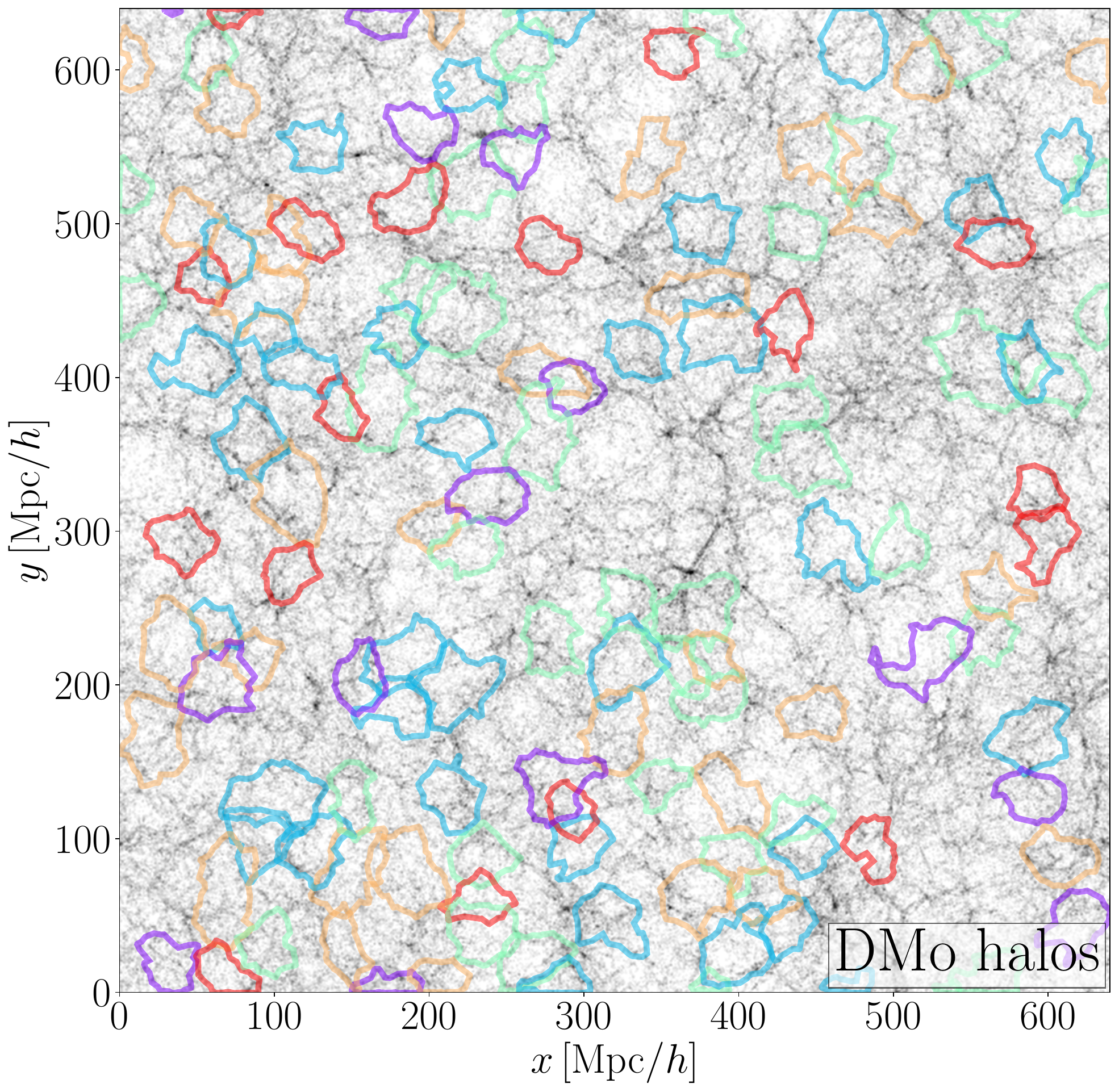}}

               \caption{Projected density field of a $50\, \hMpc$ deep slice in the \HR{} simulations, including projected boundaries of voids with $r_\void \geq 15\, \hMpc$, whose spherical equivalent resides completely within the slice. Presented for the density field and therein identified voids in CDM (top left), baryons (top right), as well as halos in the \hydro{} (bottom left) and \DMo{} (bottom right) simulations. For the projected matter density fields, subsamplings with more particles than used for the void finding are depicted. Colors of void boundaries indicate the depth of the void barycenters within the slice.}

               \label{fig_projected_density_hr}

\end{figure}

To showcase similarities in both matter and halo voids, figure~\ref{fig_projected_density_hr} presents the projected density field across the entire \HR{} simulation box in a $50\, \hMpc$ deep slice, including projected void shapes through their boundaries. Regions of lower projected density are depicted in lighter colors, while regions of high density correspond to darker tones in color. This is presented for CDM (top left), baryons (top right), halos (bottom left) and \DMo{} halos (bottom right). For better visibility, we restrict ourselves to depicting only voids with $r_\void \geq 15\, \hMpc$, whose spherical equivalent lies completely within the projected slice, as otherwise the whole area would be filled with voids. While in the case of all halo voids (bottom row), the projected tracers are identical to the ones used for the void identification after mass cuts, the matter voids are found in subsamples of around $8.21$ million tracers, while the projected density fields use subsamples of around $50$ million tracers in order to resolve cosmic structures more accurately. Colors of void boundaries indicate the depth of the barycenter within the slice. This is done to illustrate neighboring voids and those that overlap in projection. Voids in purple/blue are closest to the beginning of the slice, while orange/red voids are the deepest. As all voids are \emph{isolated} ones, there is no direct overlap in real space, and every void occupies a unique volume. Due to periodic boundary conditions, some voids even cross from one end of the projected simulation box to the other.

Even in projection, typical structures of the cosmic web are clearly visible, namely nodes of clustered tracers in dark and extended filaments, as well as more empty regions in lighter shades. While the filaments and nodes are clearly visible, most of the visibility of sheets gets lost due to the projection. However, in all 4 plots the filaments and nodes align, as expected, since all types of matter are attracted to the gravitational potential wells and halos simply arise from the clustering of matter particles. Quite often the void boundaries align with these visible structures. However, a one-to-one congruence between plots is not quite as clear for projected voids, since only a small selection of them is depicted. Some matter voids get `merged' when comparing them with the biased tracer field of halos, and exact void shapes are highly sensitive to variations in tracer locations between the different cases. Moreover, slight variations in barycenters and radii might lead to some more or less `identical' voids being depicted in different colors or left out of the projection. Nevertheless, there is still some overlap of voids in all cases, with only minor variations in their shapes. Such examples include the blue void in CDM and baryons, located at around $x \simeq 250 \, \hMpc$ and $y \simeq 50\, \hMpc$, which is also present in halos, although with larger projected size, as well as a red void located near $x \simeq 80\, \hMpc$ and $y \simeq 20\, \hMpc$ in both halos and \DMo{} halos. Nevertheless, finding and comparing `identical' voids between different tracers is neither possible for most of them, nor is it very useful. In this work we are only interested in how void statistics are affected by baryonic physics, and not the effects on individual voids identified in multiple tracers, as stochastic variations between tracer positions are expected to be much stronger than any effects solely caused by the implemented physics.

\begin{figure}[!ht]

               \centering

               \resizebox{0.89\hsize}{!}{

                               \includegraphics[trim=7 0 0 7, clip]{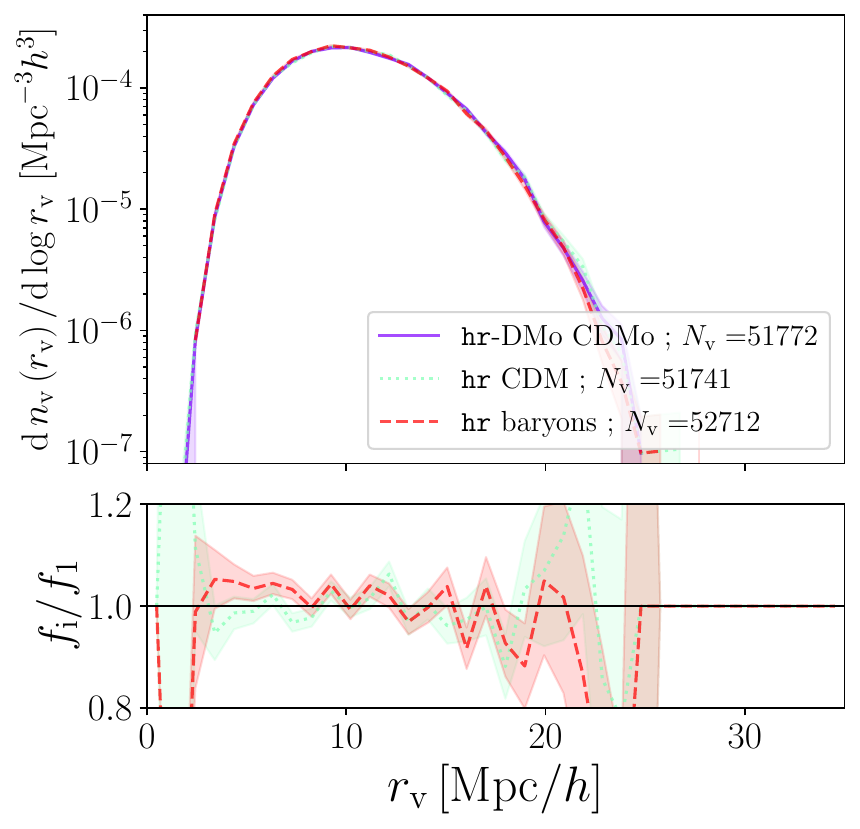}

                               \includegraphics[trim=0 0 0 7, clip]{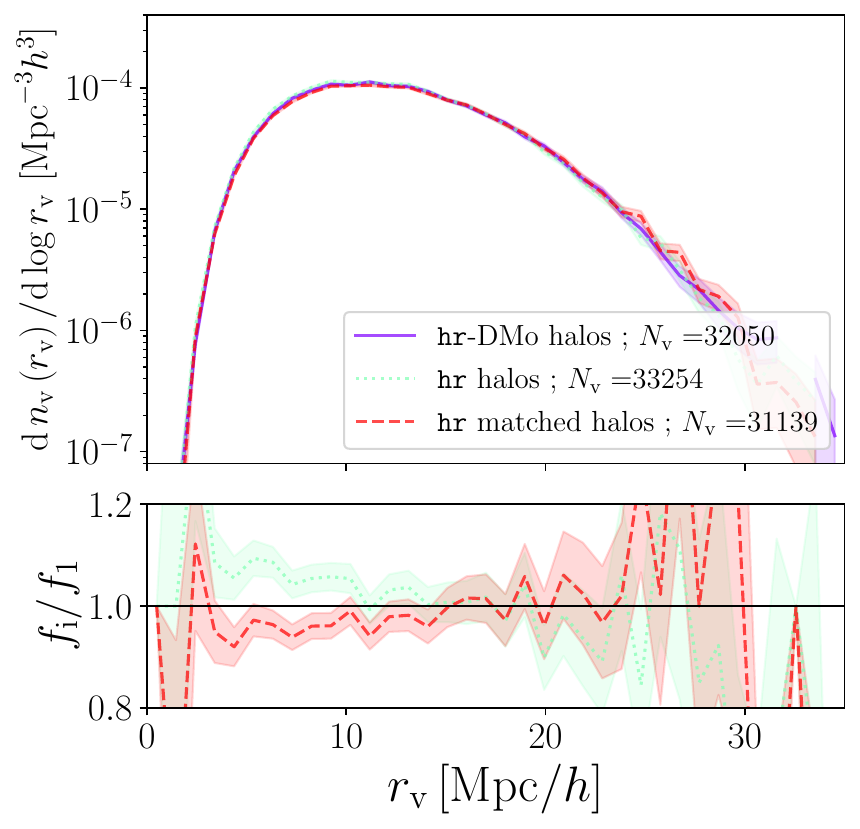}}

               \resizebox{0.89\hsize}{!}{

                               \includegraphics[trim=7 0 0 9, clip]{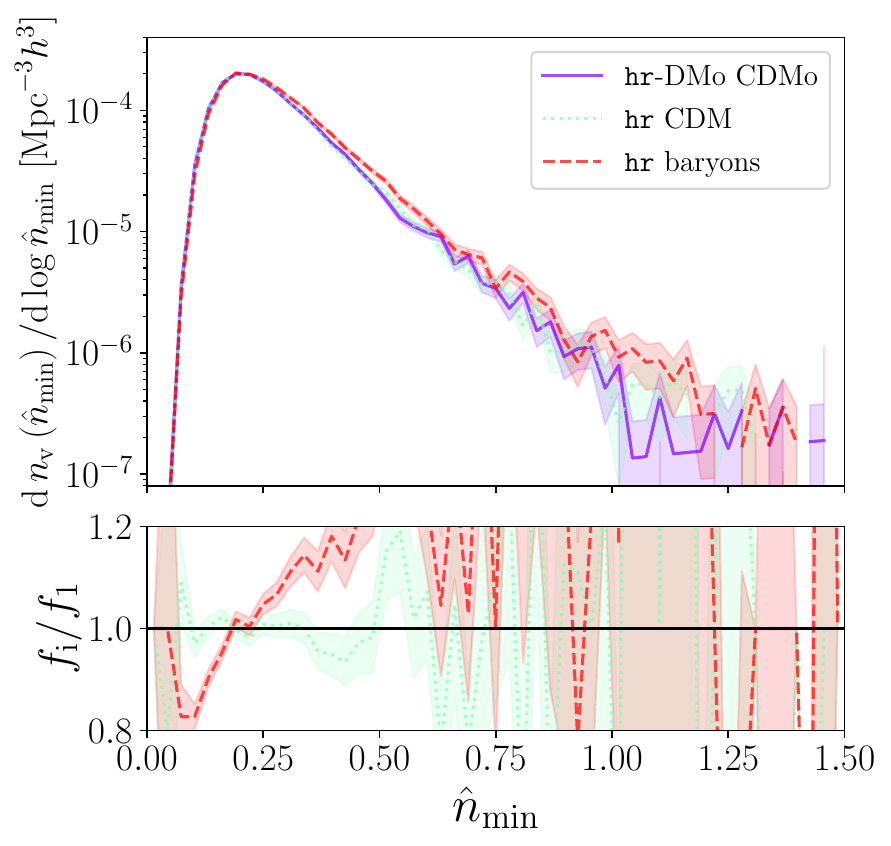}

                               \includegraphics[trim=0 0 0 9, clip]{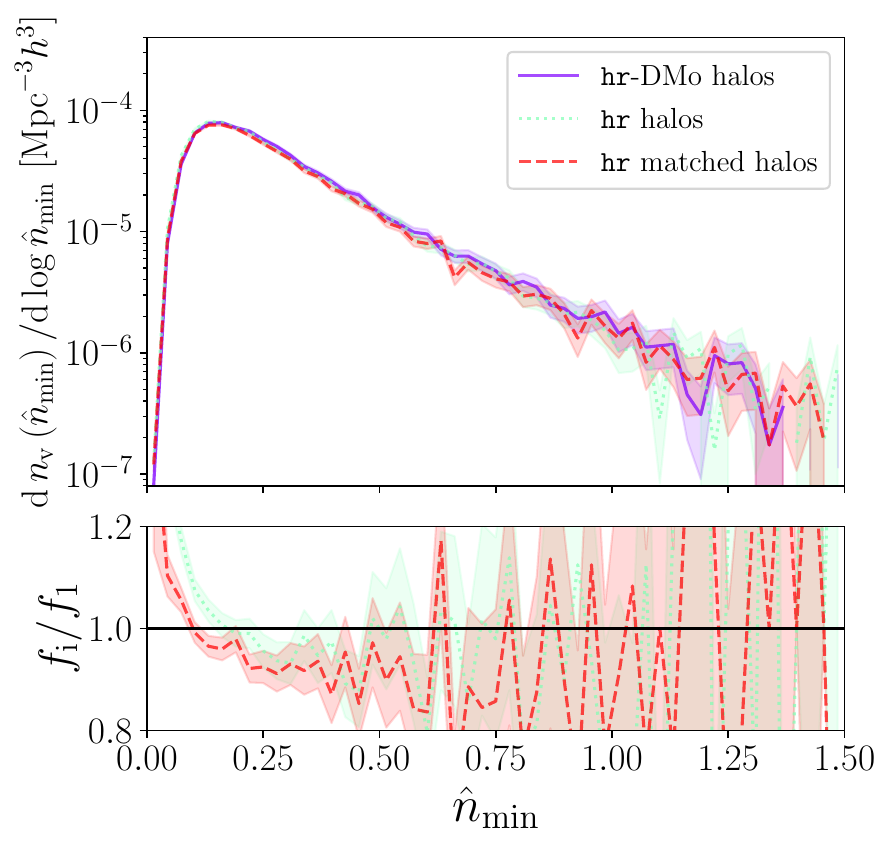}}

               \resizebox{0.89\hsize}{!}{

                               \includegraphics[trim=7 11 0 0, clip]{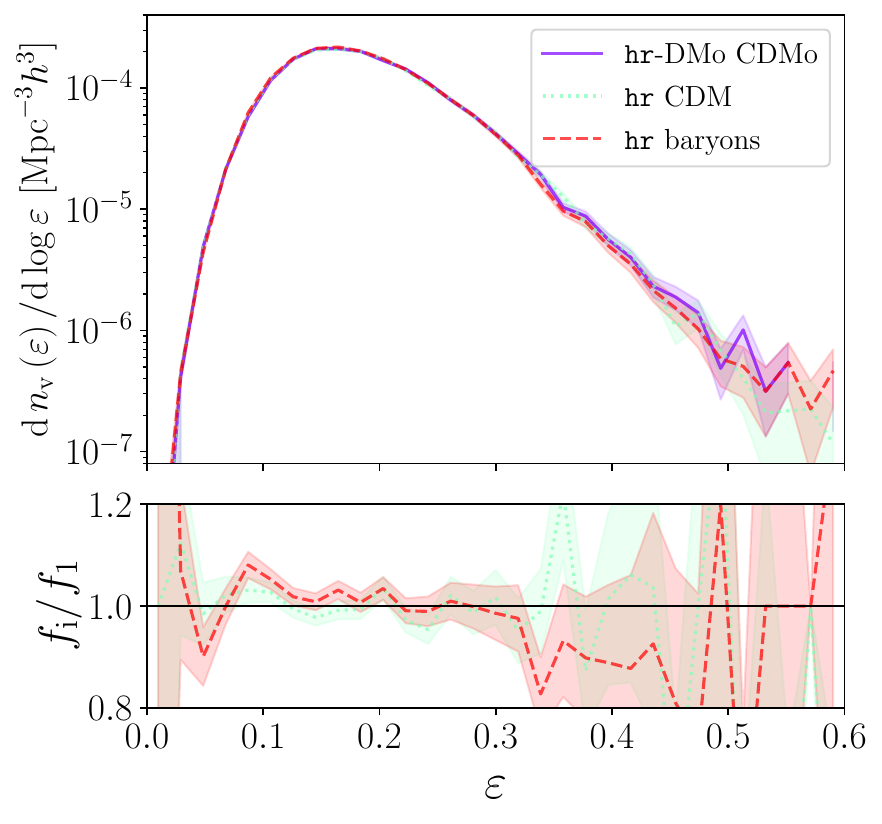}

                               \includegraphics[trim=0 11 0 0, clip]{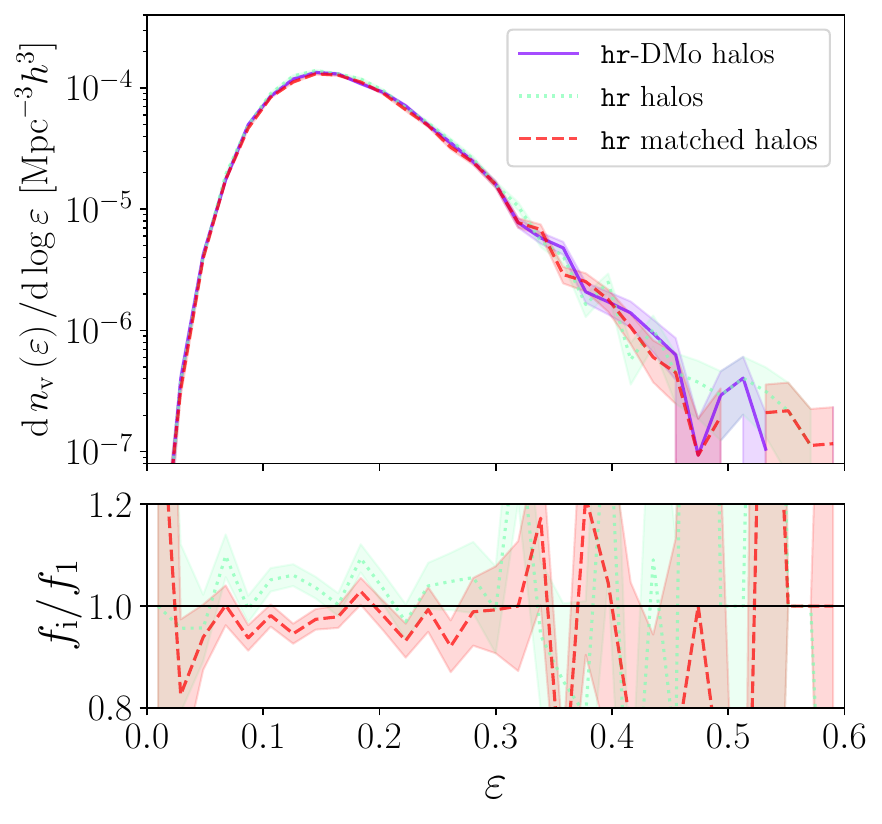}}

               \caption{Distributions in radius (top), core density (middle) and ellipticity (bottom) of \emph{isolated} voids identified in matter (left), as well as halos (right) in \HR{}. Upper panels depict the functions, whilst lower panels represent the fractional change with respect to the \DMo{} simulation. For halos, catalogs obtained from a fixed mass cut, as well as the one with matched density are presented. Errors assume Poisson statistics and are propagated for the lower ratio panels.}

               \label{fig_abundances_hr}

\end{figure}

In addition, even in projection we already observe that the distribution of baryons is more diffuse than for CDM, as previously noted in reference~\cite{Paillas2017}, which in turn indicates a higher baryon fraction inside voids~\cite{Rodriguez2022}. A striking difference between projected matter and halo voids in figure~\ref{fig_projected_density_hr} is that there are considerably fewer matter voids depicted compared to halo voids. This can be explained by looking at the void numbers found in each tracer and by investigating the void size functions in the upper plots of figure~\ref{fig_abundances_hr}. As we find significantly more matter than halo voids at similar tracer densities, matter voids necessarily become smaller than halo voids, as all voids combined have to fill the simulation volume. This shifts the void size function of matter voids towards smaller voids, and hence less projected voids are present on the top of figure~\ref{fig_projected_density_hr} due to the imposed size cut. As already discussed in reference~\cite{Schuster2023}, this higher number of matter voids happens because for halos, only the peaks in the density field are observed, while in matter subsamplings, both matter tracers that clustered in halos are present, as well as matter tracers that have not clustered, but instead reside in regions where no halos are present. Due to this, a finer structure of the cosmic web can be found inside halo voids, which results in matter-made structures `splitting up' some of the halo voids, leading to the higher number of matter voids.

Besides these universal differences in the void size functions of halo and matter voids, figure~\ref{fig_abundances_hr} presents the abundance of \emph{isolated} voids by their radii $r_\void$ for all matter tracers (top left) and for halo voids obtained from mass cuts, as well as matched densities (top right). For all cases of matter voids, their size functions are nearly indistinguishable, which can be seen in the lower panels, where abundances of either CDM or baryon voids are relative to the abundance of CDMo voids. There is a slight scatter in these ratios, although the void size functions align almost perfectly within the error bars, which could already be expected from the alike void numbers. Just a minuscule increase in small baryon voids can be observed compared to CDM voids due to the higher number of baryon voids, while CDM and CDMo voids show no significant differences.

On the other hand, the size functions of halo voids experience slight changes, caused by more significant differences in void numbers. The \DMo{} halo voids are used as the baseline and in comparison, halo voids from the \hydro{} run after mass cut experience a slight shift towards smaller voids, while after matching halo number densities, this effect reverses, and the matched halo voids are typically larger than voids identified in the \DMo{} run at identical halo density. Due to the matched halo number densities, we ascribe this effect to the implemented baryonic physics, although it is important to note that when matching densities, we cut the least massive halos from the tracer catalogs and these halos have a larger probability of residing within voids than the more massive ones~\cite{Schuster2023}. Therefore, some of the structures at low densities that connect halo voids obtained from mass cuts at their boundaries might `vanish', tracing larger, but also fewer voids in the matched density catalogs.

In contrast, void abundances as functions of their core densities (up to $\coreDens = 1.5$) in the middle of figure~\ref{fig_abundances_hr} exhibit much clearer differences. While CDM voids in both \DMo{} and \hydro{} have identical core density distributions within their error bars, voids found in the baryons experience a significant shift towards higher values in their core densities, i.e. baryon voids are typically less underdense than CDM voids due to the more diffuse distribution of baryons, as already seen in figure~\ref{fig_projected_density_hr}. When comparing halo voids obtained from mass cuts, the ones from the \hydro{} run have a slight overabundance of voids with low core densities compared to \DMo{} halo voids, while voids with higher $\coreDens$ values show no discernible differences within the scatter (see division panels). When matching halo number densities, this effect shifts towards even smaller values in core density, which generally only few voids have. There is a slight indication that in matched density catalogs, fewer voids have high core densities, although this most likely originates from reduced void numbers in the matched \hydro{} catalogs, as this shift is mostly constant over a wide range in $\coreDens$.

Finally, the distribution in void ellipticities on the bottom of figure~\ref{fig_abundances_hr} suggests that the increase in baryon void numbers happens mostly at the peak of the distribution, between $\varepsilon \simeq 0.1$ and $\varepsilon \simeq 0.2$, while there are slightly fewer highly elliptical voids. Contrary to this, voids in both CDM and CDMo have no relevant differences in their shapes. In figure~\ref{fig_abundances_hr} voids with ellipticities larger than $\varepsilon = 0.6$ were cut from the panels, as these are very few in numbers and mostly spurious. Comparing the different halo void catalogs, we notice that dividing the abundances leads to nearly constant values for each case, representing the simple shift of the distributions due to changing void numbers, while the shapes of their distributions remain the same. 

To summarize, most void property distributions remain unchanged between \hydro{} and \DMo{} runs and between different matter types, except for changes due to different void numbers. Exclusively voids identified in baryons experience a clear shift towards higher core densities than CDM voids. For a more general discussion on differences in void distributions between matter and halo voids, we refer to reference~\cite{Schuster2023}.

\section{Void profiles \label{sec:baryonic}}

In this section we focus on effects of baryonic physics on the density, as well as velocity profiles of voids. These profiles are calculated for each individual void out to five times its effective radius in bins of width $0.1 \times r_\void$, unless mentioned otherwise. Afterwards, we stack (average) these individual void profiles in bins of various void properties, although typically their radius $r_\void$ is used. We focus solely on the profiles of \emph{isolated} voids and analyze the profiles of both matter, as well as halo voids, where in the latter case we distinguish between using halos and different tracers of matter for calculating the profiles. We examine the profiles of halo voids obtained from halos after undertaking mass cuts and from matching halo densities between \hydro{} and \DMo{} simulations. Moreover, we compare the distribution of baryons and CDM around halo voids and substantiate these results with a resolution study in the \UHR{} simulation. While calculating profiles of matter around both matter and halo-defined voids, we always use subsamplings of significantly higher tracer numbers ($50$ million) compared to the the numbers used in the void finding (around $8$ million), in order reduce effects arising from sparse sampling in low density regions and to constrain baryonic effects more accurately. For the analysis of stacked velocity profiles, we always use individual stacks (see table~\ref{table_2}), as we find that the choice of velocity estimator does not impact our conclusions on baryons, except for more general effects already discussed in reference~\cite{Schuster2023}.

Figures~\ref{fig_density_matter} to~\ref{fig_velocity_halo_matter} depict the void profiles in stacked bins of their void radii in the upper plots, whereas plots on the bottom use core density bins, with the void numbers and their mean values of each bin given in the legends. While the radius of a void and its core density are slightly correlated~\cite{Schuster2023}, bins in $\coreDens$ cover a wider range of radii, and we find stronger effects from the implemented baryonic physics than in $r_\void$ bins. Stacks in more void properties, namely compensation $\Delta_\tracer$ and ellipticity $\varepsilon$, will be depicted solely in figures~\ref{fig_density_halo_BARCDM} to~\ref{fig_velocity_halo_BARCDM_Mcut12}, as effects between \DMo{} and \hydro{} simulations, as well as between CDM and baryons in the \hydro{} simulation are most severe in core density and void radius bins. This is due to compensation and ellipticity bins covering an even wider range in radii, which we discuss in more detail in section~\ref{subsec:baryon_BAR_CDM_voids}.

\subsection{Matter voids \label{subsec:baryon_matter_voids}}

\begin{figure}[t]

               \centering

               \resizebox{\hsize}{!}{

                               \includegraphics[trim=0 5 0 5, clip]{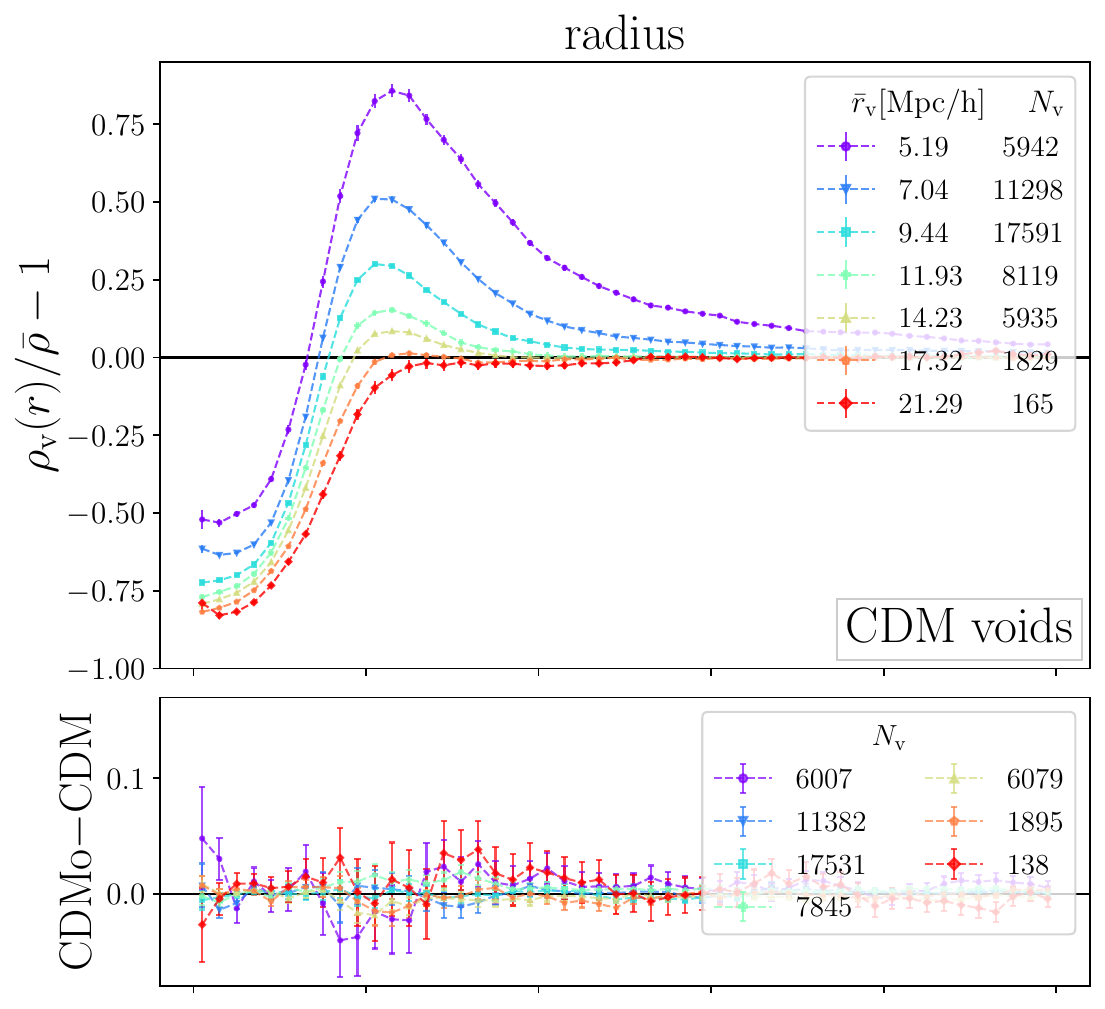}

                               \includegraphics[trim=0 5 0 5, clip]{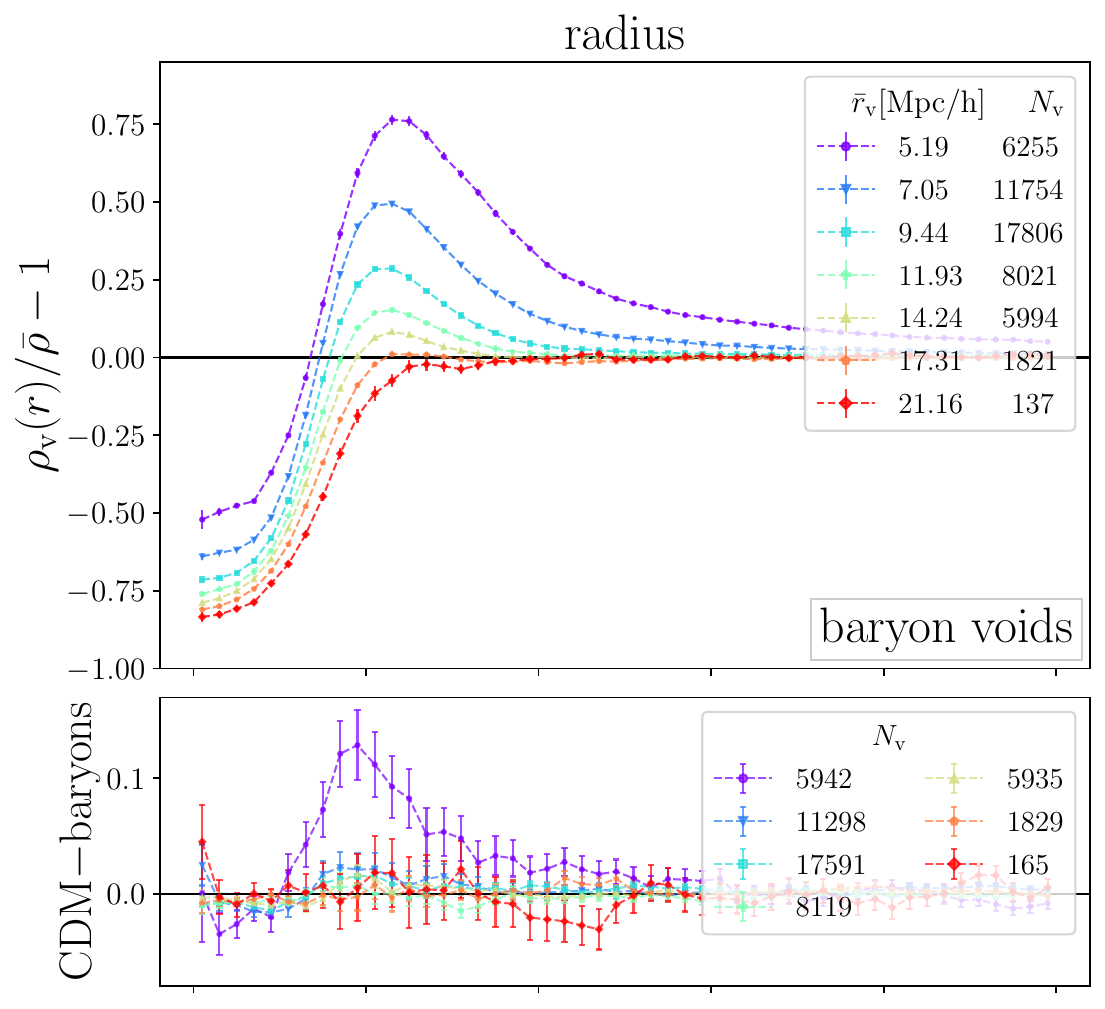}}

               \resizebox{\hsize}{!}{

                               \includegraphics[trim=0 10 0 5, clip]{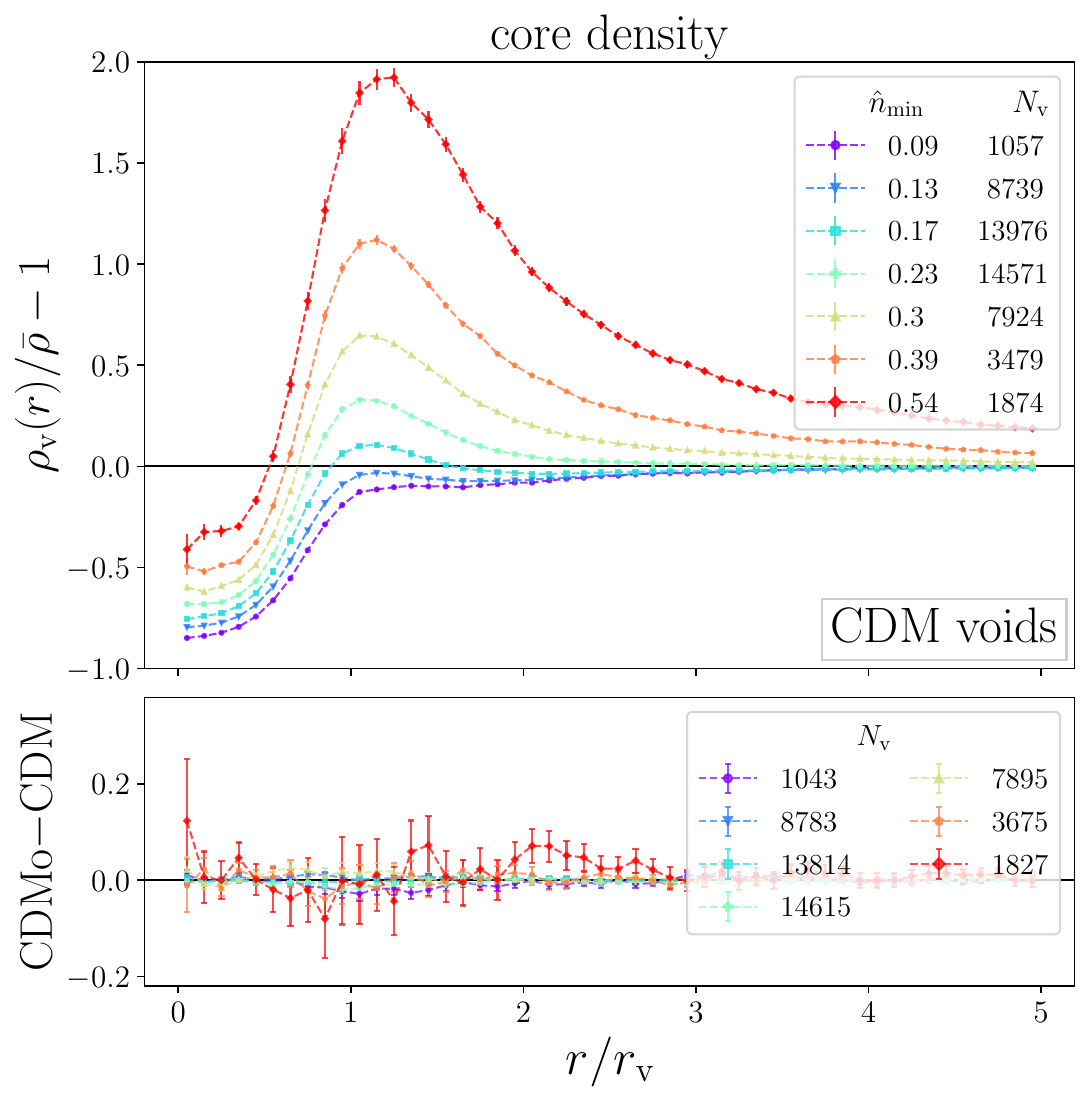}

                               \includegraphics[trim=0 10 0 5, clip]{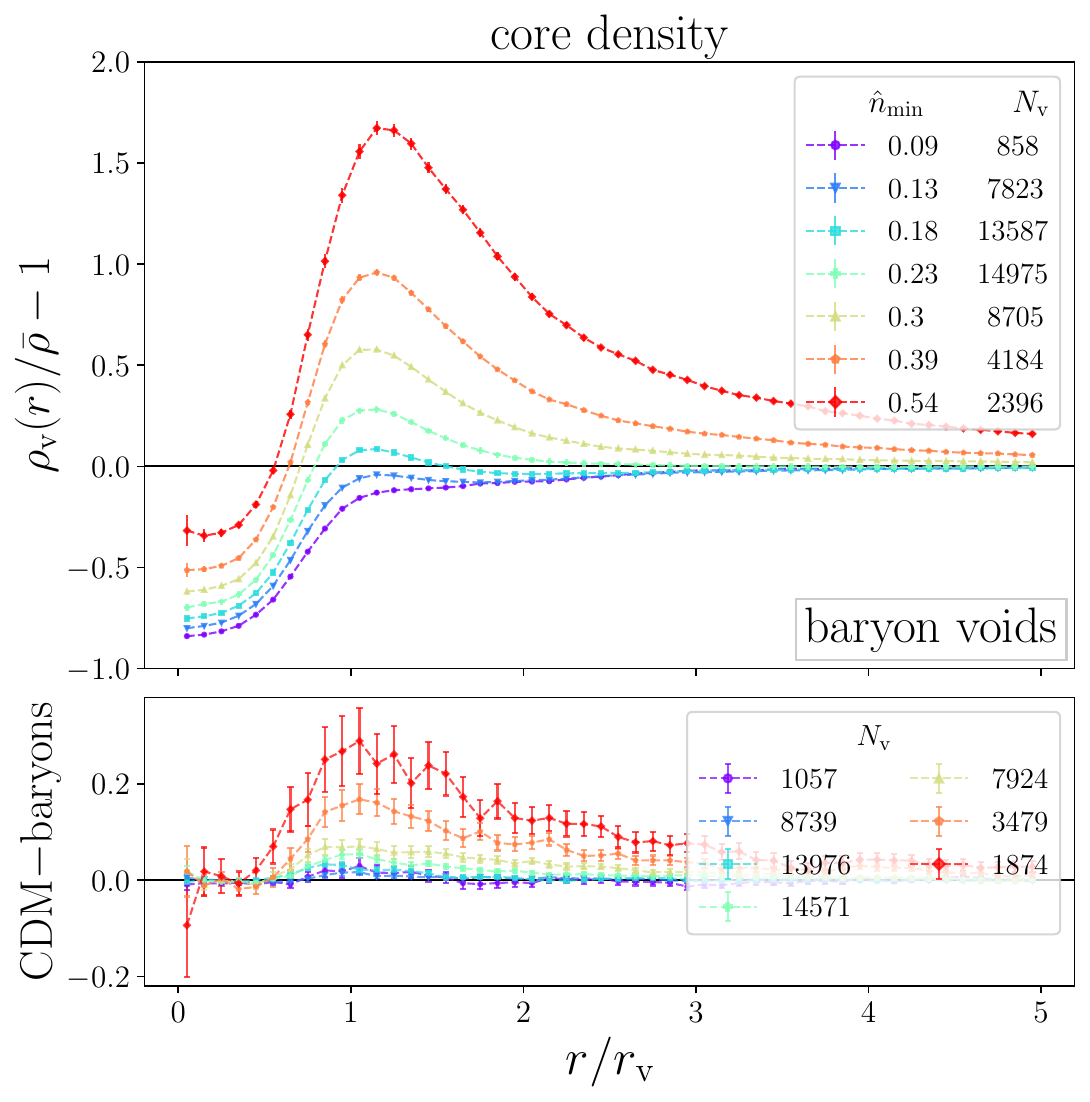}}

               \caption{Stacked density profiles of matter voids in the \HR{} simulations, in bins of void radius (top) and core density (bottom). Plots on the left depict profiles of CDM voids, with differences compared to CDMo voids in the \DMo{} simulations in the lower panels, whilst plots on the right show profiles of baryon voids, with differences between CDM voids and baryon voids in the lower panels, both from the \hydro{} simulation.}

               \label{fig_density_matter}

\end{figure}

\begin{figure}[t]

               \centering

               \resizebox{\hsize}{!}{

                               \includegraphics[trim=0 5 0 5, clip]{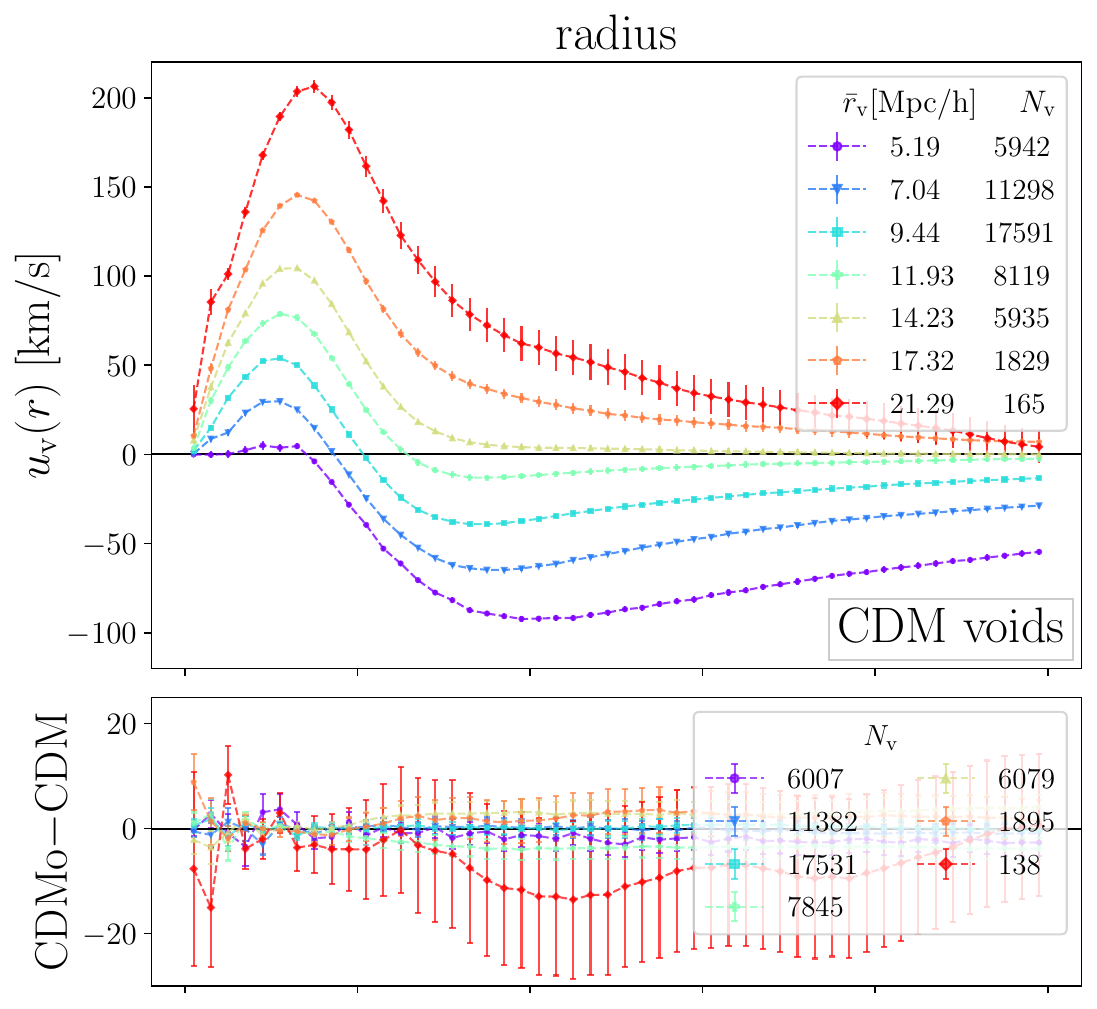}

                               \includegraphics[trim=0 5 0 5, clip]{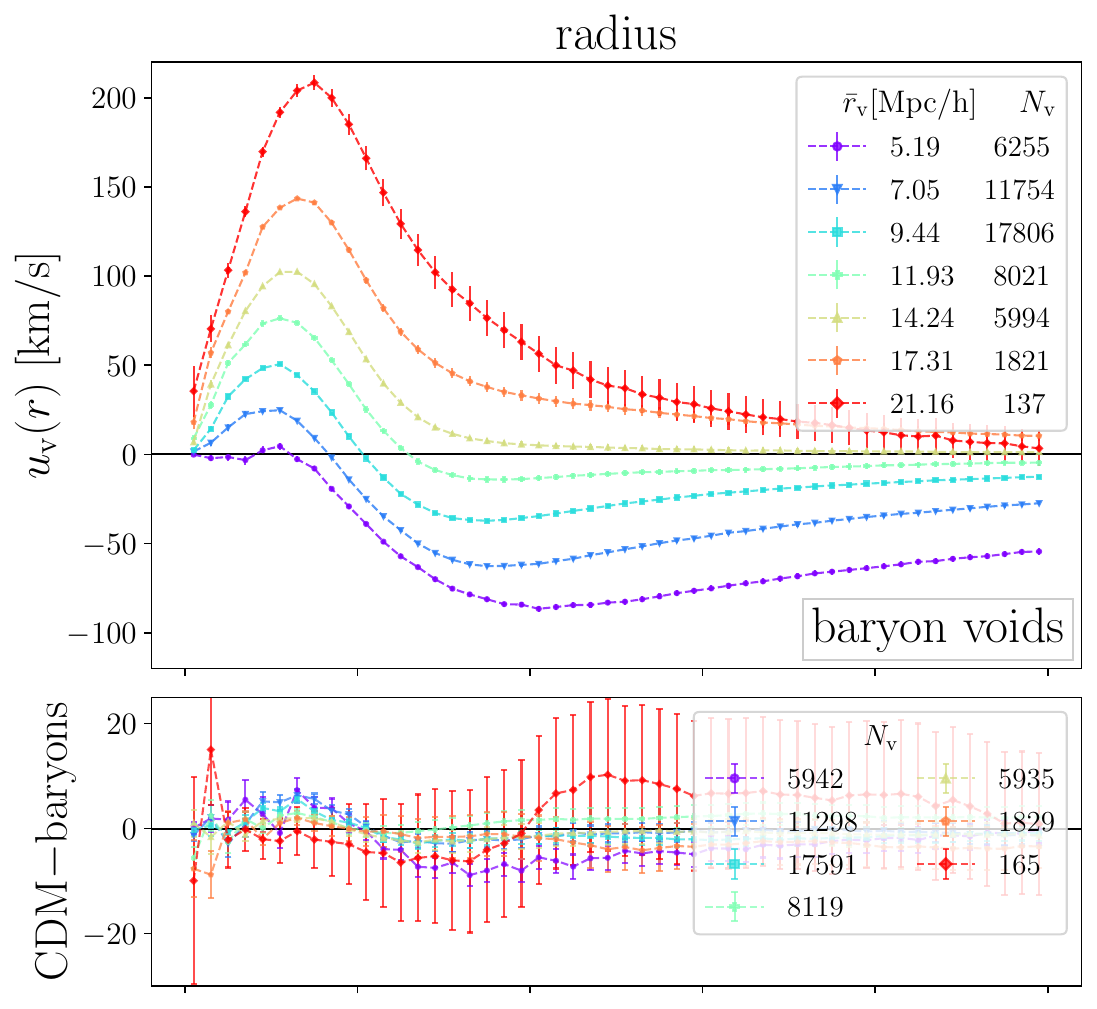}}

               \resizebox{\hsize}{!}{

                               \includegraphics[trim=0 10 0 5, clip]{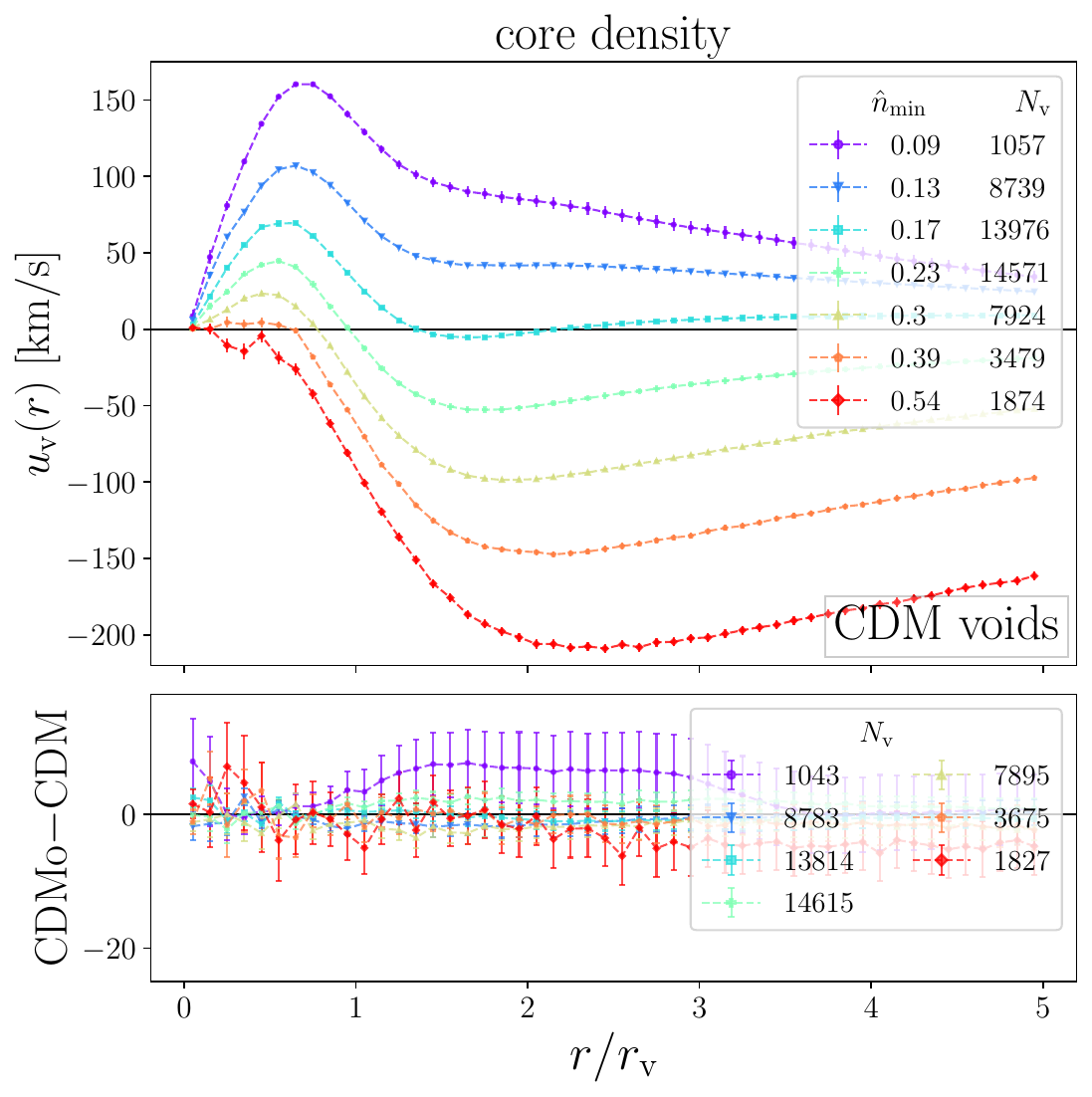}

                               \includegraphics[trim=0 10 0 5, clip]{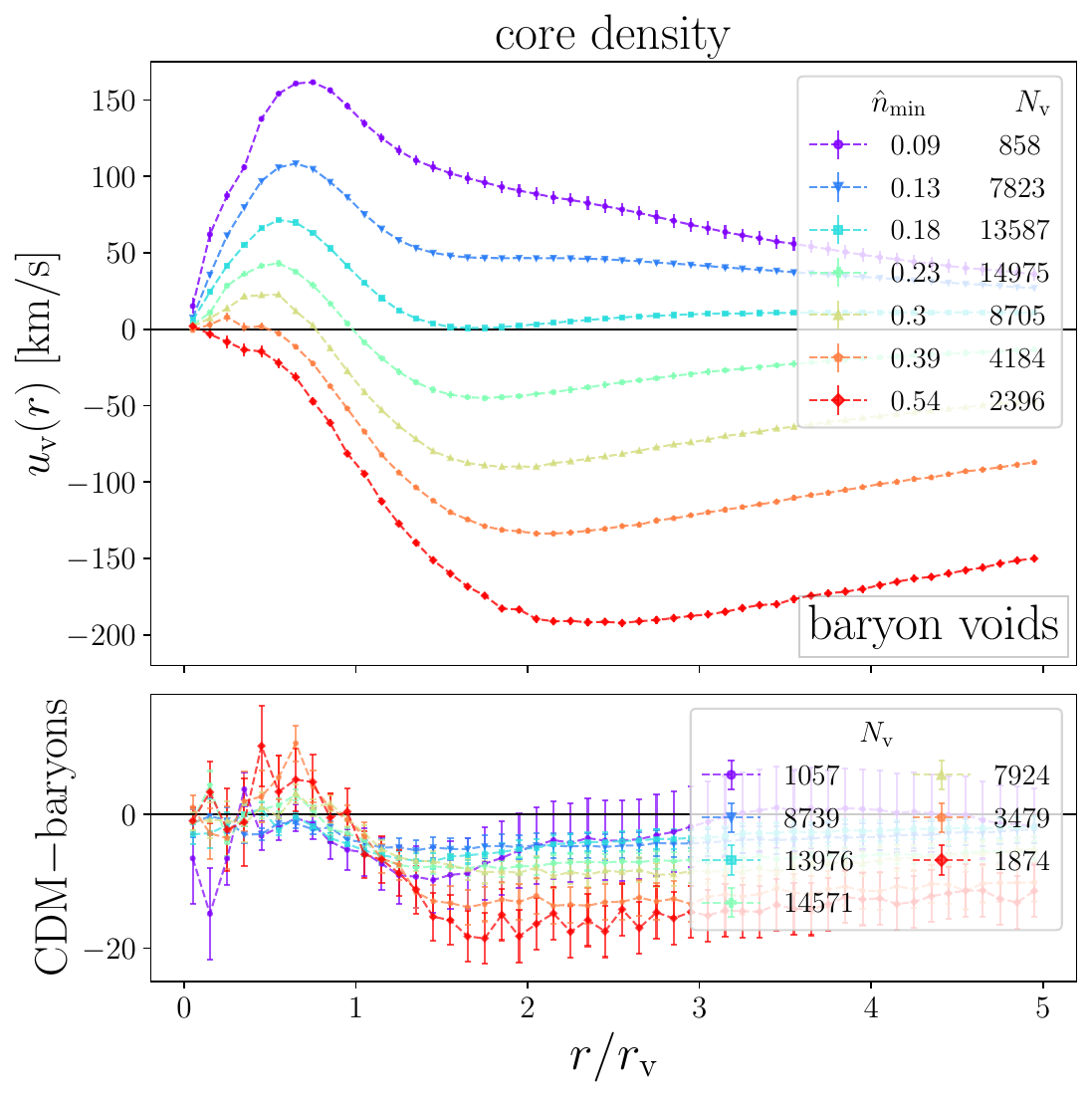}}

               \caption{Same as figure~\ref{fig_density_matter}, but for stacked velocity profiles. }

               \label{fig_velocity_matter}

\end{figure}

Figure~\ref{fig_density_matter} depicts the matter density profiles of CDM voids from the \hydro{} run on the left, with differences compared to CDMo voids from the \DMo{} simulation in the lower panels, while profiles of baryon voids are depicted on the right, with differences compared to CDM voids below, both from the \HR{} \hydro{} simulation. In each case, densities increase outward from the void center, with clear compensation walls present around $r = r_\void$. The largest voids, as well as voids with lowest core densities, have more shallow compensation walls, though a sharp decline in the density increase clearly marks the void boundaries, with densities close to the mean $\Bar{\rho}$. A correlation between core density and void radius is evident, since profiles of small voids closely resemble profiles of voids with high core density, and vice versa.

When comparing CDM voids between \hydro{} and \DMo{} simulations on the left of figure~\ref{fig_density_matter}, we observe that profiles and void numbers of each bin closely match. Most dissimilarities are still within error bars, namely in the bins of smallest and largest voids, where in the latter, the sample is quite sparse. Further differences in medium sized voids, e.g. the bin with $\Bar{r}_\void = 11.93$, are only on the order of $\lvert \Delta \rho \rvert \simeq 0.02$. In core density bins we notice the most substantial differences in bins of highest core density, with deviations on the order of $\lvert \Delta \rho \rvert \simeq 0.1$, although with large errors, and effects only arise past the compensation wall, outside of voids. Additional differences are present in the subset of voids with lowest core density, although still minor. 

More substantial effects are present in the comparison of CDM and baryon voids on the right of figure~\ref{fig_density_matter}. We note distinct profiles near the compensation walls of smallest voids, where CDM voids have noticeably higher walls, and near the centers of these small voids, where baryon voids exhibit higher densities. The void abundances from figure~\ref{fig_abundances_hr} support these conclusions, as we found baryon voids to have higher core densities, hence on average higher densities inside. Our findings from small voids indicate that baryons are distributed more evenly compared to CDM, supporting previous results~\cite{Paillas2017,Rodriguez2022}, which we investigate further in sections~\ref{subsec:baryon_BAR_CDM_voids} and~\ref{subsec:uhr}. These deviations quickly decrease with increasing void size and are almost nonexistent in the largest voids. Selecting voids by their core densities instead reveals more obvious differences, starting at a distance of around $r \geq 0.5 \, r_\void$. As expected, we do not observe major variations close to void centers, as bins in $\coreDens$ automatically limit the range of innermost densities. Voids of lowest core density have nearly indistinguishable profiles between CDM and baryon voids, but variations in profiles increase with increasing $\coreDens$, reaching values up to $\lvert \Delta \rho \rvert \simeq 0.25$ near the compensation walls of voids with highest core density, where once again, baryon voids have lower walls. These smaller compensation walls cover wide distances from void centers and deviations only vanish once the density is close to the mean $\Bar{\rho}$.

In the velocity profiles, presented in figure~\ref{fig_velocity_matter}, we see how CDM and CDMo voids have nearly identical profiles, except for deviations of order $\lvert \Delta u_\void \rvert \simeq 4 \, \kms$ for medium sized voids. Stronger deviations are only hinted at in the bin of largest voids, although even with the increased number of tracers used in the profiles calculation, error bars are still significant due to the sparse void sample, and the large variation of individual void profiles~\cite{Schuster2023}. This follows expectations, since the density profiles of CDM and CDMo are almost identical and linear mass conservation holds~\cite{Schuster2023}, which ultimately results in insignificant differences. Stacked bins in core density present similar results. Only voids of smallest $\coreDens$, typically amongst the largest voids, hint at significant deviations. However, error bars still reach $\Delta u_\void  \simeq 0$, although now with a larger void sample compared to the largest $r_\void$ bin. 

In contrast, in the baryon void profiles of figure~\ref{fig_velocity_matter} we observe stronger effects in core density bins, while bins in $r_\void$ behave similar to the ones of CDM and CDMo voids, except for smallest voids, with $\lvert \Delta u_\void \rvert \simeq 10 \, \kms$, and for largest voids due to the small sample. The former result is as expected, since these small voids also experience small variations between the density profiles of CDM and baryons. Since matter tracers should still obey linear mass conservation (see reference~\cite{Schuster2023}), differences in density profiles predict small deviations in their velocity profiles. Contrary to bins in void radius, the core density bins exhibit more significant shifts between CDM and baryons at a given value of $\coreDens$. Within voids the velocities behave similarly, but once past $r = r_\void$, deviations arise in all bins. As in the density profiles, the variation is most substantial for voids of highest $\coreDens$ and decrease with decreasing core density. While these might be physical effects, it is also likely that they are spurious. Since we are stacking velocity profiles of voids with a larger distribution in $r_\void$ when selecting voids by their core density, and baryon voids typically have higher core densities, this can lead to deviations due to the correlation between both void properties~\cite{Schuster2023} and the strong dependence of velocity profiles on void radii. 
 
 In conclusion, the density, as well as velocity profiles of CDM and CDMo voids are extremely similar between \hydro{} and \DMo{} simulations, while substantial differences arise in the comparison of CDM and baryon voids. In this latter case we note that the core density determines the magnitude of deviations more strongly than the void radius, even though both are correlated.

\subsection{Halo voids \label{subsec:baryon_halo_voids}}

\begin{figure}[t]

               \centering

               \resizebox{\hsize}{!}{

                               \includegraphics[trim=7 5 0 5, clip]{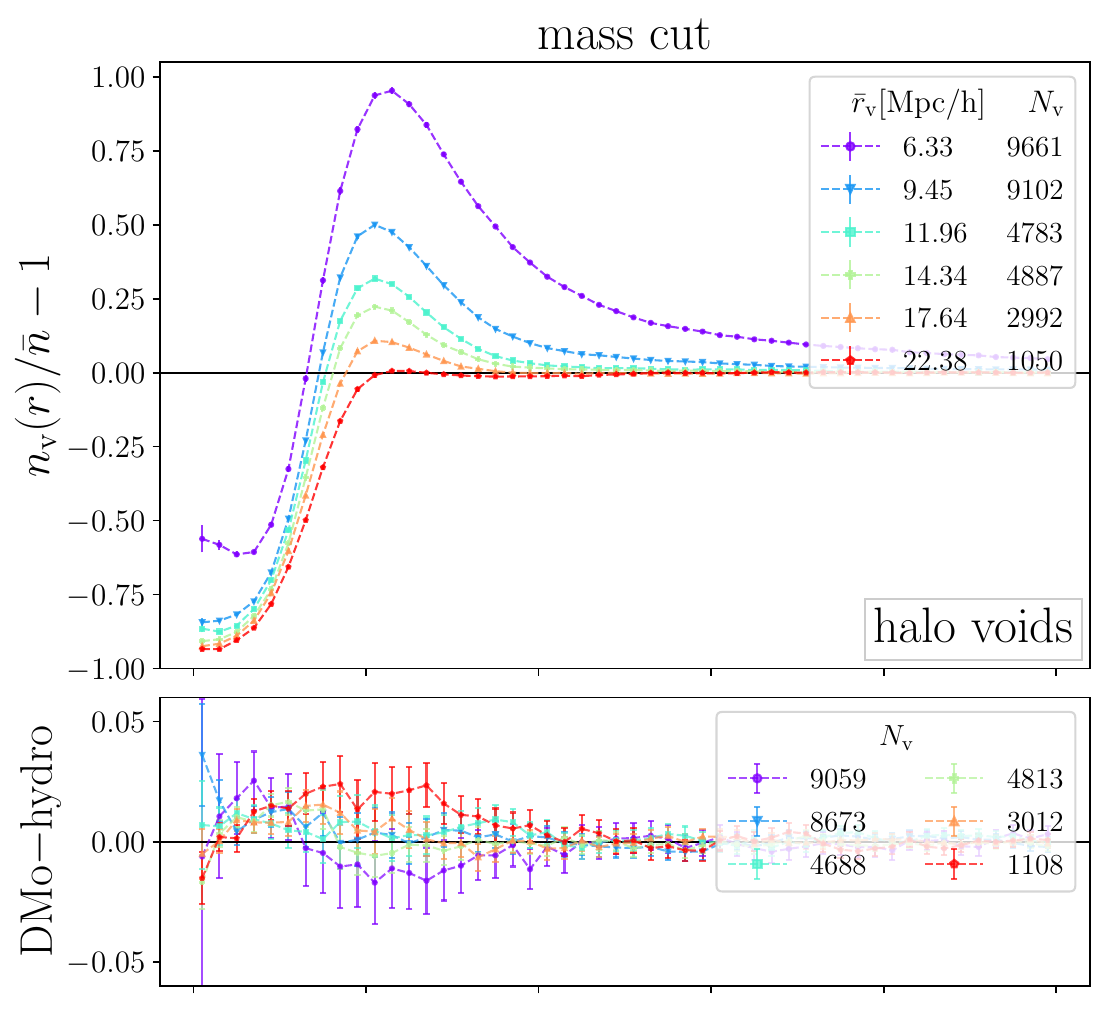}

                               \includegraphics[trim=0 5 0 5, clip]{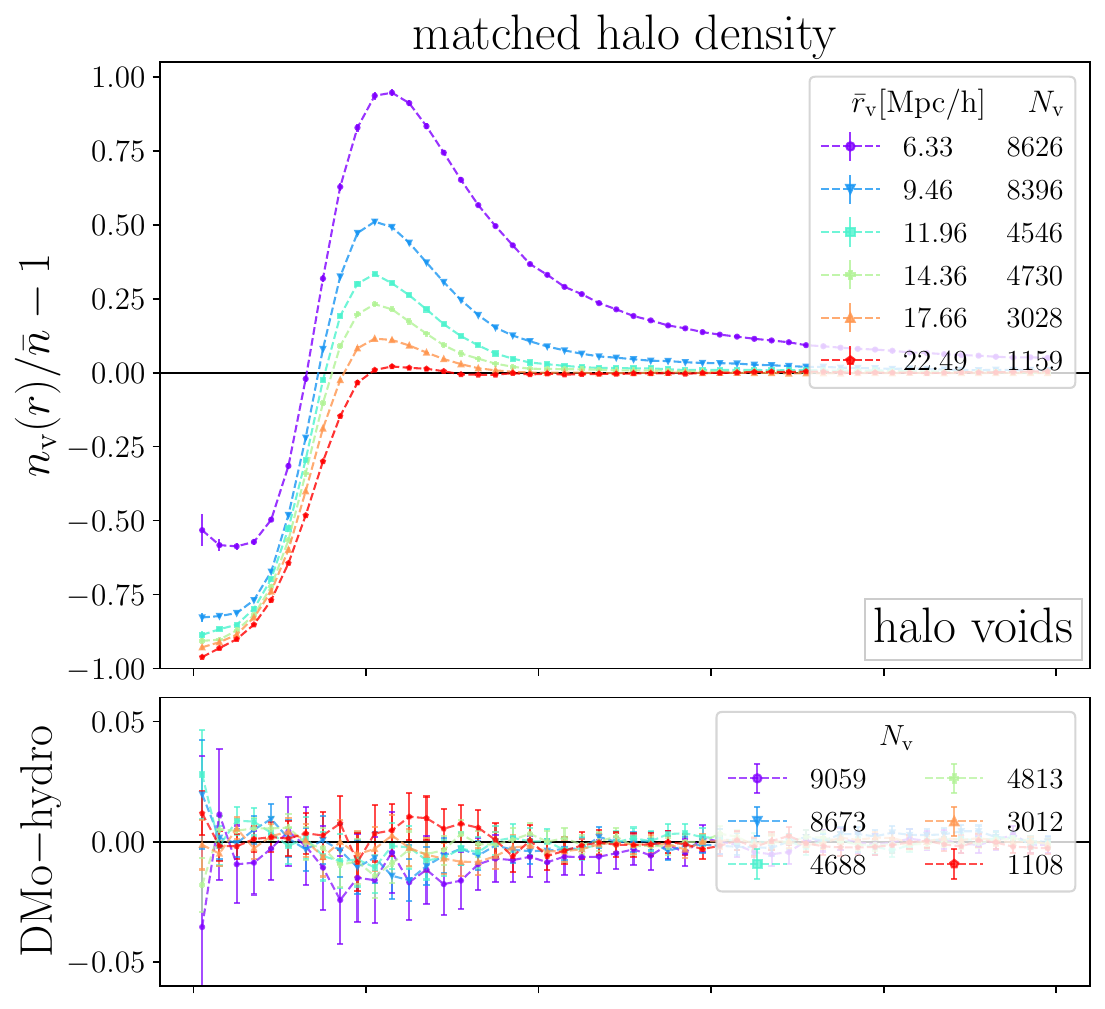}}

               \resizebox{\hsize}{!}{

                               \includegraphics[trim=0 10 0 5, clip]{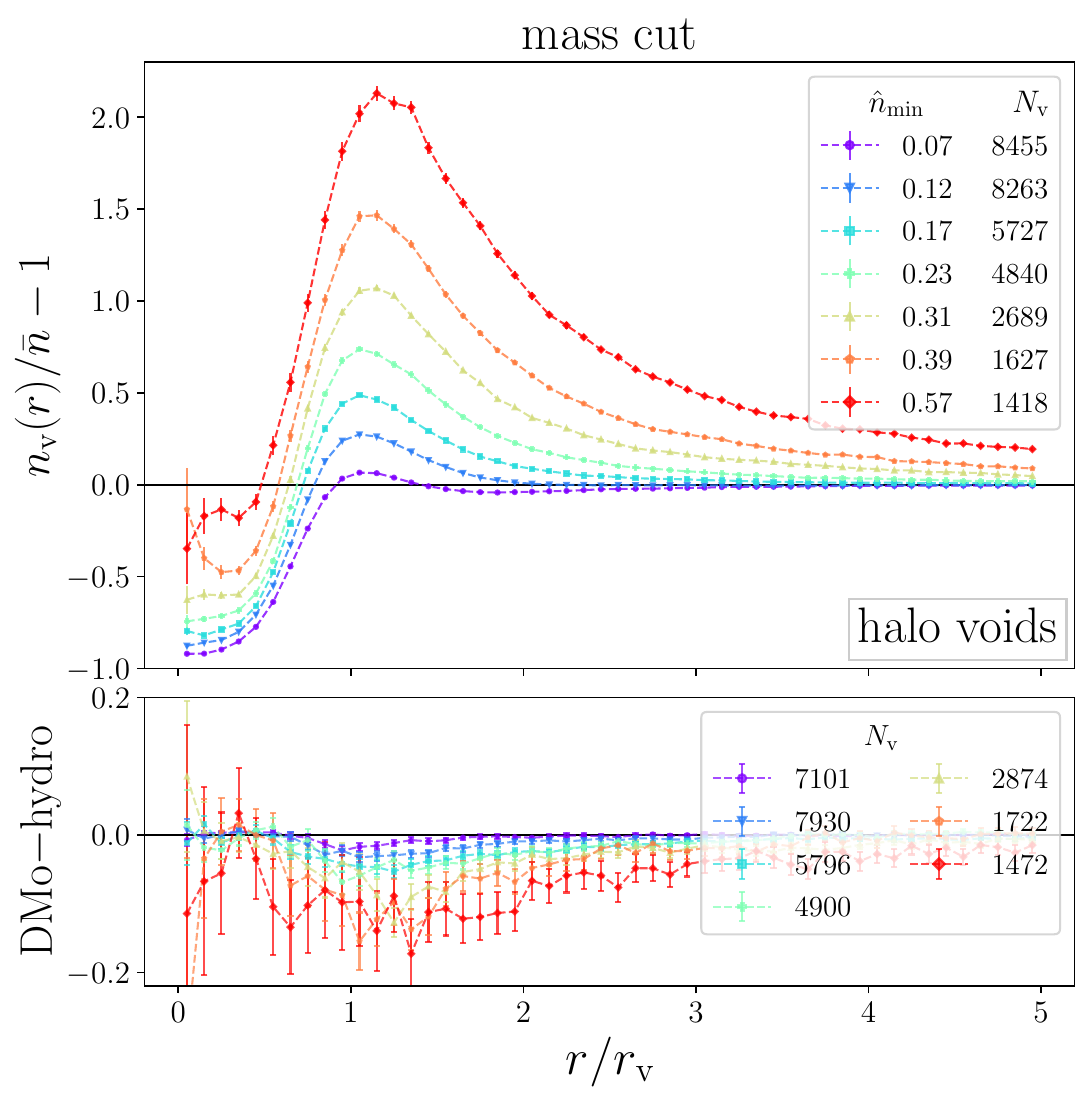}

                               \includegraphics[trim=0 10 0 5, clip]{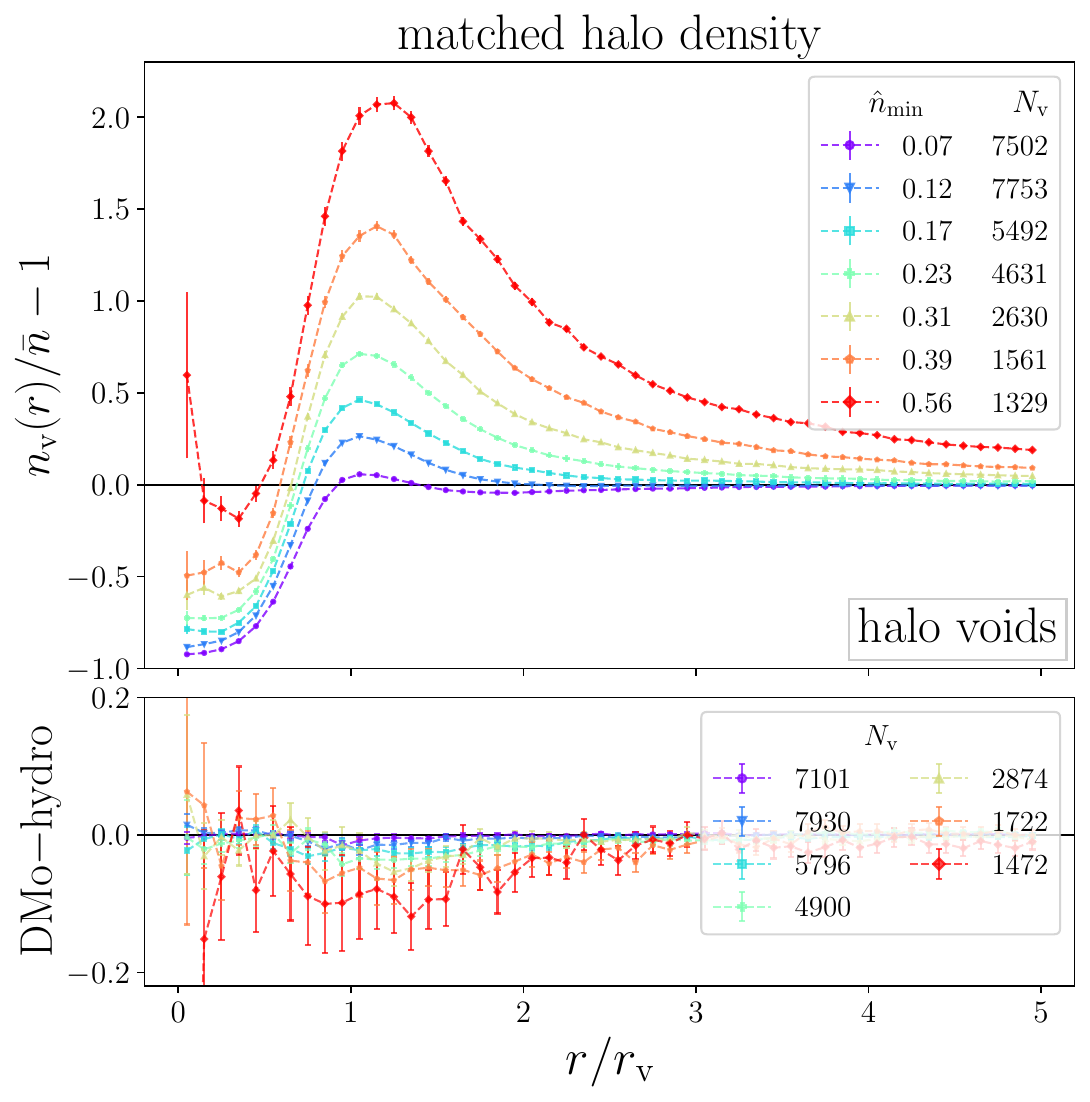}}

               \caption{Stacked density profiles of halo voids in the \HR{} simulations, for voids identified in tracers with mass cuts (left) and matched halo densities (right), in bins of void radius (top) and core density (bottom). Upper panels depict the profiles from the \hydro{} simulation and lower panels the differences between \DMo{} and \hydro{} runs.}

               \label{fig_density_halos}

\end{figure}

\begin{figure}[t]

               \centering

               \resizebox{\hsize}{!}{

                               \includegraphics[trim=7 5 0 5, clip]{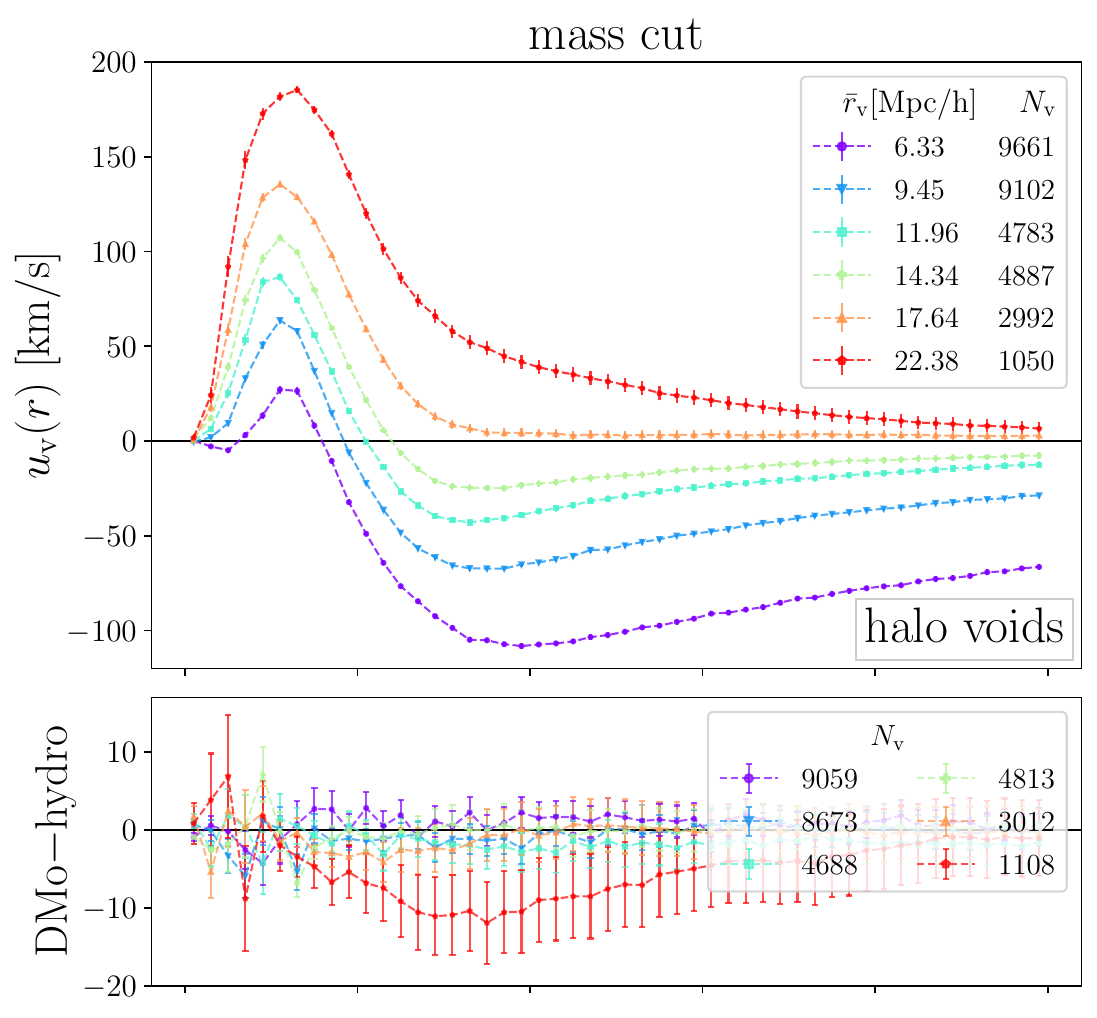}

                               \includegraphics[trim=0 5 0 5, clip]{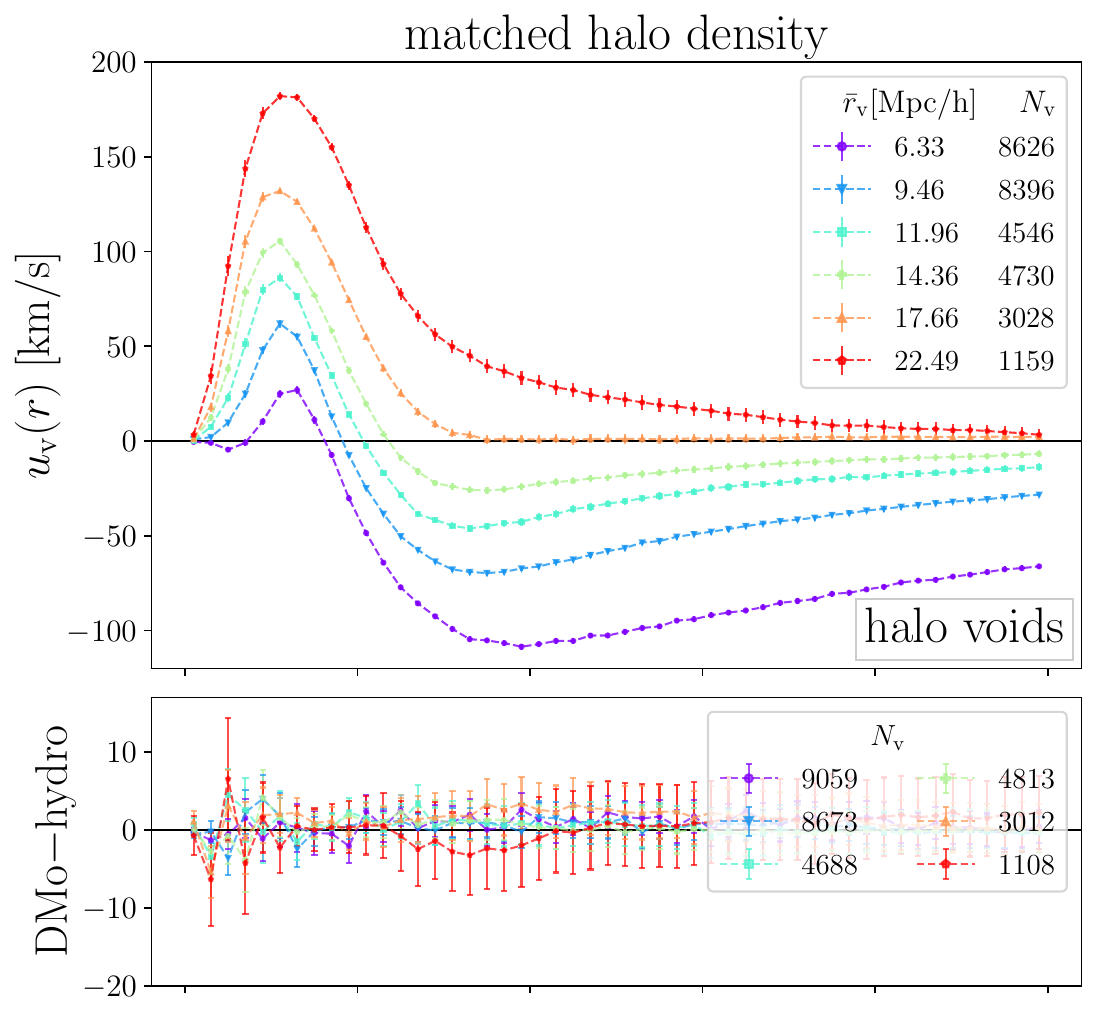}}

               \resizebox{\hsize}{!}{

                               \includegraphics[trim=0 10 0 5, clip]{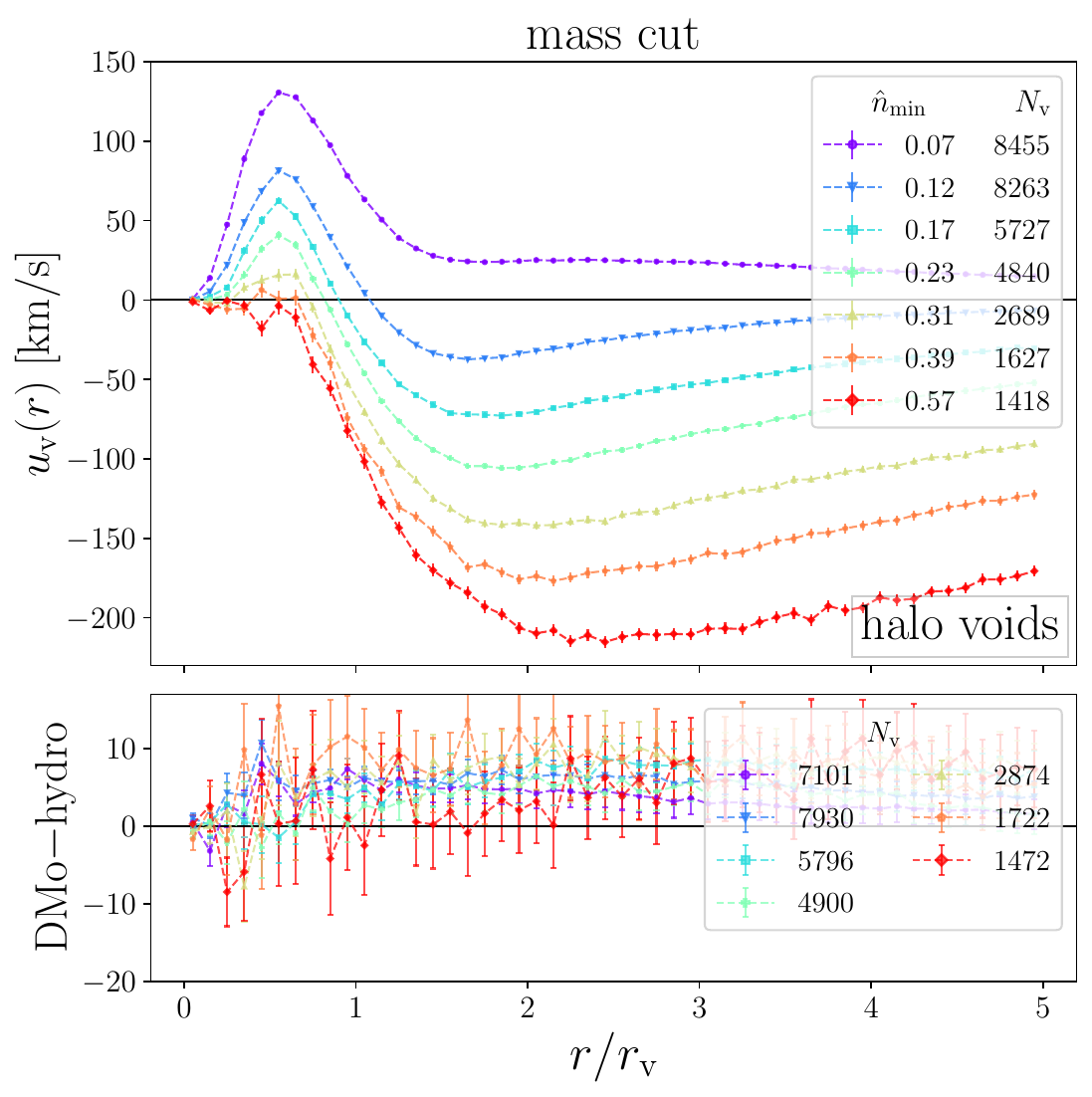}

                               \includegraphics[trim=0 10 0 5, clip]{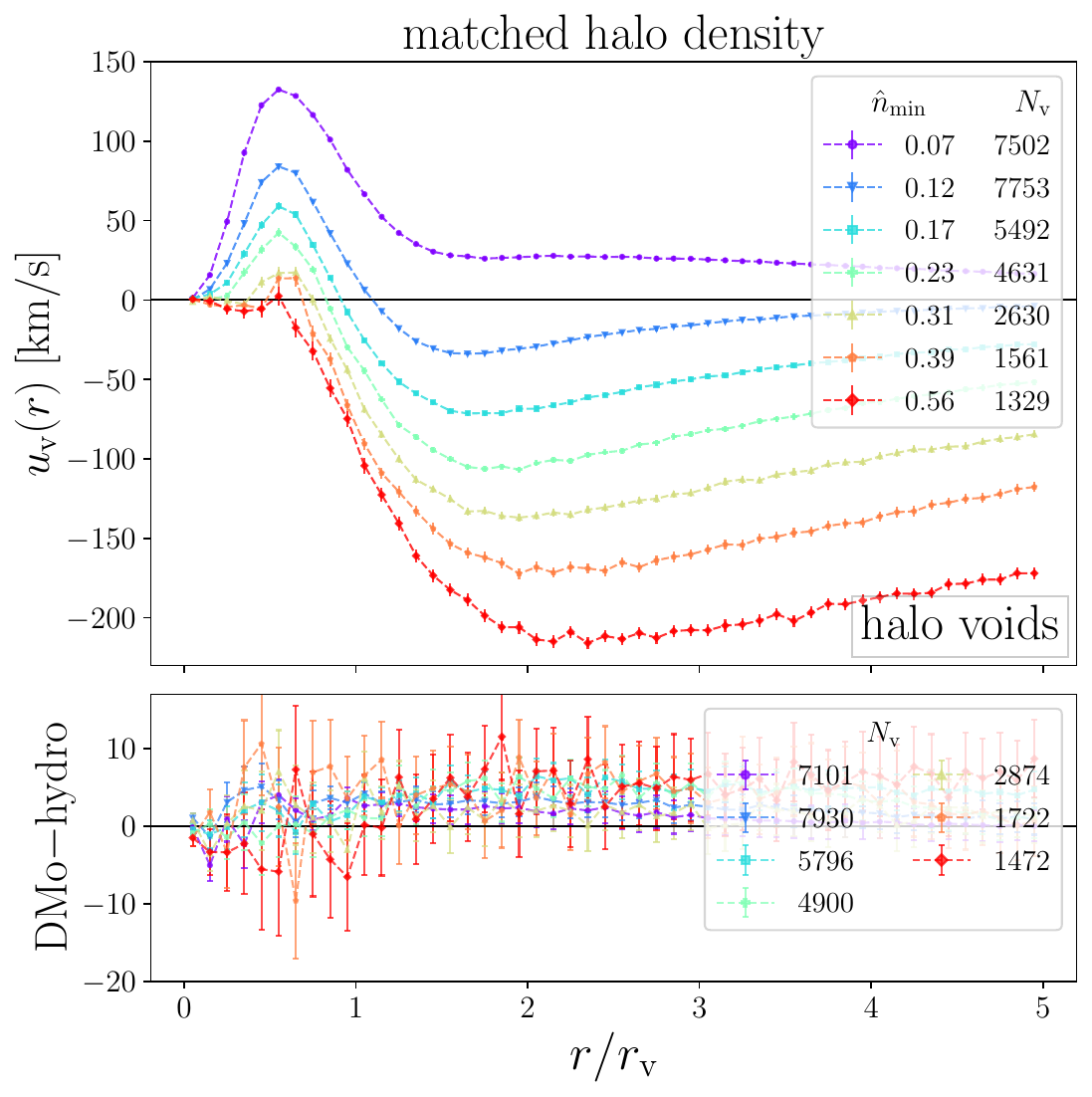}}

               \caption{Same as figure~\ref{fig_density_halos}, but for velocity profiles.}

               \label{fig_velocity_halos}

\end{figure}

After examining matter voids in section~\ref{subsec:baryon_matter_voids}, we now focus on voids more relevant for observations. Figures~\ref{fig_density_halos} and~\ref{fig_velocity_halos} depict the number density and velocity profiles of halo voids, for those obtained from halo mass cuts (left), as well as matched halo densities (right) between the \DMo{}/\hydro{} simulations. We present profiles of halo voids from the \HR{} \hydro{} run in the upper panels, while depicting differences between \DMo{} and \hydro{} runs in the lower panels. The bins in void radius are identical to those used for matter voids in section~\ref{subsec:baryon_matter_voids}, except for the two bins of smallest matter voids that are combined into one for halo voids, due to their larger size and to reduce sparse sampling effects, while core density bins are identical.

The density profiles presented in void radius bins (top) in figure~\ref{fig_density_halos} reveal small deviations near the compensation wall, where small voids from the \hydro{} run have higher walls, but the opposite is true for large voids. However, the most pronounced differences are only on the order of $\lvert \Delta n_\void \rvert  \lesssim 0.025$, and decrease even further for matched densities. We additionally note that the largest bin of $r_\void$ exhibits the most substantial difference between mass cuts and matched density profiles near the void center, with lower densities for the latter. This follows expectations, as matched densities cut halos of lowest mass, which have a higher probability of residing inside voids compared to more massive halos~\cite{Schuster2023}.

Similar to matter void profiles we observe that deviations between \DMo{} and \hydro{} runs become most substantial when stacking halo voids by their core densities (bottom). Once more these deviations decrease after matching halo densities, although less severe than for $r_\void$ bins. Halo voids in the \hydro{} run exhibit up to $ \lvert \Delta n_\void \rvert \simeq 0.1 $ higher densities near their walls, but their inner densities are almost indistinguishable, except for the innermost bin from $0$ to $ 0.1 r_\void $, where errors are huge due to sparse sampling below the mean tracer separation. As for matter voids, baryonic effects in the density profiles of halo voids increase with increasing core density, reaffirming our previous result that a voids' core density is a stronger indication on the relevance of baryonic effects than its radius. Nevertheless, binning voids in stacks of their radii is more common in cosmological applications. Hence, some effects in $r_\void$ bins would have to be accounted for when comparing statistics from dark-matter-only simulations with observations. Although in practice, the most severe deviations in bins of $r_\void$ of order $\lvert \Delta n_\void \rvert \lesssim 0.025 $ are negligible, and halo void profiles from general dark-matter-only simulations can still be of use in comparisons with observations.

In the void radius bins of velocity profiles in figure~\ref{fig_velocity_halos} (top) we find that the largest set of halo voids in mass cut catalogs exhibit the strongest deviations. Similar results were found for CDM void profiles in figure~\ref{fig_velocity_matter}, where deviations were ascribed to small void numbers in this bin and to the variability of individual profiles. However, a much larger number of halo voids is present in this bin, so we expect our results to be more solid, although small voids only experience deviations of order $\lvert \Delta u_\void \rvert \lesssim 4 \, \kms$. Once again, matching halo densities eliminates deviations and most error bars are centered around zero differences. In contrast, when binning in core density (bottom), almost every bin of both mass cut and matched density catalogs experiences small deviations of order $10 \, \kms$, though with comparably large errors. Hence, we conclude that while small differences in velocities might be present, they are insignificant, especially in $r_\void$ bins of matched density voids.

\begin{figure}[t!]

               \centering

               \resizebox{\hsize}{!}{

                               \includegraphics[trim=7 5 0 5, clip]{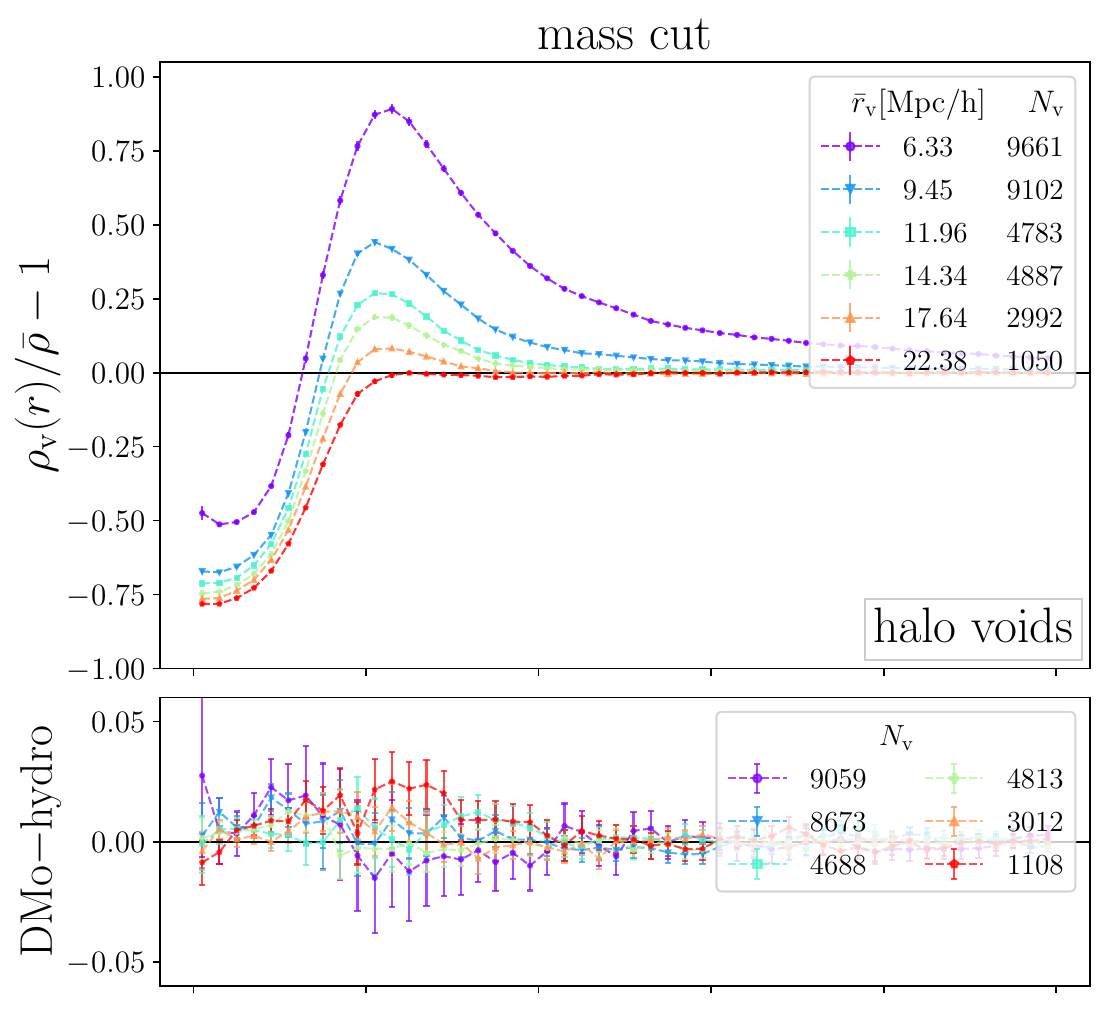}

                               \includegraphics[trim=0 5 0 5, clip]{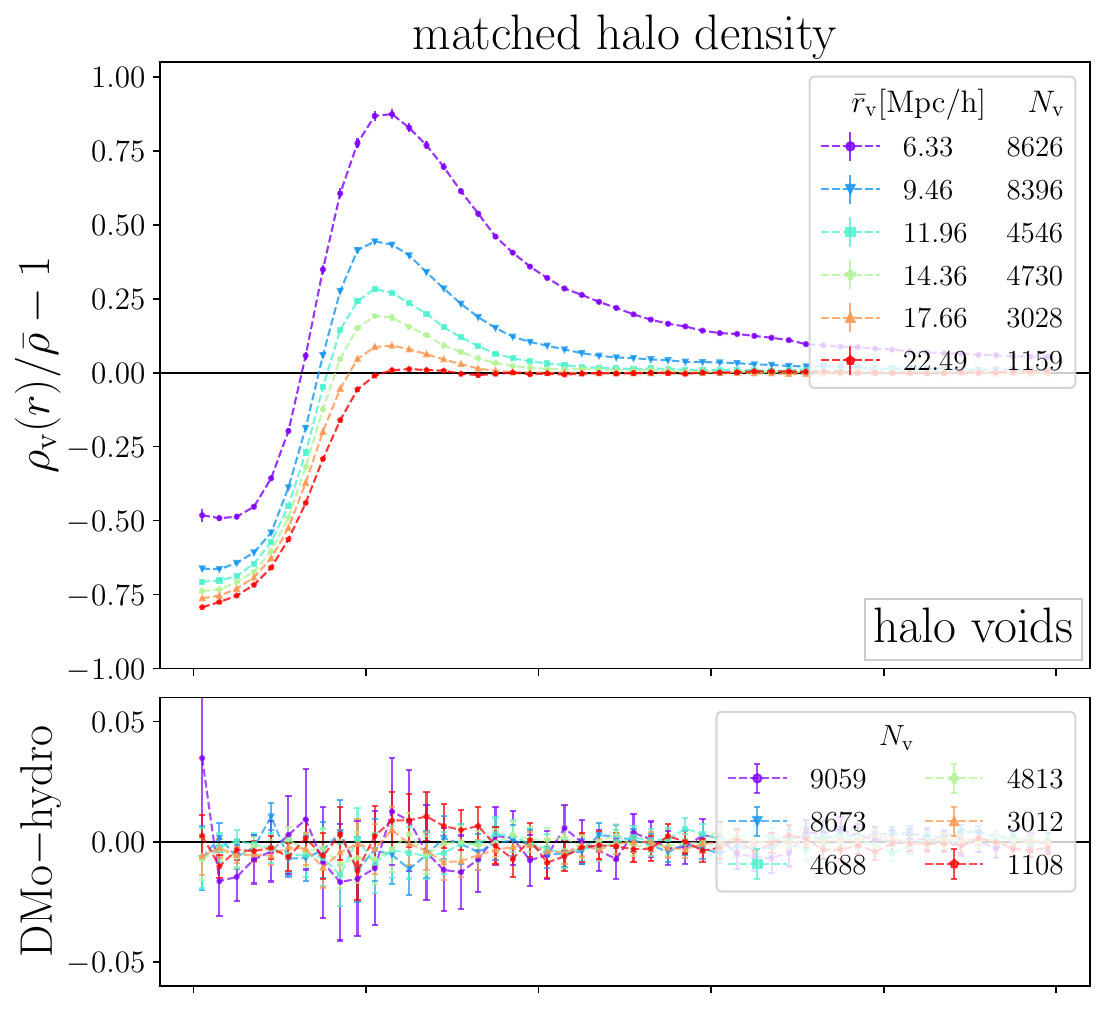}}

               \resizebox{\hsize}{!}{

                               \includegraphics[trim=0 10 0 5, clip]{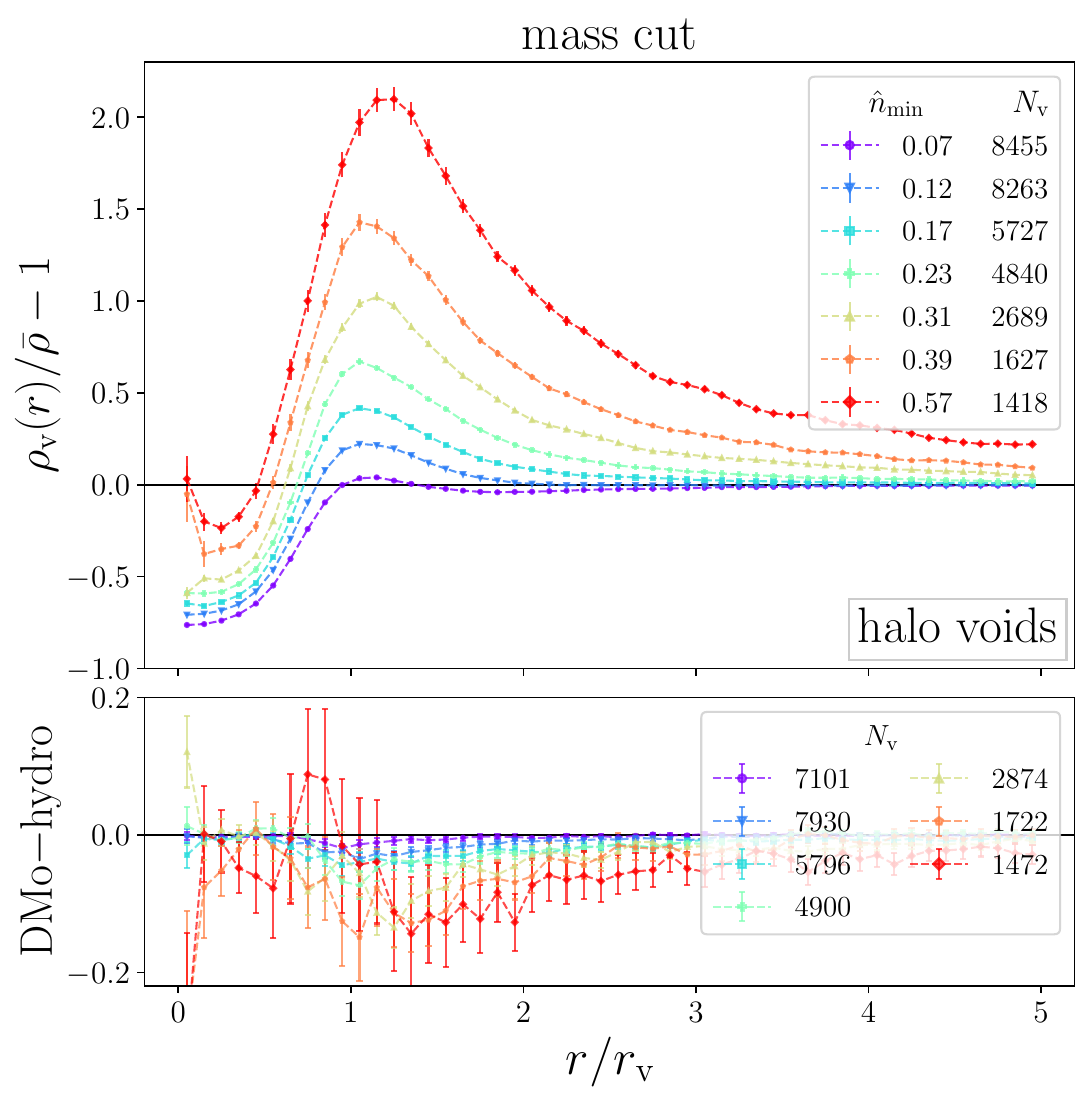}

                               \includegraphics[trim=0 10 0 5, clip]{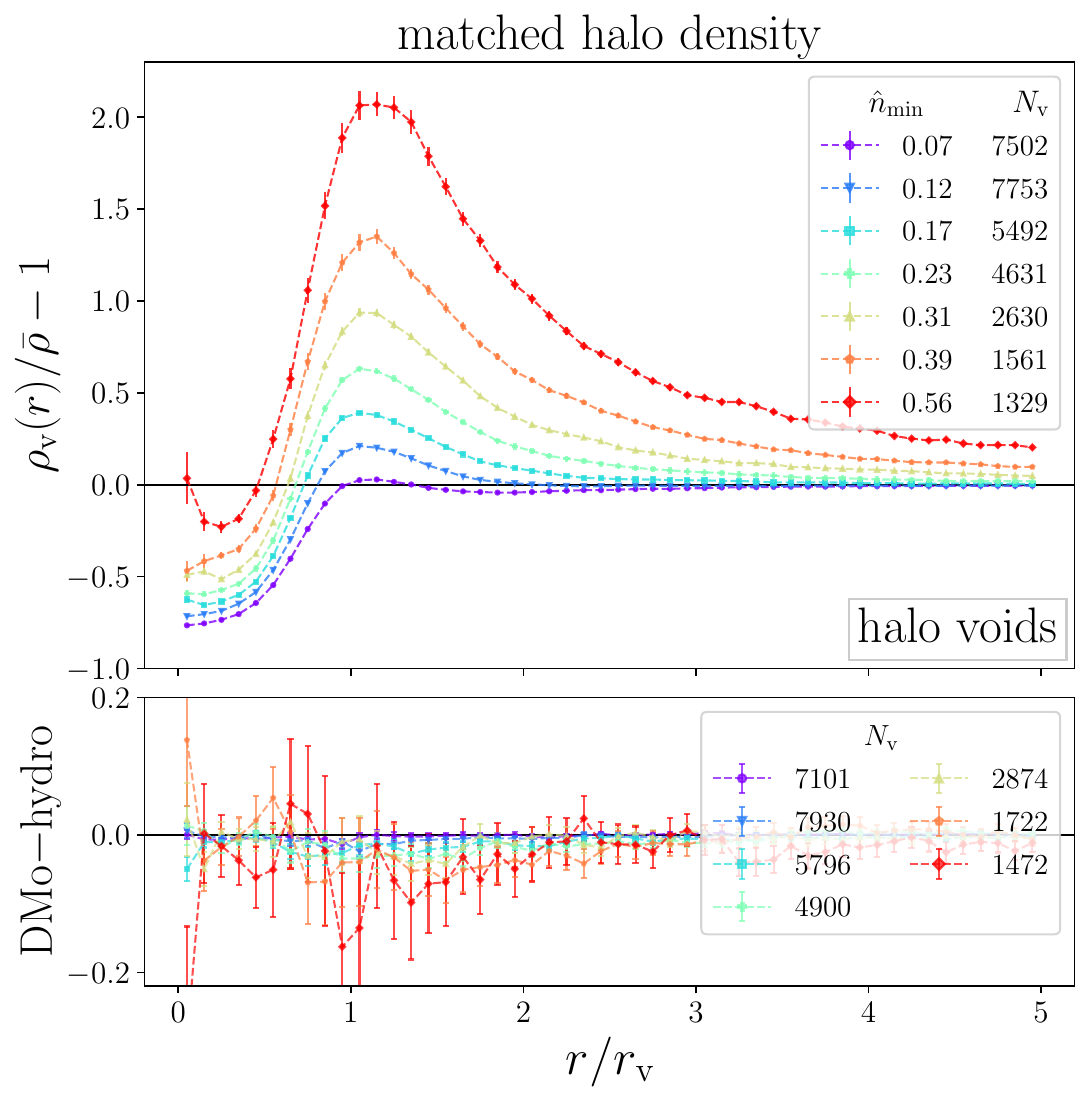}}

               \caption{Stacked CDM density profiles around halo voids in the \HR{} simulations, for voids found in tracers with mass cuts (left) and matched halo densities (right), in bins of void radius (top) and core density (bottom). Upper panels present the distribution of CDM around halo voids in the \hydro{} simulation, with  differences between \DMo{} and \hydro{} simulations in the lower panels.  }
 
               \label{fig_density_halo_matter}

\end{figure}

Additional tests at higher halo mass cuts ($M_\halo \geq 10^{12} \, \Msun$, as in \MR{}) for the void identification revealed no significant effects in halo void profiles, though in \HR{}, statistics are much sparser at this mass cut. For current state-of-the-art galaxy surveys, such as the Euclid mission~\cite{Hamaus2022}, we do not see the need in accounting for baryonic effects in the profiles of halo voids, as the \MR{} simulation is the closest match in the expected tracer densities and neither at this resolution, nor at identical mass cut in \HR{} do we find significant effects. We expect this to also hold true for voids identified in observed galaxies.

\begin{figure}[t!]

               \centering

               \resizebox{\hsize}{!}{

                               \includegraphics[trim=7 5 0 5, clip]{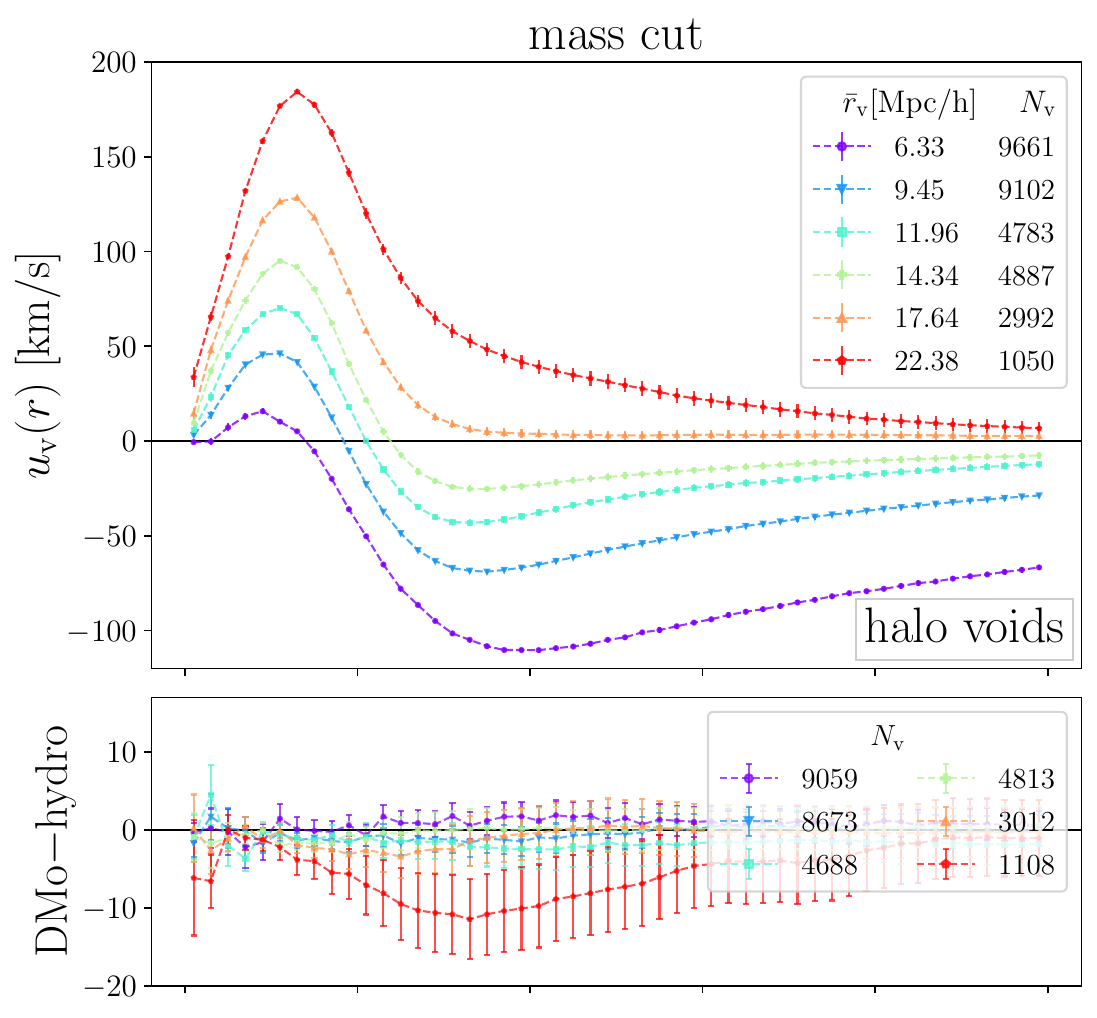}

                               \includegraphics[trim=0 5 0 5, clip]{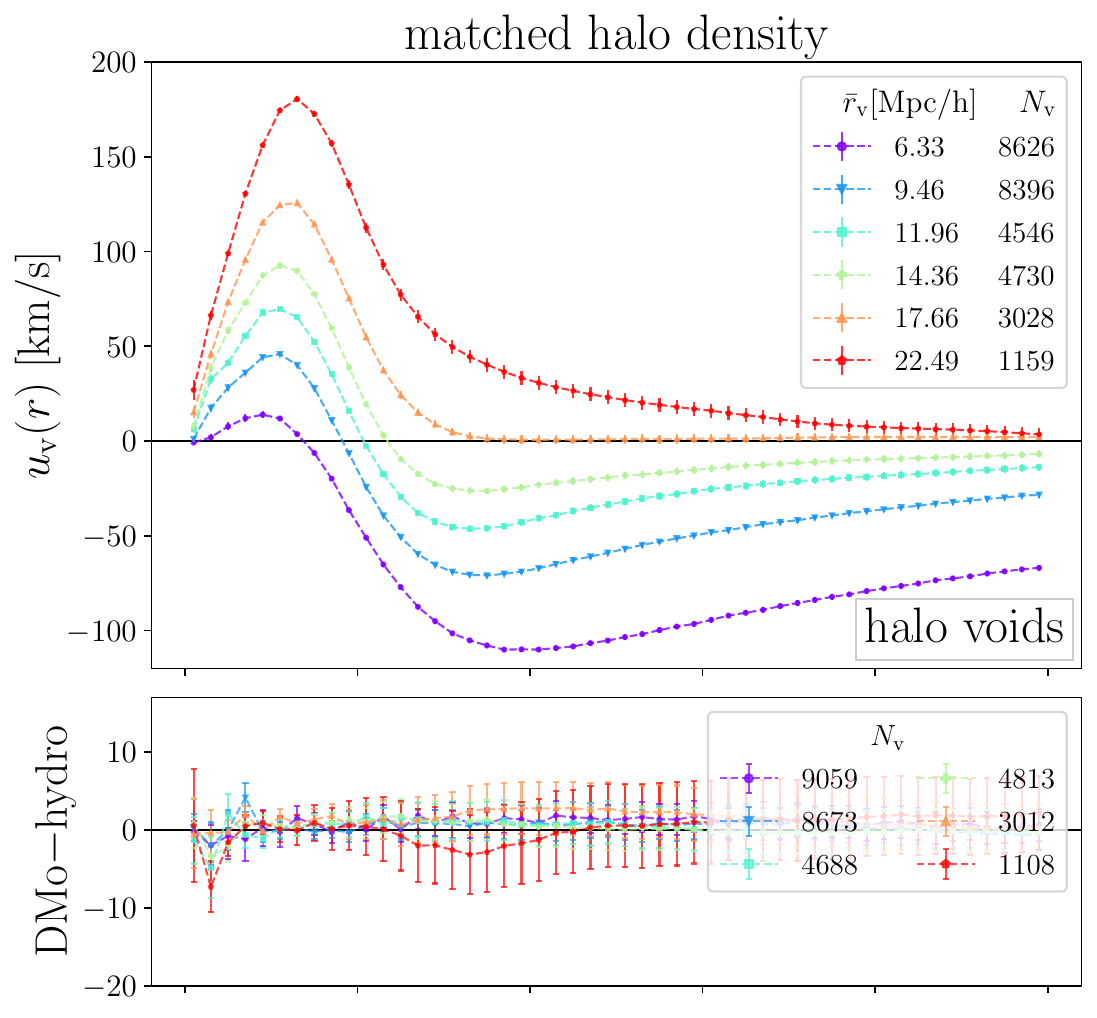}}

               \resizebox{\hsize}{!}{

                               \includegraphics[trim=0 10 0 5, clip]{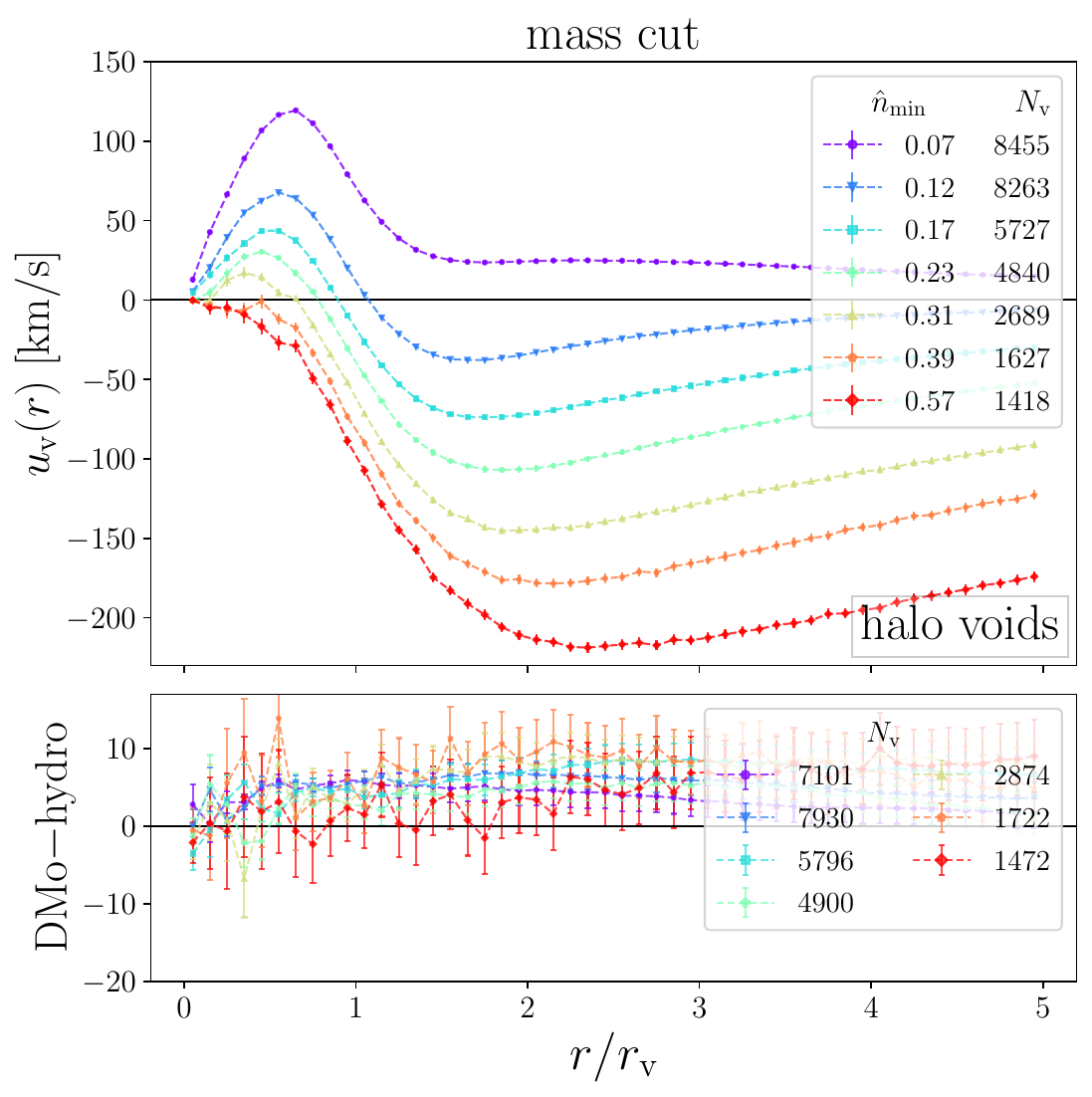}

                               \includegraphics[trim=0 10 0 5, clip]{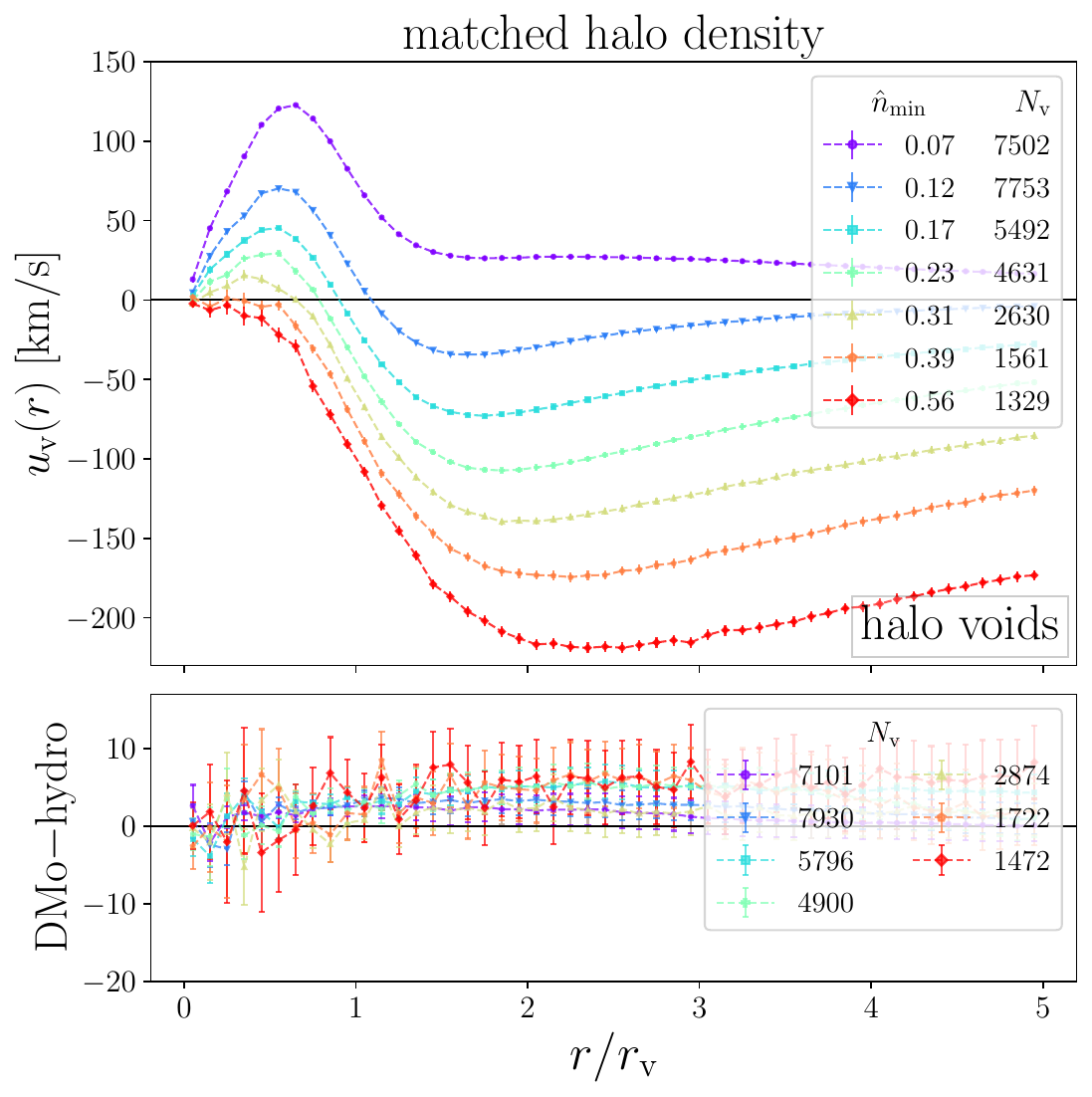}}

               \caption{Same as figure~\ref{fig_density_halo_matter}, but for velocity profiles. }
 
               \label{fig_velocity_halo_matter}

\end{figure}

\subsection{CDM profiles around halo voids \label{subsec:baryon_halo_voids_CDM}}

In order to analyse the impact of baryons on weak lensing studies around halo-defined voids, we investigate density, as well as velocity profiles of CDM around halo voids in figures~\ref{fig_density_halo_matter} and~\ref{fig_velocity_halo_matter} and compare statistics between \hydro{} and \DMo{} runs. As for matter voids, we calculate profiles using higher CDM density subsamplings. Comparing these profiles with previous halo void profiles from figures~\ref{fig_density_halos} and~\ref{fig_velocity_halos}, we find higher CDM densities within voids, while compensation walls exhibit lower densities. This effect is most obvious in the smallest voids and is caused by the halo bias~\cite{Sutter2014a,Pollina2017}. 

However, deviations between \hydro{} and \DMo{} runs in void radius bins (top) are extremely similar to those in figure~\ref{fig_density_halos} and once more vanish almost completely after matching halo densities. We find conforming results in core density bins (bottom). Even though the CDM- and halo density around halo voids differ (see above), deviations between \hydro{} and \DMo{} runs match in both cases and remain significant even for matched densities in bins of $\coreDens$.

The velocity profiles, depicted in figure~\ref{fig_velocity_halo_matter}, reveal almost identical velocities of CDM compared to halos (figure~\ref{fig_velocity_halos}), except for slightly higher CDM velocities in the innermost part of voids. We expect that this is not a genuine deviation in halo and CDM velocities, but instead a result from the much higher sampling rate of CDM. As noted in reference~\cite{Schuster2023}, individually stacked velocity profiles are biased towards profiles of small (more numerous) voids, of which many have small or zero velocities within due to sparse sampling, evident in figure~\ref{fig_velocity_halos}. Due to the higher CDM sampling, less profiles have almost zero velocity within voids, and we are able to retrieve the `true' velocities in void centers. In addition, these CDM velocities behave more linear inside, as expected from linear mass conservation~\cite{Schuster2023}. Similar to density profiles, deviations in velocity between \DMo{} and \hydro{} runs in figure~\ref{fig_velocity_halo_matter} are similar to those in figure~\ref{fig_velocity_halos}. We assess that baryonic effects seem to mostly impact the velocities of large voids and only when using mass cut catalogs, while matched halo densities reveal no substantial effects, except small deviations of $\lvert \Delta u_\void \rvert \lesssim 6 \, \kms$ in $\coreDens$ bins.

\subsection{Baryons and CDM around halo voids \label{subsec:baryon_BAR_CDM_voids}}

\begin{figure}[t]

               \centering

               \resizebox{\hsize}{!}{

                               \includegraphics[trim=7 5 0 5, clip]{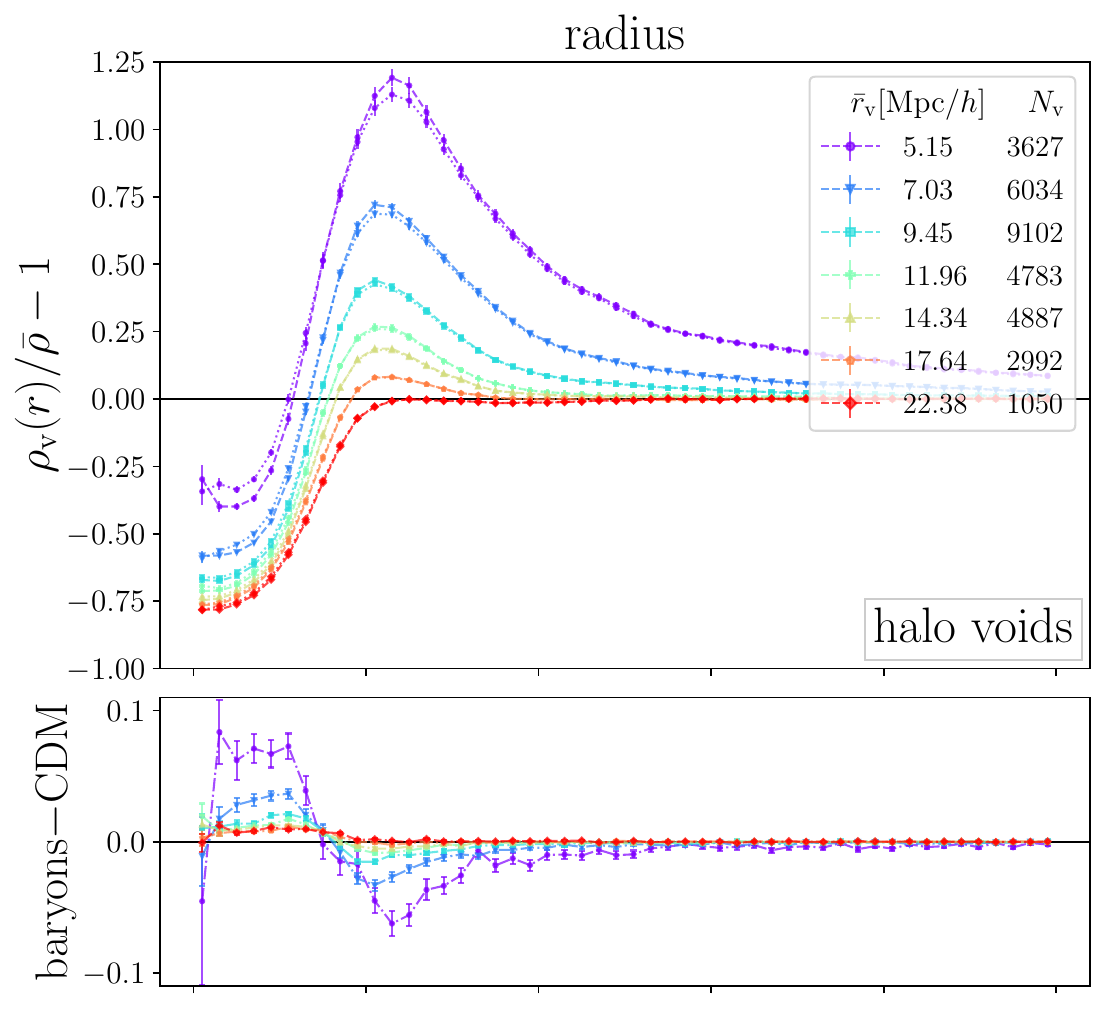}

                               \includegraphics[trim=0 5 0 5, clip]{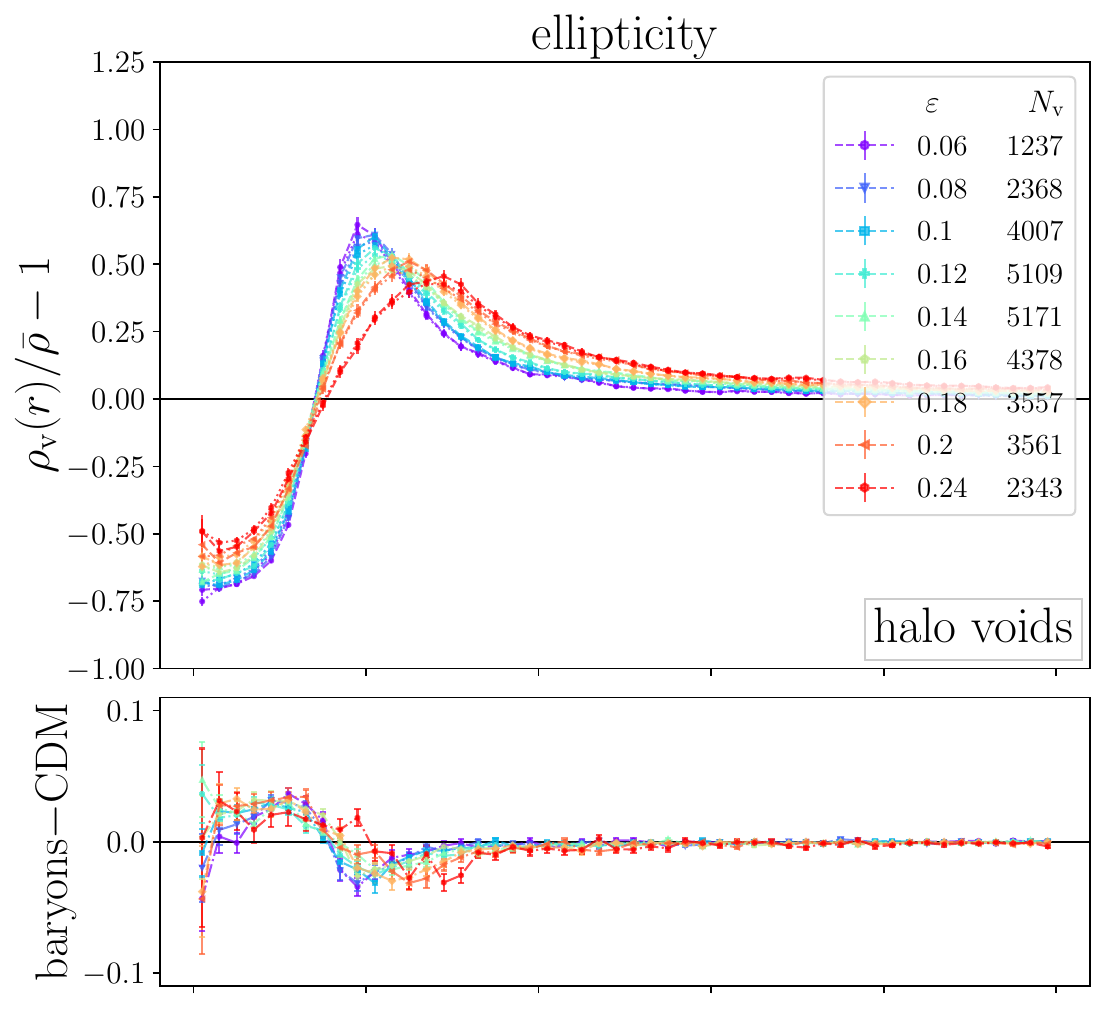}}

               \resizebox{\hsize}{!}{

                               \includegraphics[trim=0 10 0 5, clip]{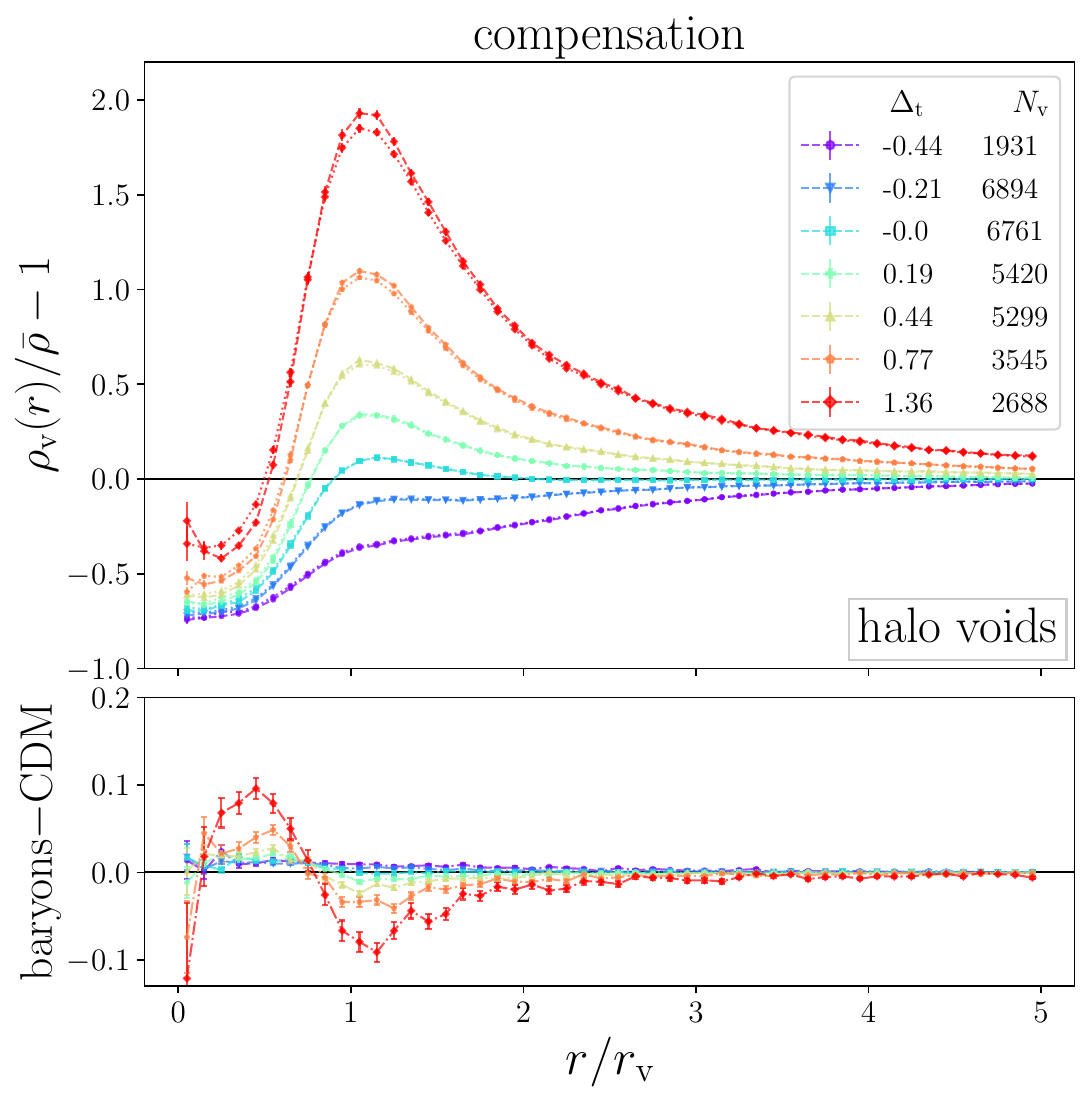}

                               \includegraphics[trim=0 10 0 5, clip]{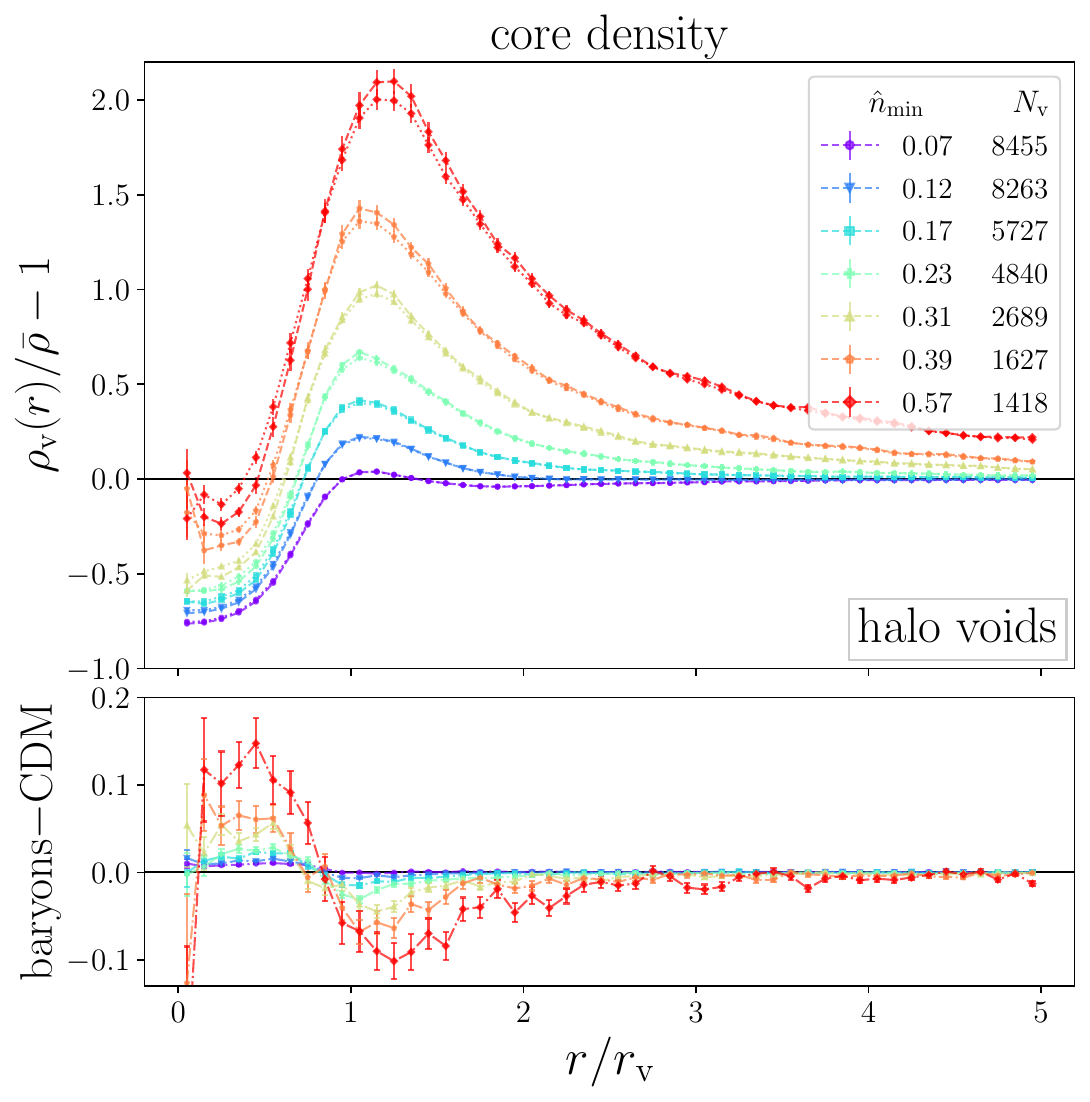}}

               \caption{Stacked matter density profiles around halo voids identified at `standard' mass cut ($M_\halo \geq 10^{11} \Msun{}$ ) in the \HR{} \hydro{} simulation in bins of radius (top left), ellipticity (top right), core density (bottom left), and compensation (bottom right). Upper panels depict stacked baryon (dotted) and CDM (dashed) density profiles, while lower panels present stacks of differences between baryons and CDM around individual voids.}

               \label{fig_density_halo_BARCDM}

\end{figure}

\begin{figure}[t]

               \centering

               \resizebox{\hsize}{!}{

                               \includegraphics[trim=7 5 0 5, clip]{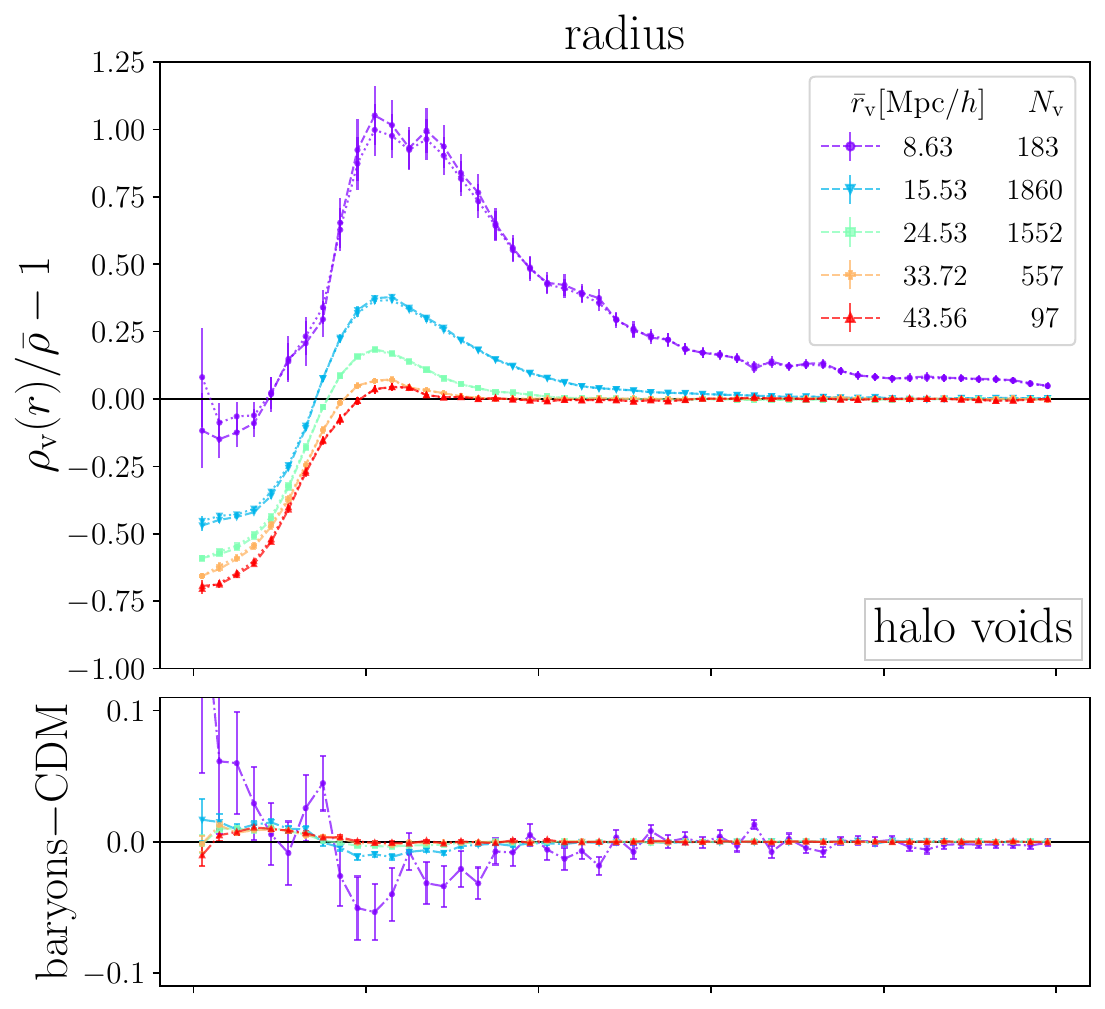}

                               \includegraphics[trim=0 5 0 5, clip]{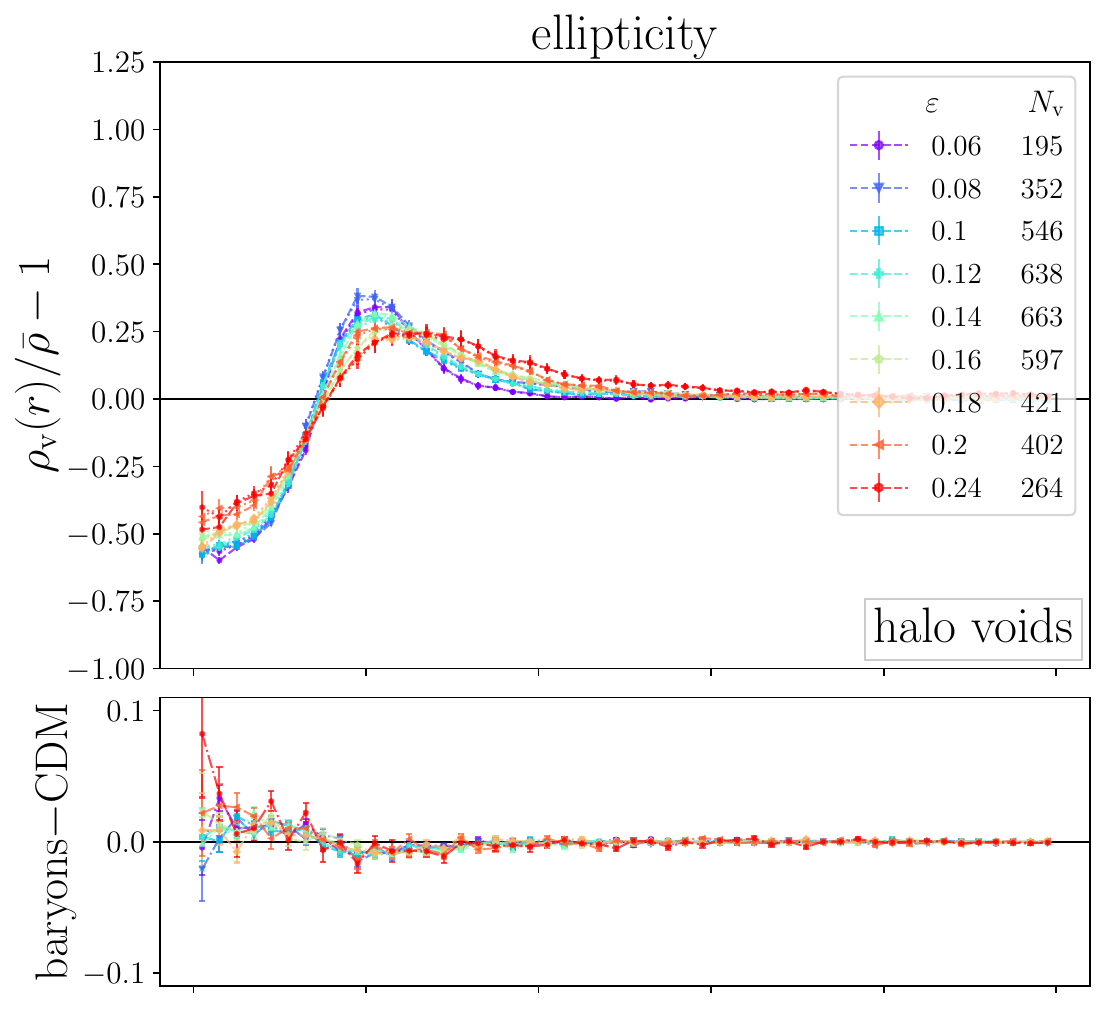}}

               \resizebox{\hsize}{!}{

                               \includegraphics[trim=0 10 0 5, clip]{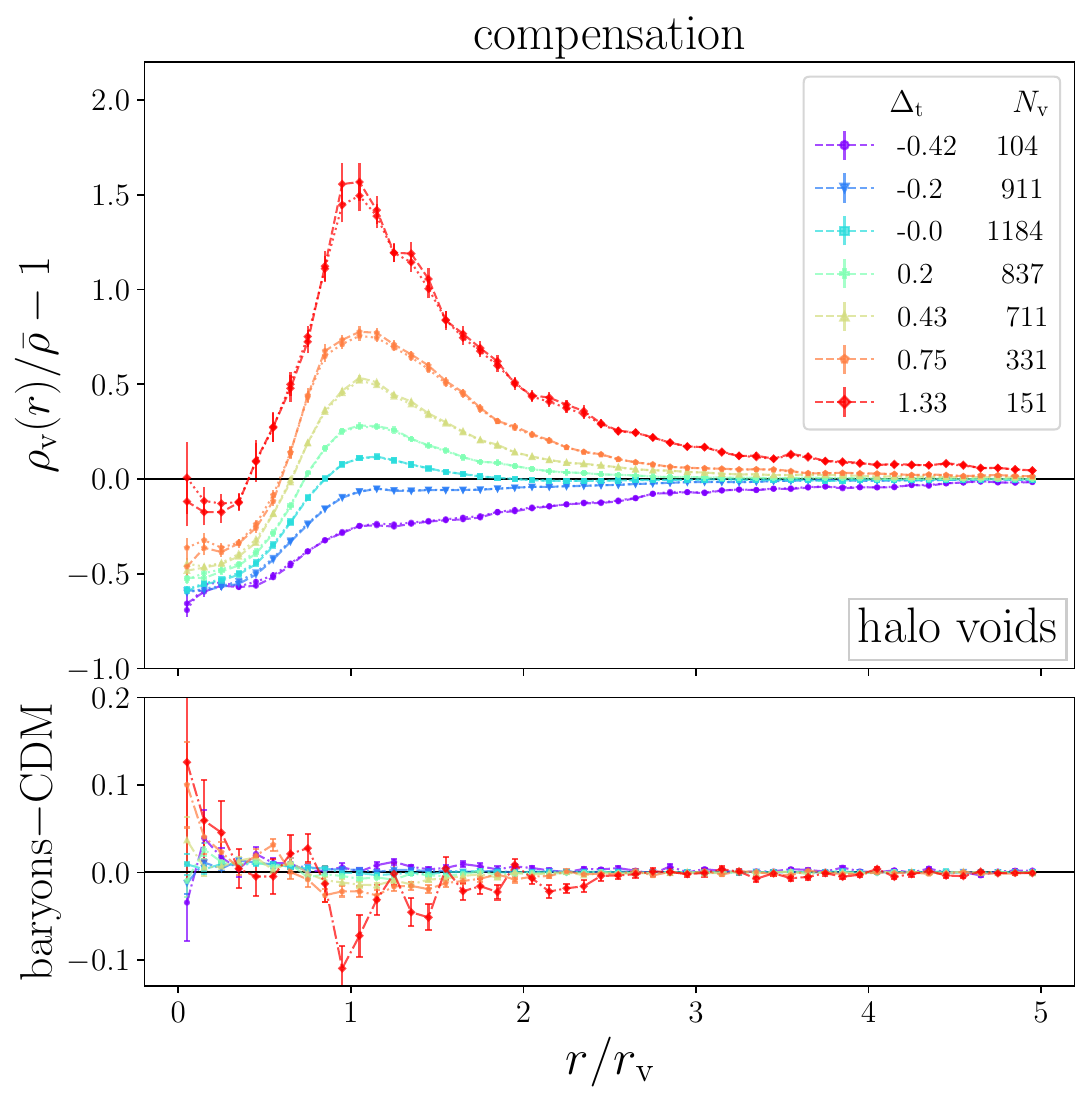}

                               \includegraphics[trim=0 10 0 5, clip]{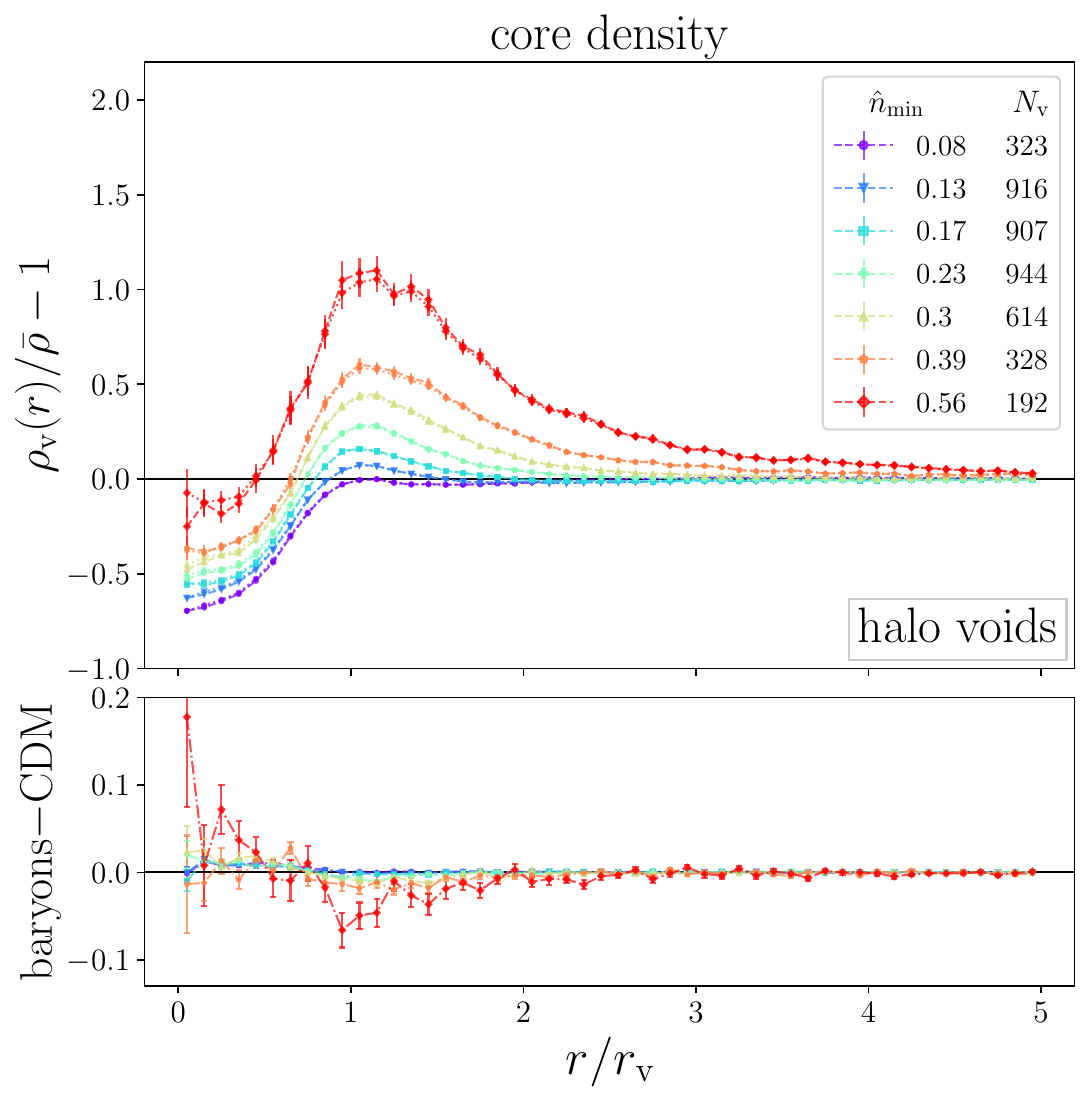}}

               \caption{Same as figure~\ref{fig_density_halo_BARCDM}, but for halo voids identified at mass cut $M_\halo \geq 10^{12} \Msun{}$.}

               \label{fig_density_halo_BARCDM_Mcut12}

\end{figure}

\begin{figure}[t]

               \centering

               \resizebox{\hsize}{!}{

                               \includegraphics[trim=7 5 0 5, clip]{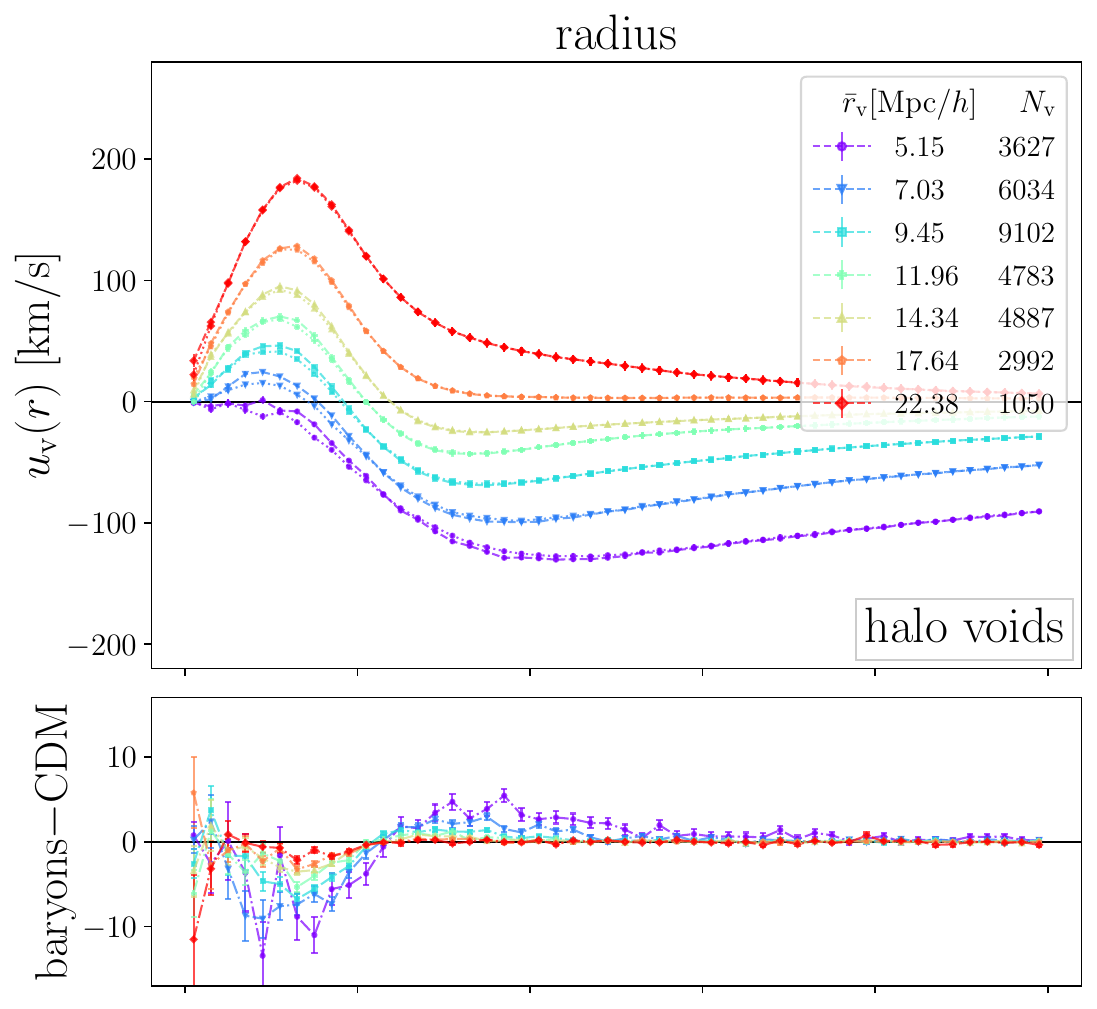}

                               \includegraphics[trim=0 5 0 5, clip]{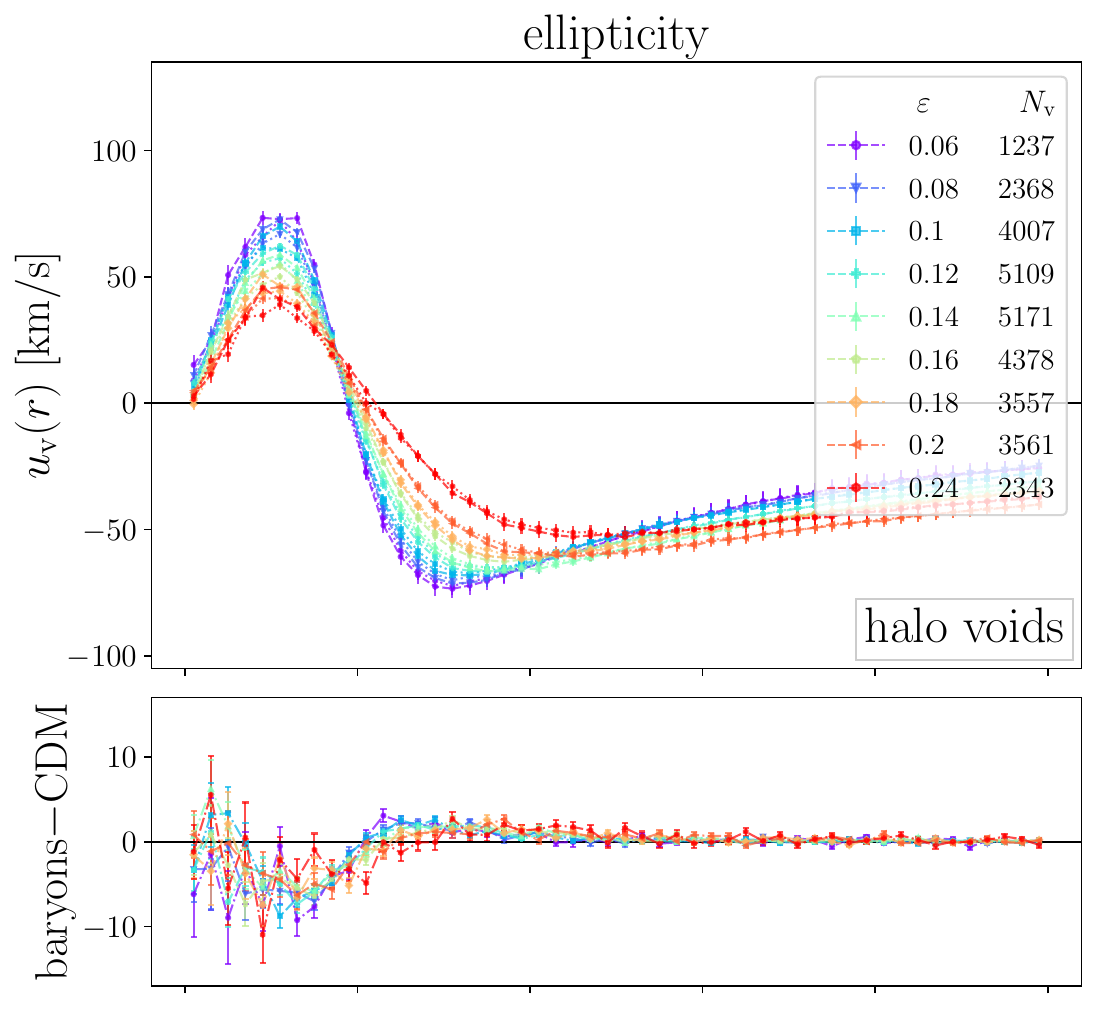}}

               \resizebox{\hsize}{!}{

                               \includegraphics[trim=0 10 0 5, clip]{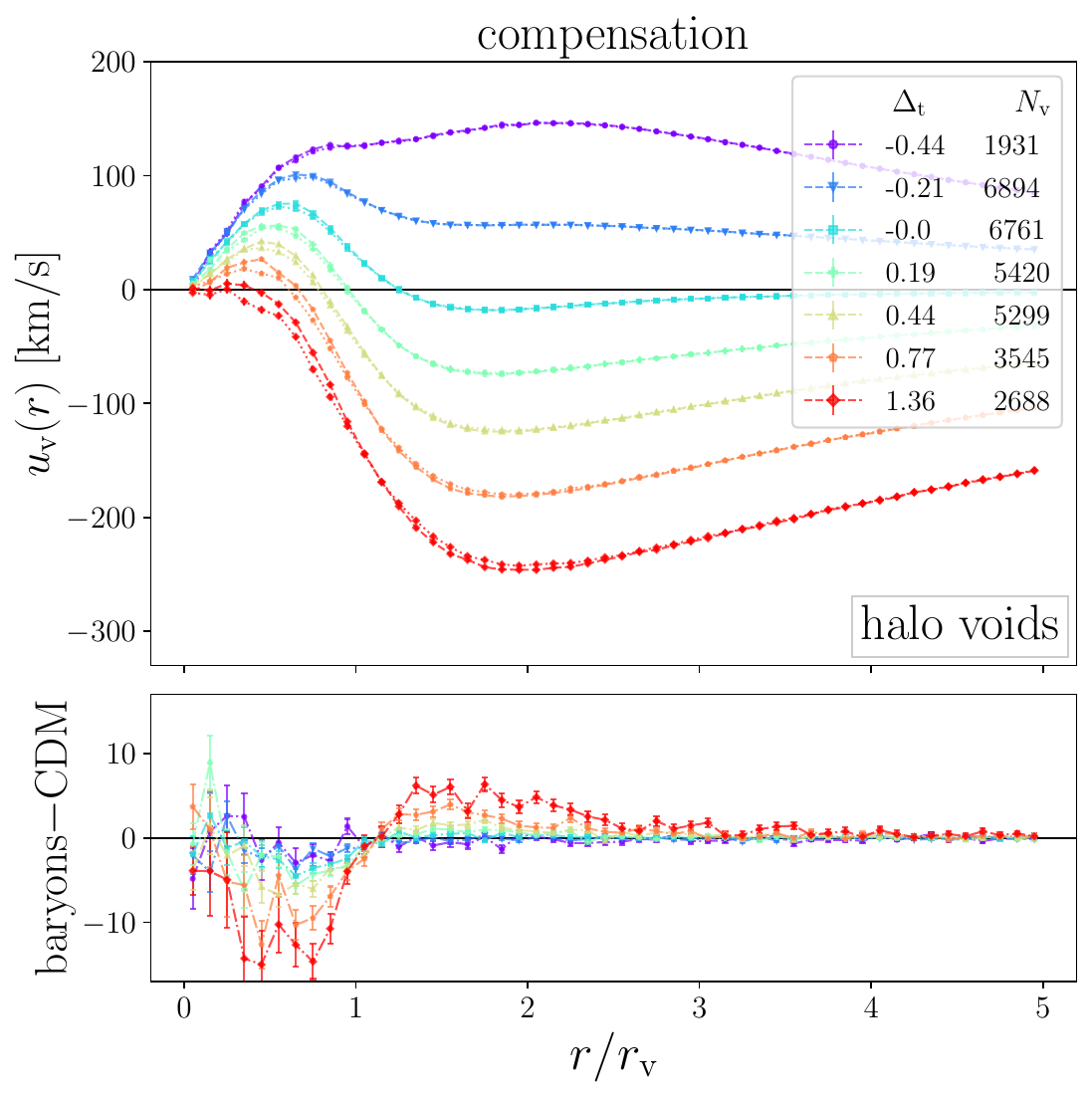}               

                               \includegraphics[trim=0 10 0 5, clip]{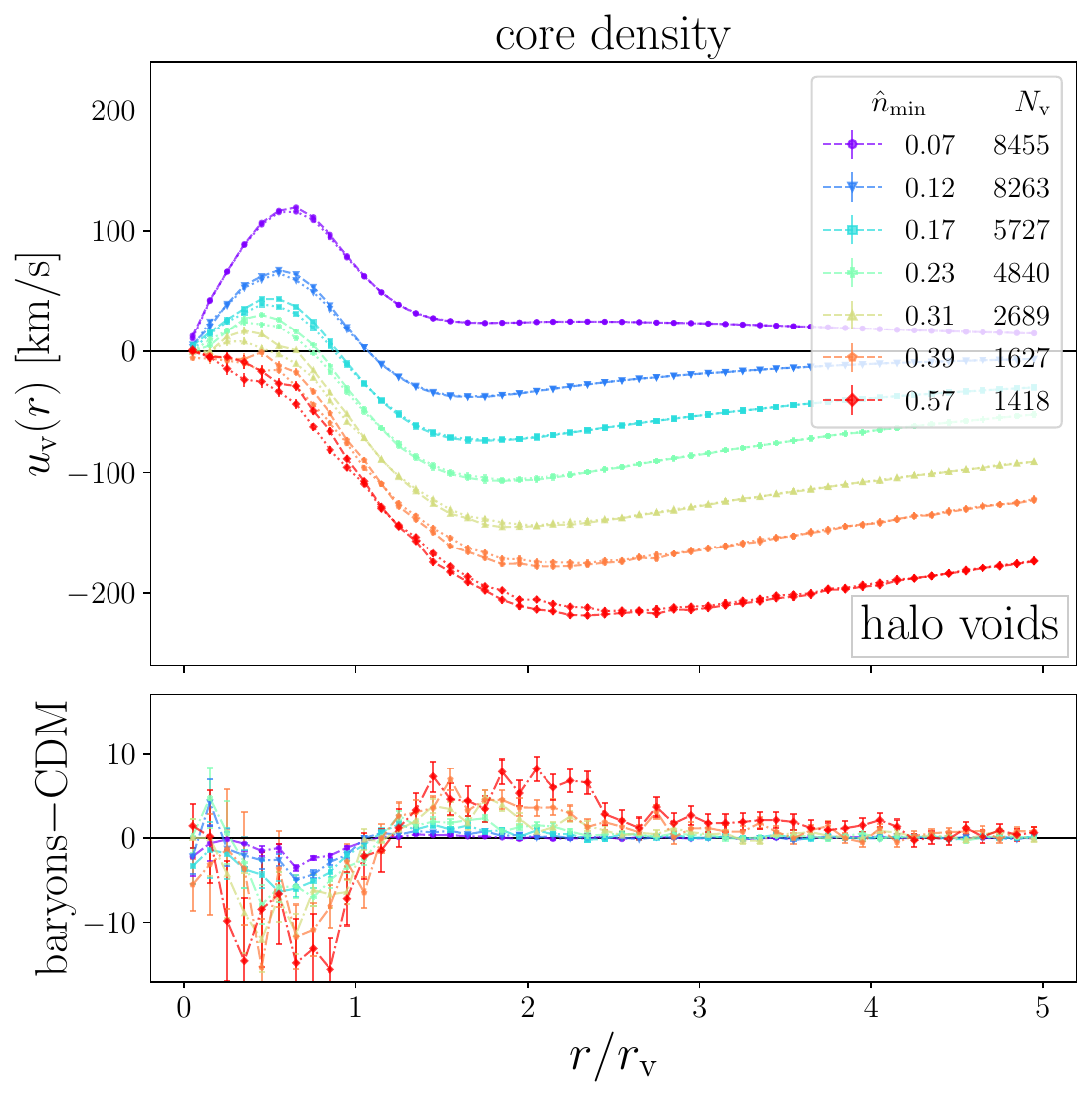}}

               \caption{Same as figure~\ref{fig_density_halo_BARCDM}, but for velocity profiles ($M_\halo \geq 10^{11} \Msun{}$ ).}

               \label{fig_velocity_halo_BARCDM}

\end{figure}

\begin{figure}[t]

               \centering

               \resizebox{\hsize}{!}{

                               \includegraphics[trim=7 5 0 5, clip]{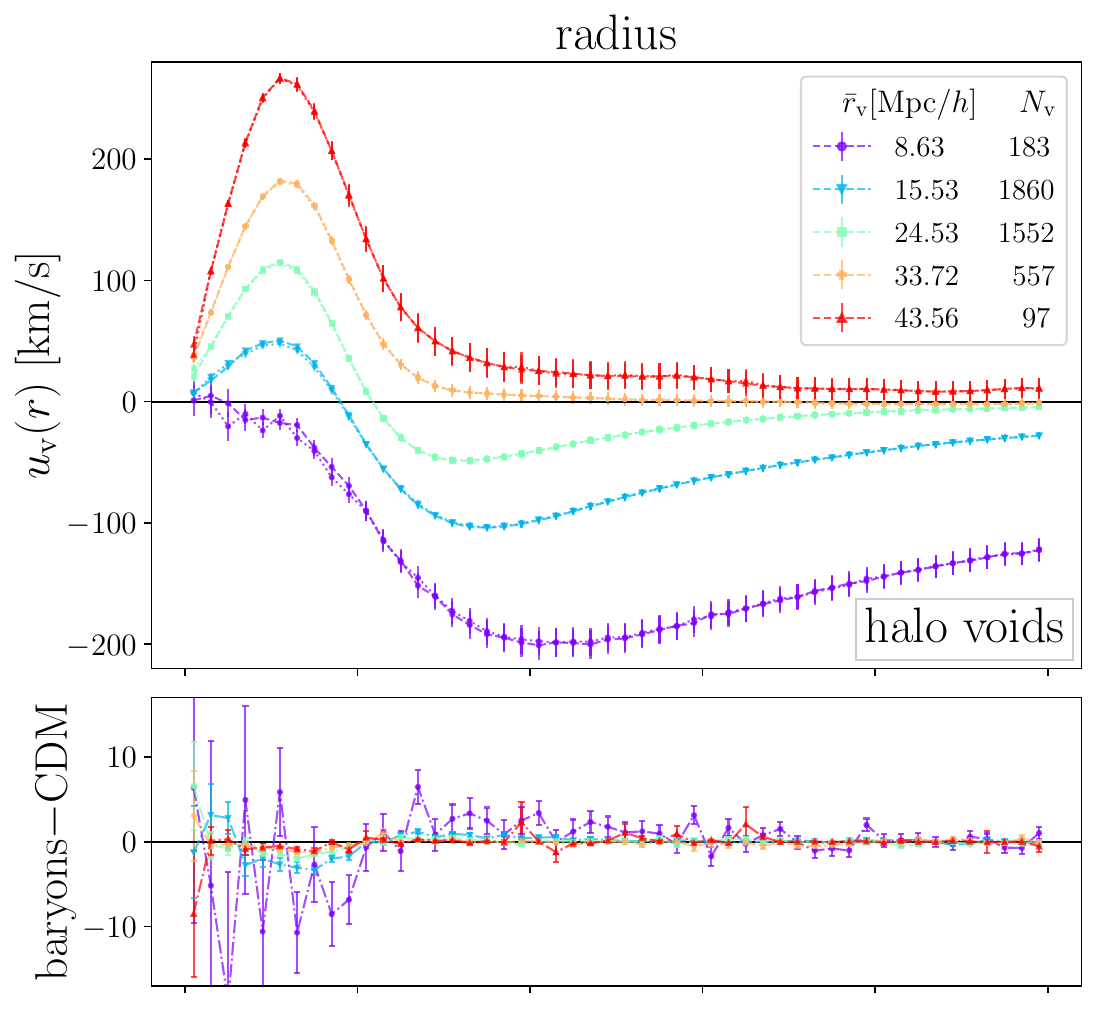}

                               \includegraphics[trim=0 5 0 5, clip]{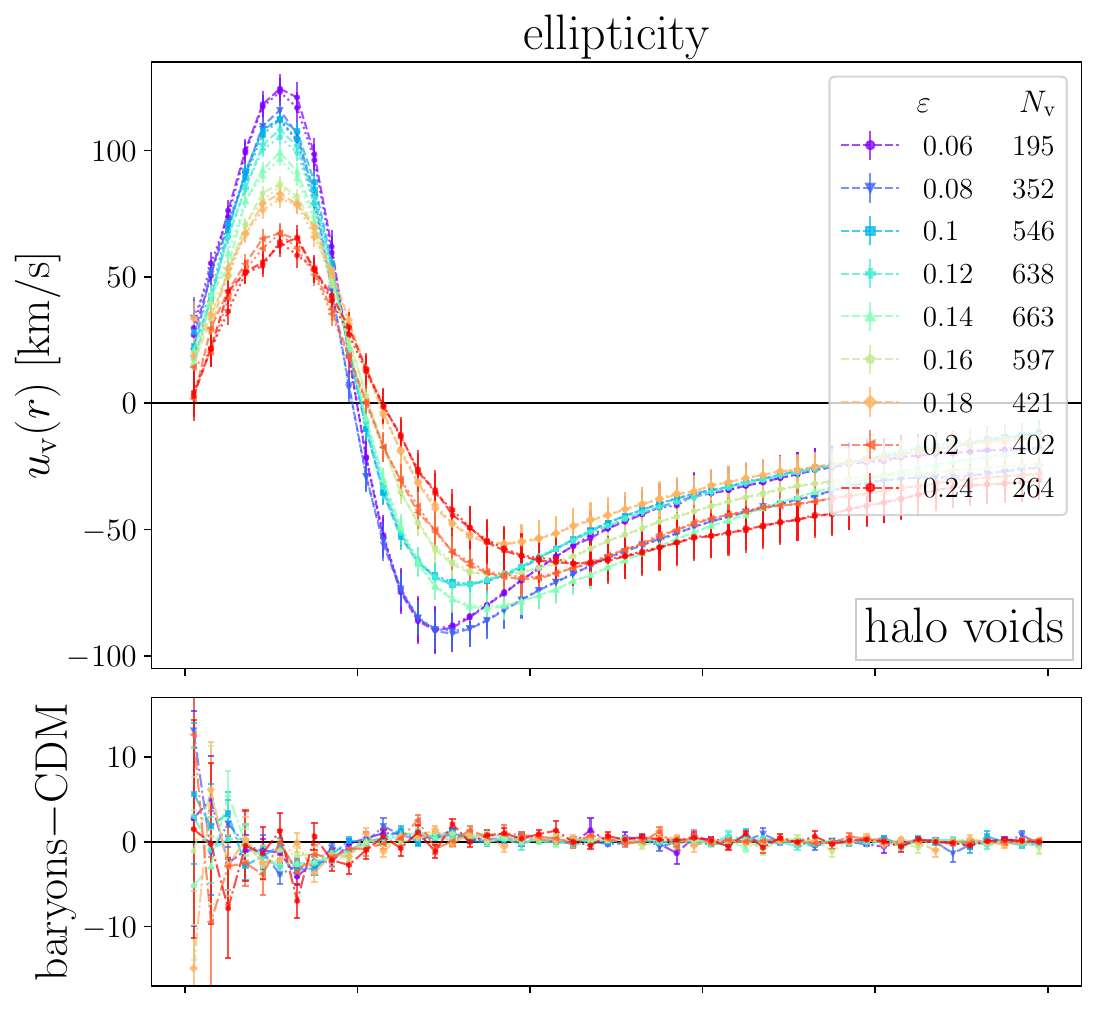}}

               \resizebox{\hsize}{!}{

                               \includegraphics[trim=0 10 0 5, clip]{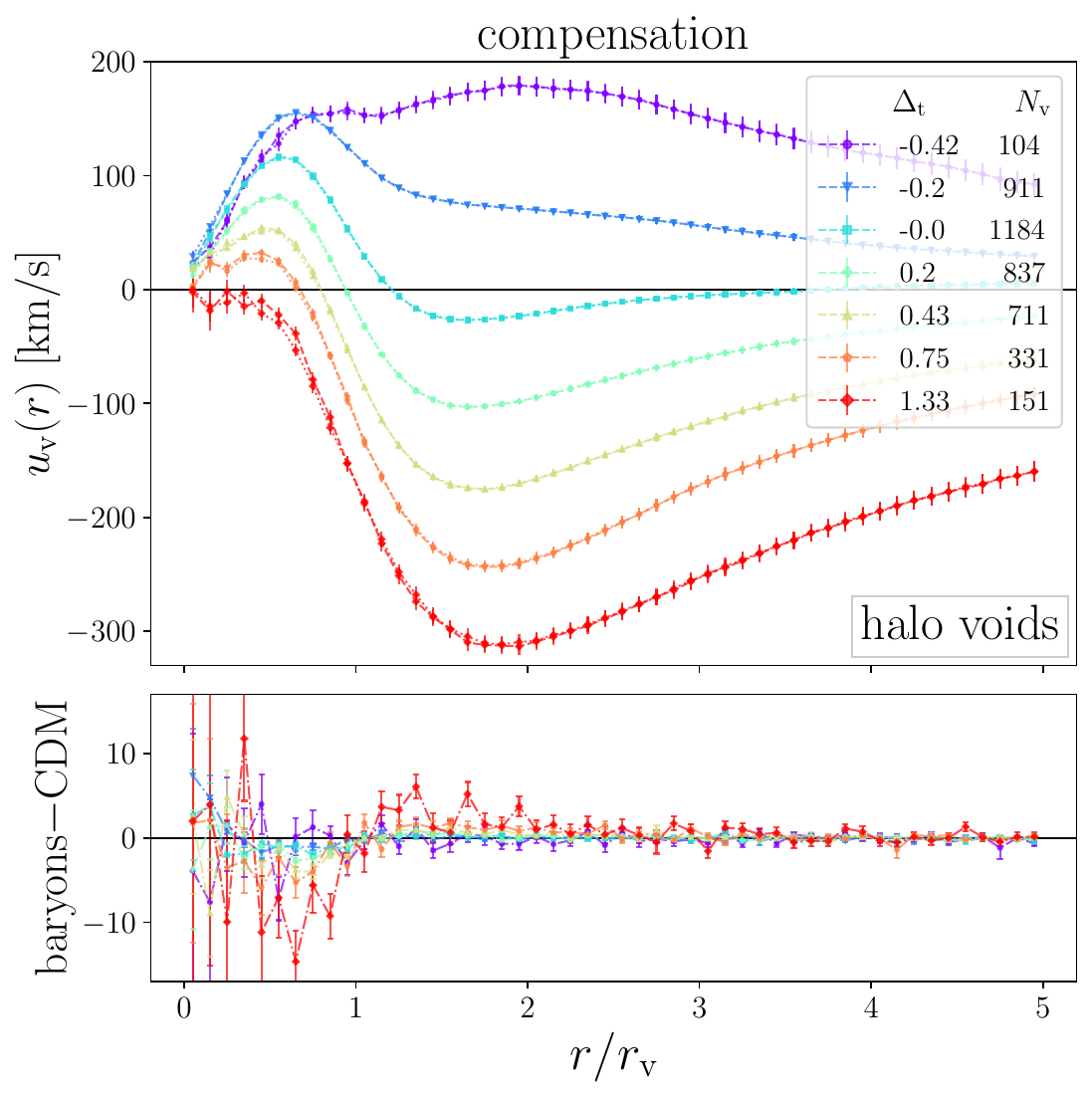}

                               \includegraphics[trim=0 10 0 5, clip]{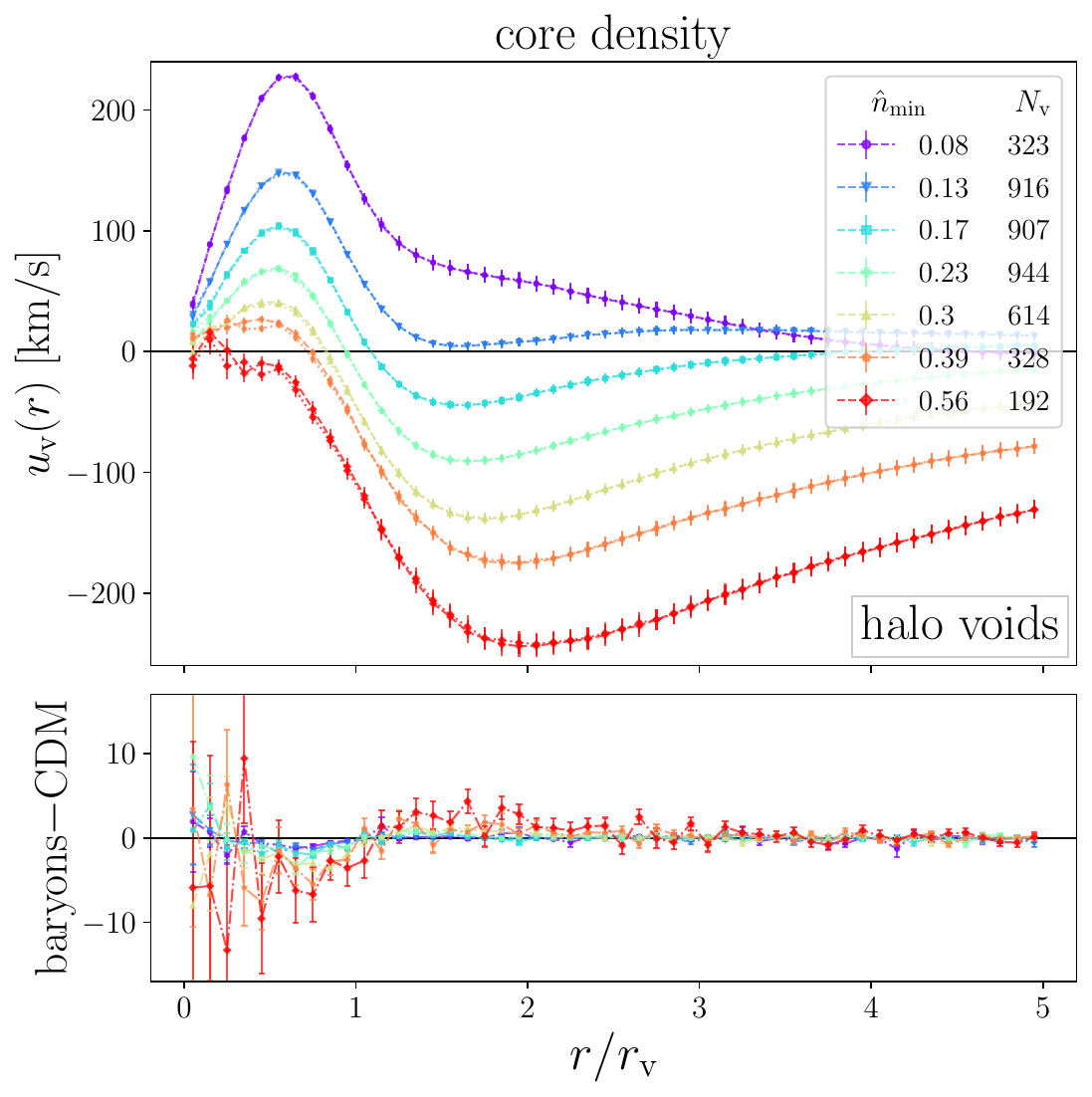}}

               \caption{Same as figure~\ref{fig_velocity_halo_BARCDM}, but for halo voids identified at mass cut $M_\halo \geq 10^{12} \Msun{}$.}

               \label{fig_velocity_halo_BARCDM_Mcut12}

\end{figure}

For our last test of baryonic effects in \HR{} we focus on the distribution, as well as movement of baryons (dotted) and CDM (dashed) around halo voids in figures~\ref{fig_density_halo_BARCDM} to~\ref{fig_velocity_halo_BARCDM_Mcut12}. We use voids from mass cuts in the \hydro{} run, since matched halo densities are of no relevance here due to the lack of a one-to-one correspondence between individual voids from \hydro{} and \DMo{} simulations (see figure~\ref{fig_projected_density_hr}). As previously mentioned, we find no significant effects in the \MR{} simulations, where a halo mass cut of $ M_\halo \geq 10^{12} \Msun{}$ for void finding is used. We now test this mass cut, along with the previously used $10^{11} \Msun{}$, in the \HR{} simulation. This enables us to test whether baryonic effects are only significant in voids identified in lower mass halos, or if the absence of effects in~\MR{} is due to the lower resolution.

Next to stacks in void radius (always top left) and core density (bottom right), we present bins in ellipticity $\varepsilon$ (top right), as well as compensation $\Delta_\tracer$ (bottom left), which allows us to argue why core density is the strongest discriminator of effects from baryonic physics. As before, we use around $50$ million tracers of either CDM or baryons to reduce sampling effects. Contrary to prior sections, we now compute differences between CDM and baryons around individual halo voids and average these `individual differences', depicted in the lower panels. The magnitude of stacked differences is identical to the subtraction of stacked CDM and baryon profiles around halo voids. However, if baryonic physics is relevant even for (most) individual voids, errors in differences can be reduced.

Figures~\ref{fig_density_halo_BARCDM} and~\ref{fig_density_halo_BARCDM_Mcut12} depict the number densities of baryons (dotted) and CDM (dashed) around voids identified in halos of mass $M_\halo \geq 10^{11} \Msun{}$ and $10^{12} \Msun{}$, respectively. Mean densities are always calculated for the respective tracer. Contrary to previous sections and our~\MR{} analysis, we find deviations from zero between baryon and CDM densities around voids of all sizes, and identified in both mass cuts. Nevertheless, their magnitude is highly dependent on the mass cut and scale. Typically baryons are distributed more evenly than CDM, with higher densities inside voids and lower compensation walls~\cite{Paillas2017,Rodriguez2022}. Figure~\ref{fig_abundances_hr} already hinted towards this effect, as baryon voids exhibit higher core densities than CDM voids. Once voids are large enough (bin of $\Bar{r}_\void = 14.34 \, \Mpch$), density differences between CDM and baryon decrease to an almost constant offset $\lvert \Delta \rho \rvert \simeq 0.01$ inside voids in both mass cuts. While small, it shows that to identify baryonic effects around voids in simulations, a high enough resolution is necessary, and particle masses in~\MR{} (table~\ref{table_1}) are not sufficiently small. Section~\ref{subsec:uhr} investigates this further at even higher resolution. While it might seem that the amplitude of profiles in bins of $r_\void$ is simply modulated for baryons, we find that a multiplicative amplitude on the profiles of baryons, including an optional offset, does not accurately reproduce the profiles of CDM.

In bins of ellipticity (top right) we observe an almost identical offset in densities in all bins, both inside and around voids, although smaller for the higher mass cut. Since the ellipticity of voids only slightly depends on their size~\cite{Schuster2023}, this offset can be expected, as effects from the more numerous small voids dominate the stacks. Hence, bins in $\varepsilon$ hint at the average deviation in density at a given resolution in halo mass and are not suited for describing scale-dependent effects.

On the contrary, stacking voids in bins of their compensation (bottom left) results in notably similar deviations to bins in $\coreDens$. Undercompensated voids experience the smallest variations in CDM and baryon densities, which continuously increase with $\Delta_\tracer$. This follows expectations, as the compensation is a measure of a voids' environment, of which its center is part as well. A correlation between overcompensated voids and voids with high $\coreDens$ is evident from their density profiles. Undercompensated voids do not exhibit this strong correlation, as these voids with both small inner densities and a slow increase in density are rather rare compared to the total number of voids with small core densities, many of which are still surrounded by large compensation walls.

As seen in the density profiles of previous chapters, we observe the most substantial effects when binning in core densities. An additional bin with even higher $\coreDens$ values is still possible. We refrain from presenting it, as only around $230$ voids have higher values than those in the bins of figure~\ref{fig_density_halo_BARCDM}, which already depict inner densities close to the mean. Nevertheless, this omitted bin experiences deviations in baryon and CDM densities of around $\lvert \Delta \rho \rvert \simeq 0.5$.

These substantial differences in $\coreDens$ bins reinforce our conclusion that the magnitude of baryonic effects most strongly depend on a voids' core density, even though they appear size-dependent at first glance. A significant reason for the latter is the strong anticorrelation between $\coreDens$ and $r_\void$, which we confirmed in additional tests of binning voids in $r_\void$ with a selection on voids with either high or low core density. When voids are small and have low inner densities, we observe almost no effects, except for the constant offset discussed above, while for small voids with high $\coreDens$, effects increase significantly. However, in the higher mass cut in figure~\ref{fig_density_halo_BARCDM_Mcut12} the void sample is of larger size, and we note that differences in identical $\coreDens$ bins are substantially smaller than in figure~\ref{fig_density_halo_BARCDM}. Moreover, restricting voids in core density bins at $10^{11} \, \Msun$ mass cut to solely voids of radii that are used in the $r_\void$ bins reduces the effects, yet deviations are still larger than in the bin of smallest $r_\void$. These cases suggest a more complex interplay between the inner density of a void and its size that ultimately determines the magnitude of baryonic effects. Furthermore, as deviations remain even in the $ 10^{12} \Msun{}$ mass cut that was originally used in~\MR{}, the resolution dependence of baryonic effects is reaffirmed.

While the magnitude of deviations is similar to those in other void profiles from previous chapters, errors are significantly smaller, which is noticeable in $r_\void$ bins. As these deviations were first calculated around individual voids and only stacked afterwards, this indicates that baryonic physics fundamentally influences individual voids and causes effects that do not merely arise in the statistics of many stacked voids. We explore this in more detail in section~\ref{subsec:uhr}, using individual void profiles

The velocity profiles, depicted in figures~\ref{fig_velocity_halo_BARCDM} ($10^{11} \, \Msun{}$ mass cut ) and~\ref{fig_velocity_halo_BARCDM_Mcut12} ($10^{12} \, \Msun{}$), mirror and support the results from density profiles. Differences in the lower mass cut are non-zero in the inner void regions and outside near the compensation wall in all depicted stacks, independent of void property, with a sign-change slightly outside $r = r_\void$. As deviations are at most on the order of $\lvert \Delta u_\void \rvert \simeq  12 \, \kms$, this suggest that while the shape of velocity profiles depends highly on the selected void property for binning, CDM and baryons always move with closely matching velocities around each individual void.

Furthermore, differences in CDM/baryon densities around halo voids in bins of $r_\void$ are intuitively supported by the corresponding velocity profiles. Inside the smallest voids, baryon velocities are lower than for CDM, i.e. baryons have a stronger tendency to stream towards the inner regions as they are pushed out of compensation walls in high density regions due to pressure and shocks, which yields higher baryon densities inside voids. Past the walls, CDM has higher inflowing velocities, i.e. CDM is strongly attracted to the wall around small voids, while baryons are either pushed outwards or are decelerated when moving inwards, which counters some particles' inward velocity. This yields smaller net velocities of baryons, as well as higher CDM densities at the compensation wall.

Ellipticity stacks once more highlight the averaging of effects over vast ranges in $r_\void$, with a decrease in $\lvert \Delta u_\void \rvert$ in the $10^{12}\, \Msun$ mass cut depicted in figure~\ref{fig_velocity_halo_BARCDM_Mcut12}, while the absolute velocities in this mass cut are higher due to the increase of the average void size. In core density and compensation stacks, voids of lowest $\coreDens$, as well as undercompensated voids experience minor deviations between matter velocities, which increase as the inner and average density of voids increases. This follows naturally from the correlation of overcompensated voids and ones with high $\coreDens$, since both are located in high density environments, which leads to a more complex interplay of matter than in low density regions. This can in turn result in stronger effects due to more frequent baryon interactions compared to the sole interaction of CDM via gravity.

\subsection{Resolution study \label{subsec:uhr}}

Previous sections revealed the magnitude of effects from baryonic physics in void profiles, as well as a complex interplay of a void's core density and its size that determines the strength of these effects. Furthermore, we note that they are highly dependent on the simulation resolution, since effects around voids identified with a halo mass cut of $10^{12} \, \Msun$ were only present in \HR{}, but not \MR{}. Hence, we investigate this further in the \UHR{} simulation, where we identify 281 voids in halos with a mass above $1.6 \times 10^9 \, \Msun$ in the \hydro{} simulation, as well as 348 and 278 halo voids in the \DMo{} run at mass cuts and matched densities, respectively. We refrain from presenting comparisons of halo void profiles between \hydro{} and \DMo{}, as we observe no deviations outside of errors due to extremely sparse statistics.

\begin{figure}[t]

               \centering

               \resizebox{\hsize}{!}{

                               \includegraphics[trim=7 5 0 5, clip]{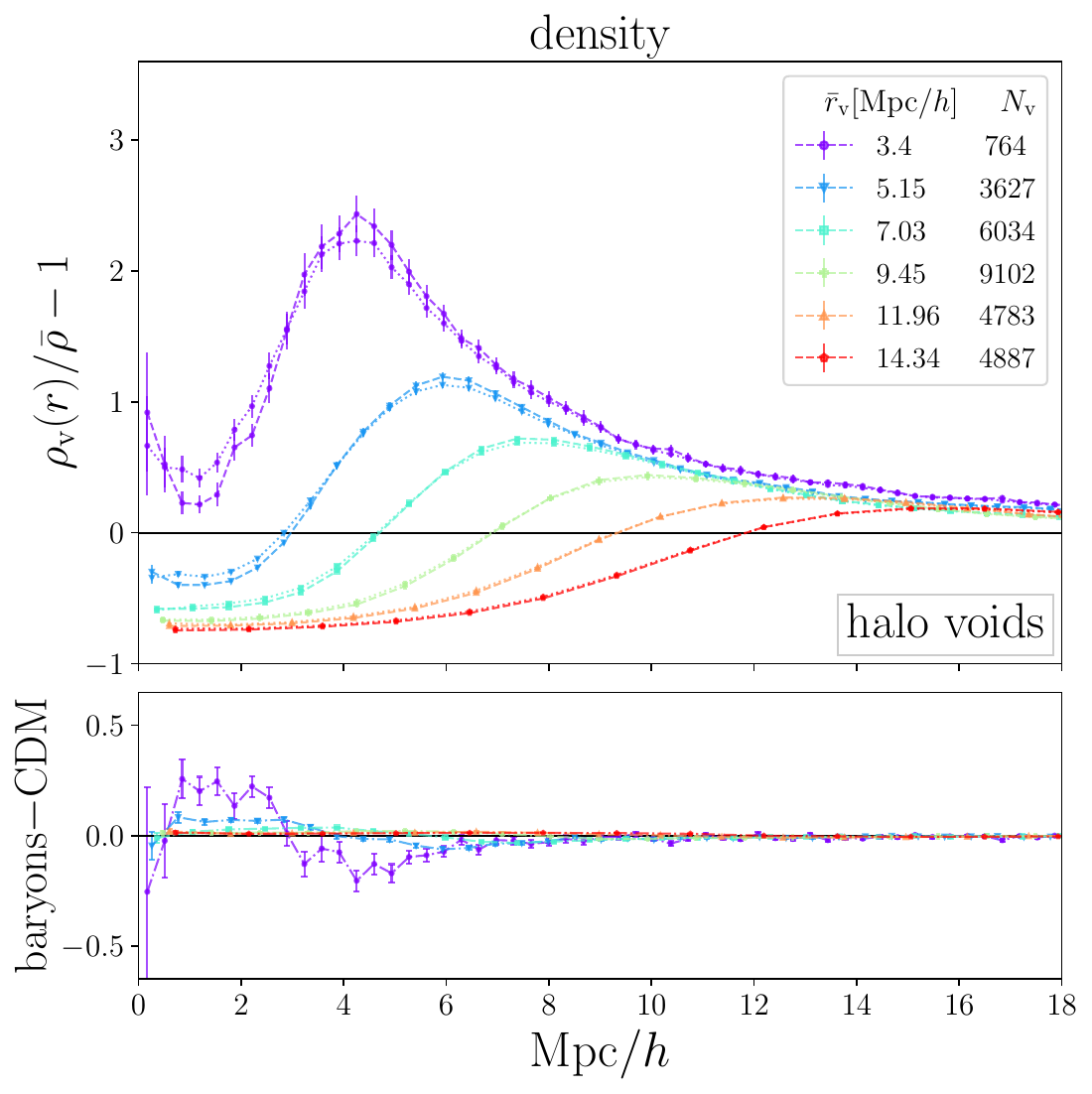}

                               \includegraphics[trim=0 5 0 5, clip]{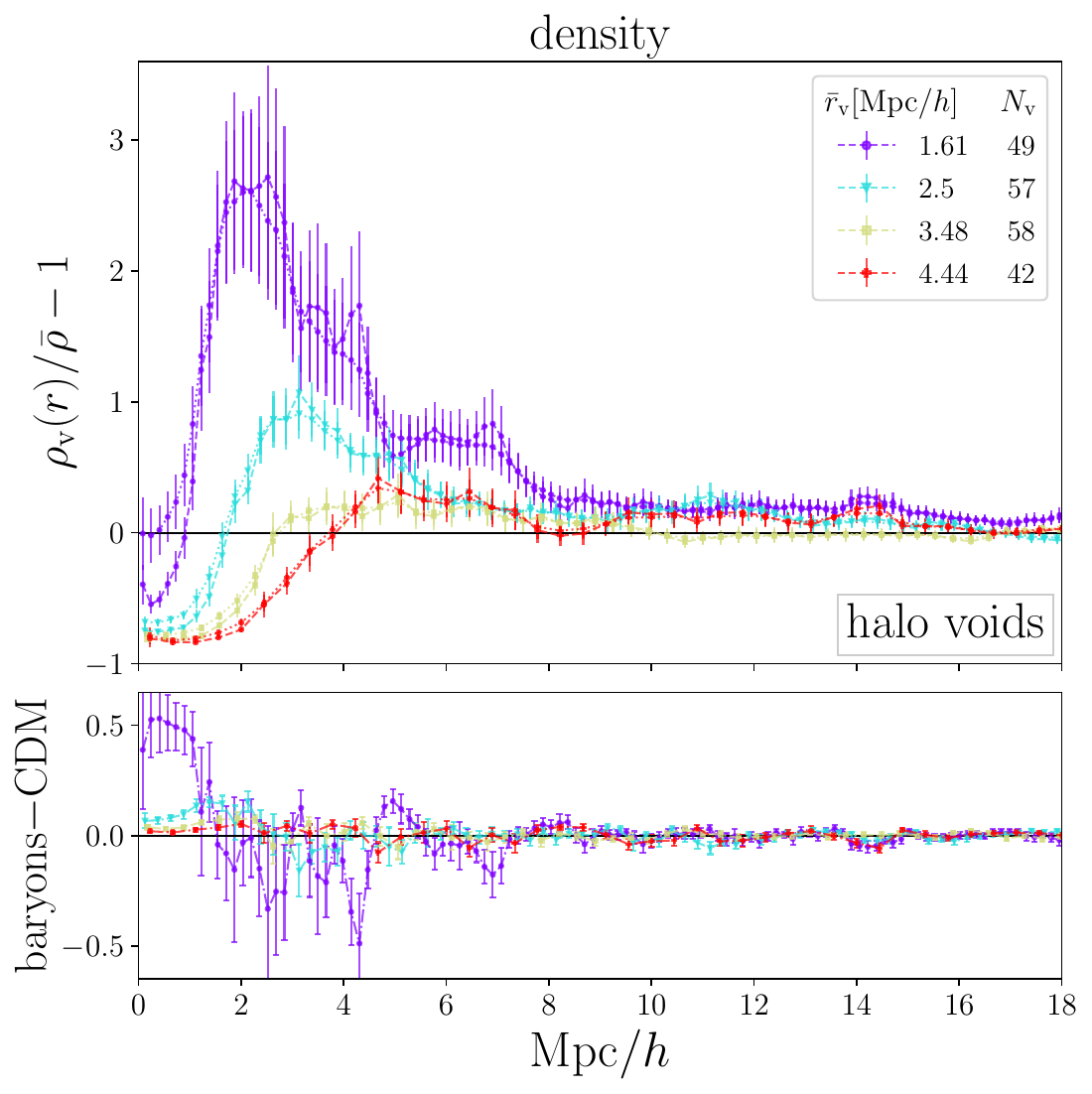}}

               \resizebox{\hsize}{!}{

                               \includegraphics[trim=0 8 0 5, clip]{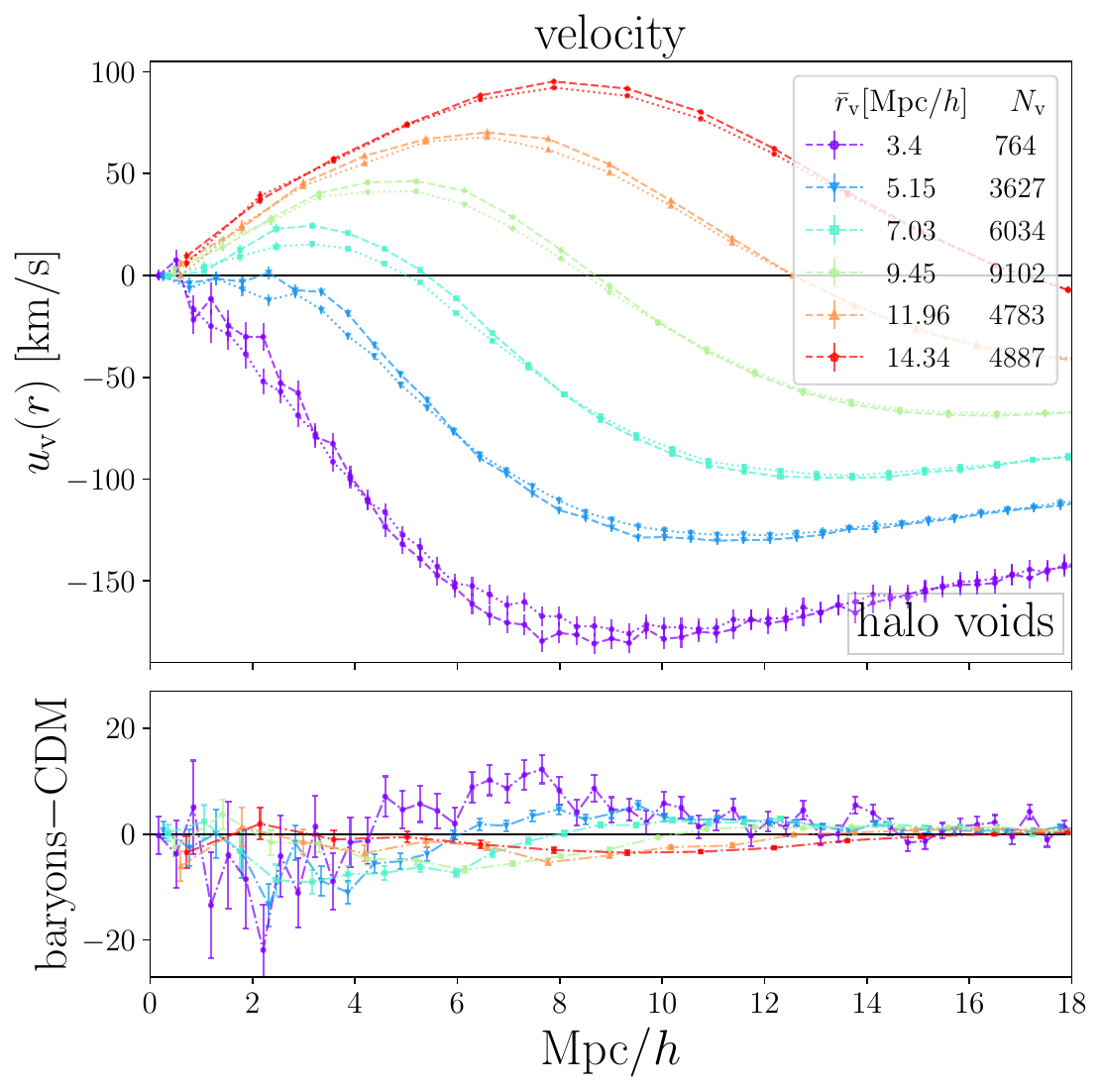}

                               \includegraphics[trim=0 8 0 5, clip]{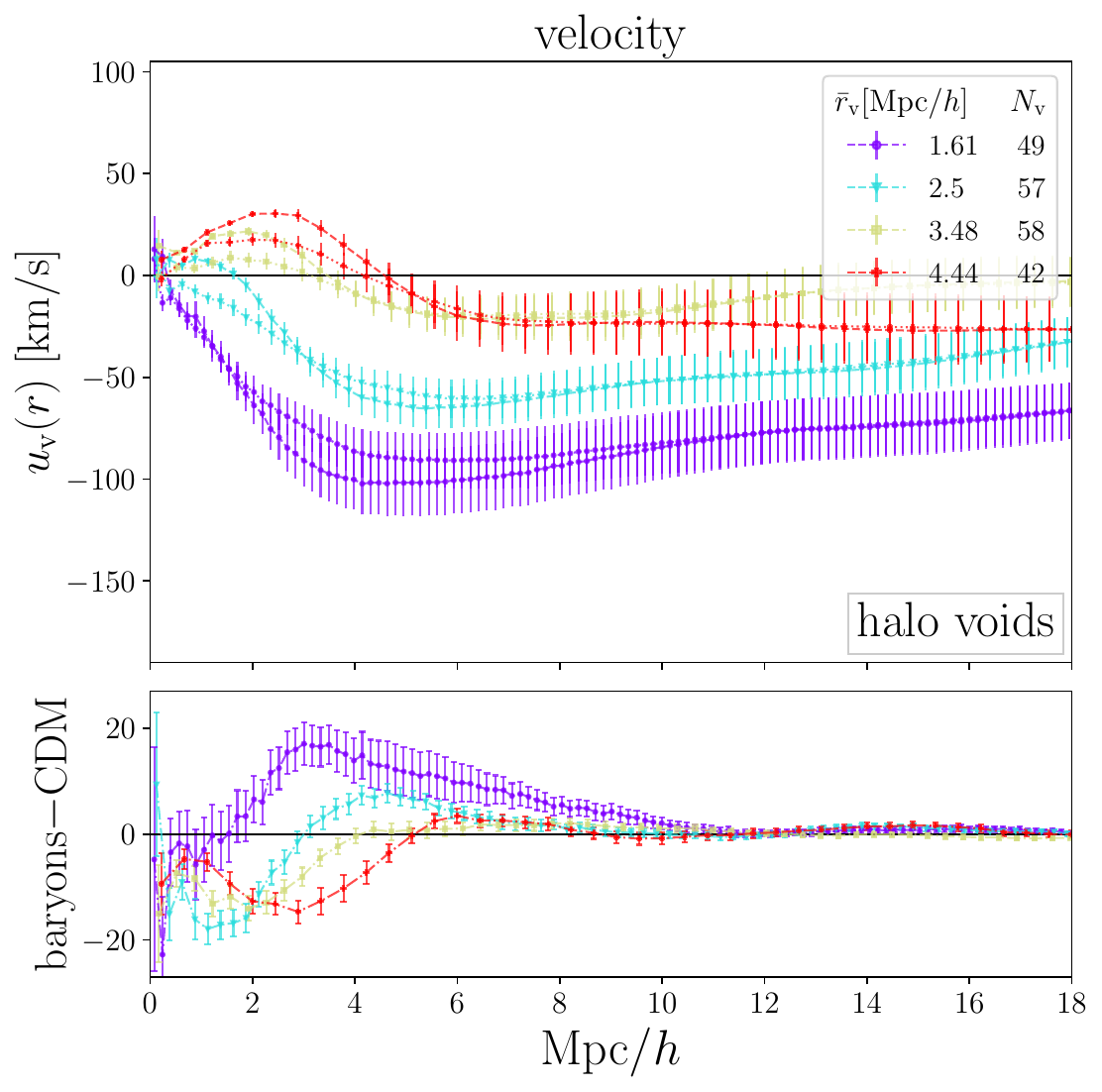}}

               \caption{Stacked density (top) and velocity (bottom) profiles of CDM (dashed) and baryons (dotted) around halo voids in the \HR{} (left) and \UHR{} (right) simulation, in bins of void radius. To showcase scale-dependent effects, profiles are depicted in comoving scale and are not rescaled in units of $r_\void$. Compared to figure~\ref{fig_density_halo_BARCDM}, profiles from the \HR{} simulation contain an additional bin in $r_\void$ of smaller voids, while large voids are omitted.}

               \label{fig_density_velocity_scale}

\end{figure}

Instead, we focus on the CDM and baryon distributions and movements around halo voids, as deviations are most distinct in this scenario. Figure~\ref{fig_density_velocity_scale} presents density (top), as well as velocity (bottom) profiles from \UHR{} (right, with $ M_\halo \geq 1.6 \times 10^9 \, \Msun$ ) and \HR{} (left, with $ M_\halo \geq 10^{11} \, \Msun$ ). In the latter, we add an additional bin in $r_\void$ of even smaller voids than depicted in figure~\ref{fig_density_halo_BARCDM}, while larger voids with $r_\void > 16 \, \Mpch$ are omitted to enhance visibility. To increase the significance of our results from sparse void numbers in \UHR{}, we use subsamples of around $\sim 31 \%$ of all tracers from each matter species for calculating profiles. To probe the scale dependence of effects, we depict profiles in comoving scales, up to $18 \, \Mpch$ from void centers. Profiles are still calculated and stacked in units of $r_\void$, but $x$-axis values are then multiplied with the mean radii $\Bar{r}_\void$ of each stack

The added smaller halo voids in \HR{} (left) have radii near the mean halo separation, and their CDM, as well as baryon densities are above the mean. Usually these voids are considered spurious, but a clear underdensity compared to their higher wall is obvious, defining them as local voids. They feature even stronger deviations between the matter densities, of order $\lvert \Delta \rho \rvert \simeq 0.2$. These represent larger differences than in core density bins, but as mentioned in section~\ref{subsec:baryon_BAR_CDM_voids}, restricting core density bins to voids of radii depicted in figure~\ref{fig_density_halo_BARCDM} reveal larger deviations than in $r_\void$ bins, hinting once more at the interplay of size and core density for baryonic effects. The same holds for the smallest depicted voids in \UHR{} (right), which exhibit the largest differences between baryons and CDM, with $\lvert \Delta \rho \rvert $ around $0.5$, of same magnitude as $\coreDens$ bins (not depicted). Given a CDM number density of $0.5 \, \Bar{\rho}$, this means that within voids of size $1-2 \, \Mpch$ there are almost twice as many baryon tracers as CDM, although in terms of mass, CDM still dominates. In larger voids $\lvert \Delta \rho \rvert $ quickly decreases and reaches the same magnitude as in \HR{}. Higher halo mass cuts for void identification in \UHR{}, namely $10^{10} \, \Msun$ and $10^{11} \, \Msun$ contain fewer and hence larger voids, around which deviations in baryon and CDM densities are once more on the order of \HR{} results. This dictates an upper limit on the deviations in matter densities, irrespective of resolution.

A prominent feature of depicting voids in comoving scales are the outer tails of their compensation walls that align almost perfectly and converge towards the mean density at similar scales from void centers, except for smallest voids, which usually reside in high density environments. Moreover, in \HR{} the deviations in matter densities near the walls align at around $7 \, \Mpch$ from void centers and vanish near $12 \, \Mpch$. However, inside voids they differ significantly. In \UHR{} this shifts towards smaller scales. Some minor deviations outside errors still exist on larger scales, however due to the sparse void sample, we question the significance of these deviations.

The velocity profiles depicted in the lower panels of figure~\ref{fig_density_velocity_scale} confirm results from section~\ref{subsec:baryon_BAR_CDM_voids}. Matter tracers exhibit even more substantial deviations around the smallest \HR{} voids (left), around $\lvert \Delta u_\void \rvert \simeq 15 \, \kms$. Deviations once more vanish at similar scales as in density, just slightly farther outside, near $15 \, \Mpch$ from the void centers. In \UHR{} (right) the scales shift towards slightly smaller scales and differences increase to $\lvert \Delta u_\void \rvert \simeq 18 \, \kms$. While the stacked profiles exhibit significant overlap due to the sparse void sample, deviations between baryon and CDM velocities persist outside errors. This once more confirms that deviations are relevant around individual voids.

\begin{figure}[t]

               \centering

               \resizebox{\hsize}{!}{

                               \includegraphics[trim=0 5 0 5, clip]{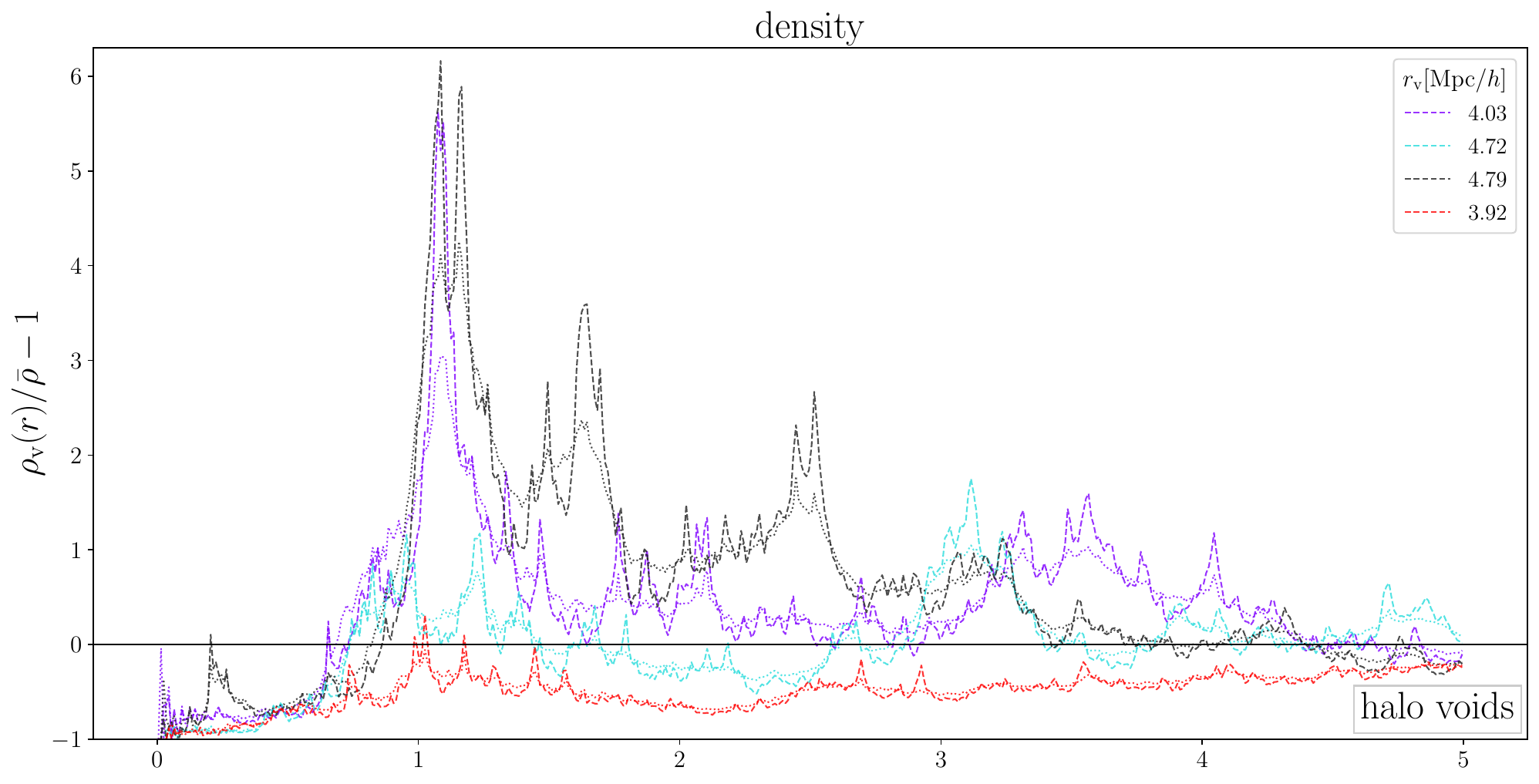}}

               \resizebox{\hsize}{!}{

                               \includegraphics[trim=0 10 0 5, clip]{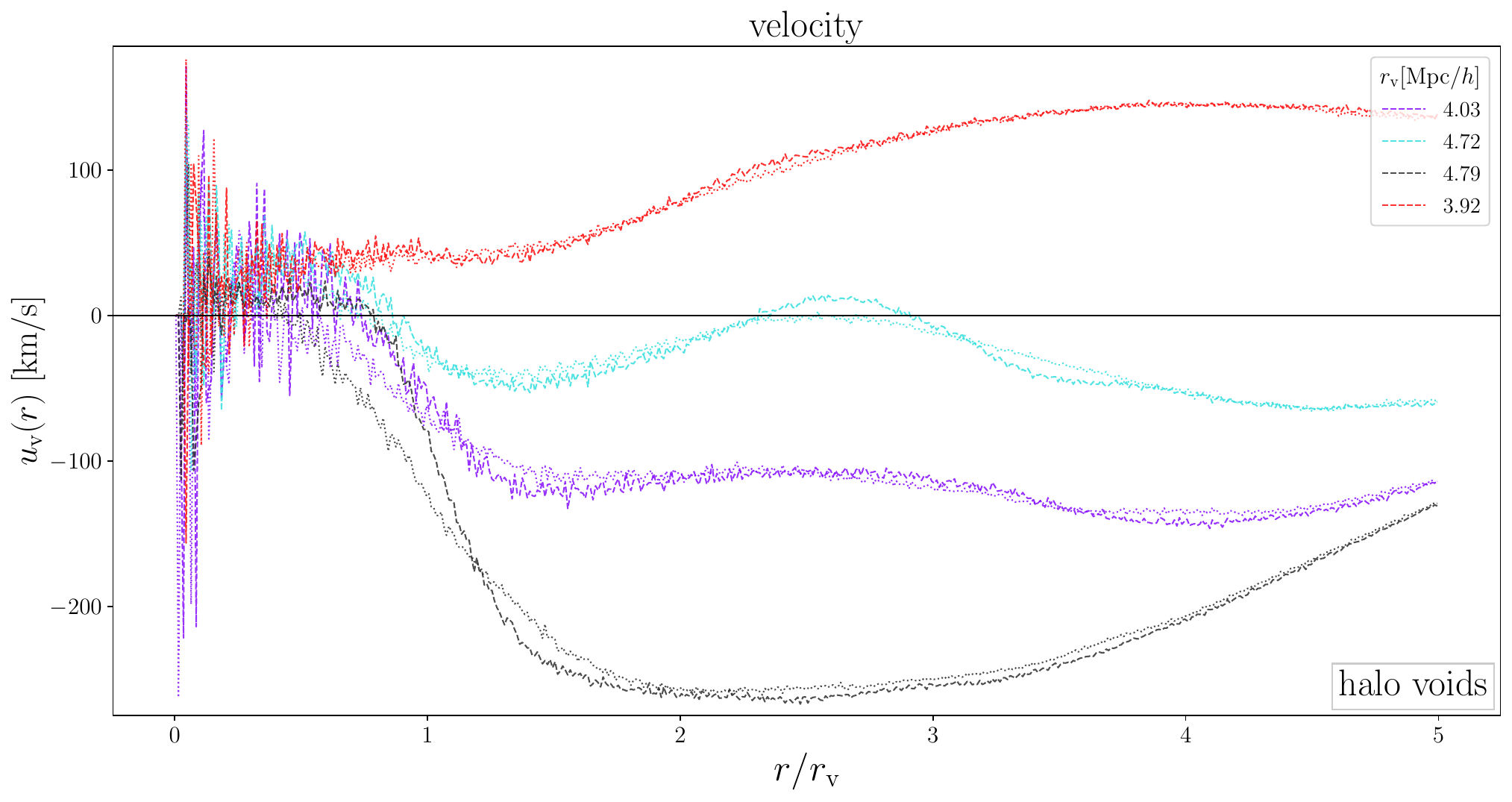}}

               \caption{Density (top) and velocity (bottom) profiles of baryons (dotted) and CDM (dashed) around individual halo voids identified in halos with a mass cut of $1.6 \times 10^9  \Msun{}$ in the \UHR{} simulation. Profiles are calculated in 100 shells per $r_\void$, corresponding to a physical shell width of $47.9 \, \hkpc$ for the largest void depicted. }

               \label{fig_uhr_individual_profiles}

\end{figure}

Before concluding, we want to investigate the structure, as well as dynamics of baryons (dotted) and CDM (dashed) around individual halo voids from the \UHR{} simulation in figure~\ref{fig_uhr_individual_profiles}, once more as functions of $r/ r_\void$. These profiles are calculated in 100 shells per $r_\void$ instead of the usual 10, in order to highlight small-scale differences. For the largest of the depicted voids, with $r_\void = 4.79 \, \Mpch$, this corresponds to a shell width of $47.9 \, \hkpc$, or equivalently $2.22 \times 10^5 $ light years ($h = 0.704$), which is close to the size of our own milky way galaxy. The four voids are nearly random draws from the catalog, with the only condition of significantly different shapes in their profiles at similar void radii.

Each individual density profile in the upper part of figure~\ref{fig_uhr_individual_profiles} exhibits a distinct peak near $r = r_\void$  and in most profiles, the density fluctuates around the mean density further outward. In contrast, the smallest void (red) is clearly undercompensated. Only near its boundary does the density cross slightly above the mean, whilst the void is embedded in an underdense environment on larger scales. 

We find that density peaks are consistently higher in CDM, while baryons are dispersed over the surrounding volume, with smaller peaks at identical locations as CDM. Not only does this hold for overdensities above the mean, but in fact any local overdensity and local underdensity experiences these effects. For global overdensities, e.g. around the two voids with highest walls, we note a large number of local peaks in CDM densities farther outside. Between peaks, CDM densities drop significantly and baryons have smaller peaks than CDM, but experience higher densities in between. Similarly, around global underdensities, e.g. the profile depicted in red, we still observe peaks of higher CDM densities, while baryons are more likely to reside around local minima. This is identical to effects in stacked profiles, where local underdensities (voids) contain a higher number of baryons, whereas CDM clusters heavily in local overdensities, such as the compensation walls. Once more, the magnitude of baryonic effects experiences a high environmental dependence, as seen by the undercompensated void that exhibits the smallest deviations compared to the other depicted voids.

 The clear impact of baryonic physics on the density around individual voids is evident, not merely arising in averaged statistics of voids. We find that CDM is always more clustered due to its lack of self-interaction (apart from gravity), while baryons are distributed more evenly due to the impact of non-gravitational forces. We interpret this as baryonic feedback caused by AGN and supernovae that pushes baryons to the outskirts of clustered regions, i.e. they are ejected from galaxies and halos into their surroundings, whilst CDM is only indirectly affected through the gravitational drag of baryons. On average this leads to higher baryon number densities inside voids and lower compensation walls. While we solely analyse these effects in simulations, an observational example of baryonic physics affecting CDM and baryon densities is found in the `Bullet cluster' (1E 0657-56)~\cite[e.g.,][]{Clowe2006,Paraficz2016}.

Individual velocity profiles on the bottom of figure~\ref{fig_uhr_individual_profiles} exhibit substantial scatter in velocities up to $r = 0.5 \, r_\void$ and only become smoother further outside. This is merely caused by the small shell size, as we find significantly reduced scatter in $0.1 \, r_\void$ wide shells, with velocities of $15-40 \, \kms$ inside these particular voids. Deviations between matter velocities happen on considerably larger scales compared to fluctuations in density and are highest near compensation walls, as previously seen in figure~\ref{fig_density_velocity_scale}. Moreover, the general shape and magnitude of these deviations looks strikingly similar to those of stacked profiles.

The particular void ($r_\void = 4.79 \, \Mpch$) with the highest infall velocities exhibits higher baryon velocities inside and up to around $r = 1.2 \, r_\void$, where baryon and CDM velocities briefly match. Further outside, CDM experiences a higher infall towards the wall. This reaffirms our interpretation that some baryons are either accelerated to regions outside of the compensation wall or are decelerated in their infall, while CDM is solely accelerated towards the wall. The undercompensated void ($r_\void = 3.92 \, \Mpch$) has matching velocities, as matter streams outwards of underdense environments. Furthermore, CDM and baryon velocities have almost insignificant deviations due to the environmental dependence of effects. An additional feature present in small shell sizes is that while velocity profiles look smooth on large scales ($0.1 \, r_\void$ shells), CDM exhibits significant scatter in velocities on small scales and experiences more `chaotic' movement than baryons, as the non-gravitational interactions of baryons smooth their velocity flow more efficiently than only gravity for CDM.

\section{Conclusion\label{sec:conclusion}}

This paper explored the effects from baryonic physics on the statistics of voids across a substantial range in scale and mass by analyzing the \Mag{} suite, a set of hydrodynamical (\hydro{}) and dark-matter-only (\DMo{}) simulations. A comparison between \hydro{} and \DMo{} simulations, as well as between voids identified in CDM and baryons enabled us to reveal the magnitude of baryonic effects around voids. Our main findings can be summarized as follows:

\begin{itemize}
\setlength\itemsep{0.33em}

    \item The magnitude of baryonic effects is highly resolution dependent. Deviations between \DMo{} and \hydro{} simulations in \MR{} exist solely within statistical errors, but we are able to identify differences between \DMo{} and \hydro{}, as well as between baryons and CDM of the \hydro{} run in the \HR{} and \UHR{} simulations, even when using halo mass cuts typically applied in \MR{}. Investigating deviations around individual voids between \DMo{} and \hydro{} runs is not feasible, as there exists no one-to-one correspondence between individual voids (figure~\ref{fig_projected_density_hr})

    \item Within their errors, abundances and profiles of CDM voids are identical between \DMo{} and \hydro{} runs. In contrast to CDM voids, voids identified in baryons exhibit somewhat higher core densities and are slightly smaller due to higher numbers identified in fixed matter subsamplings (figure~\ref{fig_abundances_hr}). In their density profiles, only the smallest subset of baryon voids features smaller compensation walls and slightly higher inner densities, while investigating profiles instead in bins of core density reveals more substantial differences in density, as well as small deviations in the velocity profiles (figures~\ref{fig_density_matter} and~\ref{fig_velocity_matter}).
    
    \item The number densities of halos at fixed mass cut change noticeably between \DMo{} and \hydro{} simulations (table~\ref{table_3}), which leads to differences in void numbers, as well as slight variations in the distributions of characteristic properties. Halo voids in the \HR{} \hydro{} simulation are slightly smaller, while ellipticity and core density distributions experience a small vertical shift due to higher void numbers compared to \DMo{}. After fixing halo number densities for the void identifications, deviations in void property distributions decrease significantly (figure~\ref{fig_abundances_hr}).

    \item The density profiles of small halo voids in the \hydro{} run feature lower inner densities and higher compensation walls compared to halo voids from the \DMo{} run, while large voids exhibit smaller walls. Void profiles of intermediate sized voids remain unchanged. Fixing halo number densities leads to almost vanishing differences. In contrast, binning voids in core density bins reveals more substantial differences, which increase as $\coreDens$ increases, and are of similar magnitude in both fixed halo densities and mass cuts (figure~\ref{fig_density_halos}). Velocity profiles feature only small variations for the largest voids, as well as in $\coreDens$ bins, which once more decrease after fixing halo number densities (figure~\ref{fig_velocity_halos}). Using CDM as tracers for profiles around halo voids results in substantially different density profiles compared to halo tracers. However, differences between \DMo{} and \hydro{} runs are almost identical to differences when halos are used as tracers (figures~\ref{fig_density_halo_matter} and~\ref{fig_velocity_halo_matter}).

    \item We find similar effects to those between baryon and CDM voids when analyzing the structure and dynamics of baryons and CDM around halo voids in the \hydro{} run. We observe a higher abundance of baryons inside voids compared to CDM, while CDM is more abundant around compensation walls. Deviations decrease with increasing void radius, as well as with decreasing core density and compensation. We find that a void's core density is the best indicator for the strength of baryonic effects (figures~\ref{fig_density_halo_BARCDM} and~\ref{fig_density_halo_BARCDM_Mcut12}). While the equivalence principle states that all tracers of matter should move at identical speed, we find slight deviations between CDM and baryon velocities, where the latter typically exhibit smaller absolute velocities. Deviations decrease similar to density differences as voids increase in size (figures~\ref{fig_velocity_halo_BARCDM} and~\ref{fig_velocity_halo_BARCDM_Mcut12}). However, this is not a violation of the equivalence principle, as the dynamics of baryons are not only governed by gravity, but also by pressure and other effects. These can cause an additional acceleration of baryons that expels them out of overdense halos near compensation walls into void interiors. This feedback can explain both the deviations in velocity, as well as higher baryon number densities inside voids, which we observe even for local underdensities around individual voids in \UHR{} down to scales of order $50 \, \hkpc$ (figure~\ref{fig_uhr_individual_profiles}).

    \item Comparing profiles in comoving scales between \UHR{} and \HR{} not only reveals the alignment of compensation walls at identical distances from void centers, but additionally allows us to place an upper limit on differences between baryons and CDM around halo voids and their relevant scales (figure~\ref{fig_density_velocity_scale}).

\end{itemize}

These results have important implications for the use of voids as cosmological probes and their observations. For the next state-of-the-art galaxy surveys such as \emph{Euclid}~\cite{Hamaus2022,Contarini2022,Bonici2023,Radinovic2023}, we expect to attain tracer densities close to those in the \MR{} simulation, as well as the \HR{} simulation with halo mass cut $M_\halo \geq 10^{12} \, \Msun$. We find that for these tracer densities, baryonic effects on void statistics are either negligible within errors (in \MR{}) or minor and only relevant on small scales (figure~\ref{fig_density_halo_BARCDM_Mcut12}). While this work focuses on effects at redshifts $z = 0$, $z = 0.25$ and $z = 0.29$, we expect them to be even less severe at higher redshifts, where structures are less developed. Hence, the use of dark-matter-only simulations with matching tracer density is still valid for comparisons with observations in such surveys. Moreover, even the signatures of massive neutrinos in simulations with a comparable tracer density as \MR{} are more significant than the baryonic effects studied in this paper, most notably in the void core~\cite{Schuster2019}. This suggests voids to be among the best targets for the hunt of neutrinos in cosmology~\cite[see also][]{Kreisch2019,Zhang2020,Bayer2021,Thiele2023}.

Our results pave the way for observing voids in deeper observations with denser tracer samples at low redshift, as planned for Roman~\cite{Spergel2015} and 4MOST~\cite{DeJong2019}. In such cases, our identified baryonic effects would either have to be accounted for, or more sophisticated hydrodynamical simulations are needed in order to use voids as probes of cosmology down to extremely small scales. This opens up the possibility to utilize voids for cosmological inference via redshift-space distortions (RSD) and the Alcock-Paczynski effect~\cite{Hamaus2016,Hamaus2020,Correa2021b,Hamaus2022} beyond current limits. In addition, our observed deviations between baryon and CDM densities in and around voids enable more extensive weak lensing studies on voids and could help in both testing gravity in new regimes and in ultimately determining the nature of dark matter~\cite{Melchior2014,Clampitt2015,SanchezC2017,Davies2018,Fang2019,Davies2021,Jeffrey2021}. Furthermore, as fast radio bursts (FRB) can be used to constrain the baryon density~\cite[e.g.,][]{Macquart2020,Hagstotz2022,Reischke2023}, the average baryonic content inside voids could potentially be probed through the use of FRB and the overabundance of baryons inside voids with respect to CDM might help in finding the missing baryons~\cite{deGraaff2019}.

While we perform our analysis solely for one set of cosmological parameters, we expect our results to be robust even in other cosmologies, especially in those with $\Omega_\matter$ and $\Omega_\mathrm{b}$ values that are relevant in observations, as effects from baryonic physics were rather small in general. Together with their linear dynamics across vast scales~\cite{Schuster2023}, the insignificance of baryonic effects on most cosmological scales shows that voids are exquisite objects for the study of our Universe, as they are governed by much simpler physics compared to other structures of the cosmic web. Upcoming data sets with an unprecedented number of voids will enable cosmologists a considerable improvement in the accurate determination of cosmological models.

\begin{acknowledgments}
We thank Steffen Hagstotz, Luisa Lucie-Smith, Alice Pisani, Ben Wandelt and Rien van de Weygaert for fruitful discussions. This research has been funded by the Deutsche Forschungsgemeinschaft (DFG, German Research Foundation) -- HA 8752/2-1 -- 669764. The authors acknowledge additional support from the Excellence Cluster ORIGINS, which is funded by the DFG under Germany's Excellence Strategy -- EXC-2094 -- 390783311. KD acknowledges funding for the COMPLEX project from the European Research Council (ERC) under the European Union’s Horizon 2020 research and innovation program grant agreement ERC-2019-AdG 882679. The calculations for the hydrodynamical simulations were carried out at the Leibniz Supercomputing Center (LRZ) under the project pr83li. We are especially grateful for the support by M. Petkova through the Computational Center for Particle and Astrophysics (C2PAP) and for the support by N. Hammer at LRZ when carrying out the {\it Box0} simulation within the Extreme Scale-Out Phase on the new SuperMUC Haswell extension system.
\end{acknowledgments}

\bibliography{ms}
\bibliographystyle{JHEP}

\end{document}